\documentclass[pdflatex,sn-mathphys-num,iicol]{sn-jnl}

\usepackage{graphicx}
\usepackage{multirow}
\usepackage{amsmath,amssymb,amsfonts}
\usepackage{amsthm}
\usepackage{mathrsfs}
\usepackage[title]{appendix}
\usepackage{xcolor}
\usepackage{textcomp}
\usepackage{manyfoot}
\usepackage{booktabs}
\usepackage{algorithm}
\usepackage{algorithmicx}
\usepackage{algpseudocode}
\usepackage{listings}
\usepackage{orcidlink}
\usepackage{url}
\usepackage{booktabs}
\usepackage{tabularx}
\usepackage{multirow}
\usepackage{enumitem}
\usepackage{rotating}
\usepackage{color}
\usepackage{soul}
\usepackage{xstring}
\usepackage{float}
\definecolor{cy}{HTML}{FFF2CC}
\definecolor{cb}{HTML}{DAE8FC}
\definecolor{cr}{HTML}{F8CECC}
\definecolor{cp}{HTML}{E1D5E7}
\newcommand{\hlyellow}[1]{{\sethlcolor{cy}\hl{#1}}}
\newcommand{\hlblue}[1]{{\sethlcolor{cb}\hl{#1}}}
\newcommand{\hlred}[1]{{\sethlcolor{cr}\hl{#1}}}
\newcommand{\hlpurple}[1]{{\sethlcolor{cp}\hl{#1}}}
\newcommand{\task}{survey item linking}
\newcommand{\taskcapital}{Survey Item Linking}
\newcommand{\taskabbrev}{SIL}
\newcommand{\taskone}{mention detection}
\newcommand{\taskonecapital}{Mention Detection}
\newcommand{\taskoneabbrev}{MD}
\newcommand{\tasktwo}{entity disambiguation}
\newcommand{\tasktwocapital}{Entity Disambiguation}
\newcommand{\tasktwoabbrev}{ED}

\newcommand{\taskthreecapital}{Sequential Model Pipeline}
\newcommand{\taskthreeabbrev}{SMP}
\newcommand{\data}{SILD}
\newcommand{\datacapital}{Survey Item Linking Dataset}
\newcommand{\datasim}{GSIM}
\newcommand{\datasimcapital}{GESIS Survey Item Metadata}
\newcommand{\datapt}{S44K}
\newcommand{\modelsoscibert}{SoSci-mBERT}
\newcommand{\modelsoscixlmr}{SoSci-XLM-R}

\begin{document}

\title[Enriching Social Science Research via Survey Item Linking]{Enriching Social Science Research via Survey Item Linking}

\author*[]{\fnm{Tornike} \sur{Tsereteli}\orcidlink{0000-0003-4298-3570}}\email{tsereteli@uni-mannheim.de}

\author[]{\fnm{Daniel} \sur{Ruffinelli}\orcidlink{0000-0002-4831-2930}}\email{druffinelli@uni-mannheim.de}

\author[]{\fnm{Simone Paolo} \sur{Ponzetto}\orcidlink{0000-0001-7484-2049}}\email{ponzetto@uni-mannheim.de}

\affil[]{\orgdiv{Data and Web Science Group}, \orgname{University of Mannheim}, \city{Mannheim}, \country{Germany}}

\abstract{Questions within surveys, called \textit{survey items}, are used in the social sciences to study latent concepts, such as the factors influencing life satisfaction.
Instead of using explicit citations, researchers paraphrase the content of the survey items they use in-text.
However, this makes it challenging to find survey items of interest when comparing related work.
Automatically parsing and linking these implicit mentions to survey items in a knowledge base can provide more fine-grained references.
We model this task, called \taskcapital{} (\taskabbrev{}), in two stages: mention detection and entity disambiguation.
Due to an imprecise definition of the task, existing datasets used for evaluating the performance for \taskabbrev{} are too small and of low-quality.
We argue that latent concepts and survey item mentions should be differentiated.
To this end, we create a high-quality and richly annotated dataset consisting of 20,454 English and German sentences.
By benchmarking deep learning systems for each of the two stages independently and sequentially, we demonstrate that the task is feasible, but observe that errors propagate from the first stage, leading to a lower overall task performance.
Moreover, mentions that require the context of multiple sentences are more challenging to identify for models in the first stage.
Modeling the entire context of a document and combining the two stages into an end-to-end system could mitigate these problems in future work, and errors could additionally be reduced by collecting more diverse data and by improving the quality of the knowledge base.}


\maketitle

\section{Introduction}
\label{section:introduction}
\begin{figure*}[!ht]
    \centering
    \includegraphics[width=\textwidth]{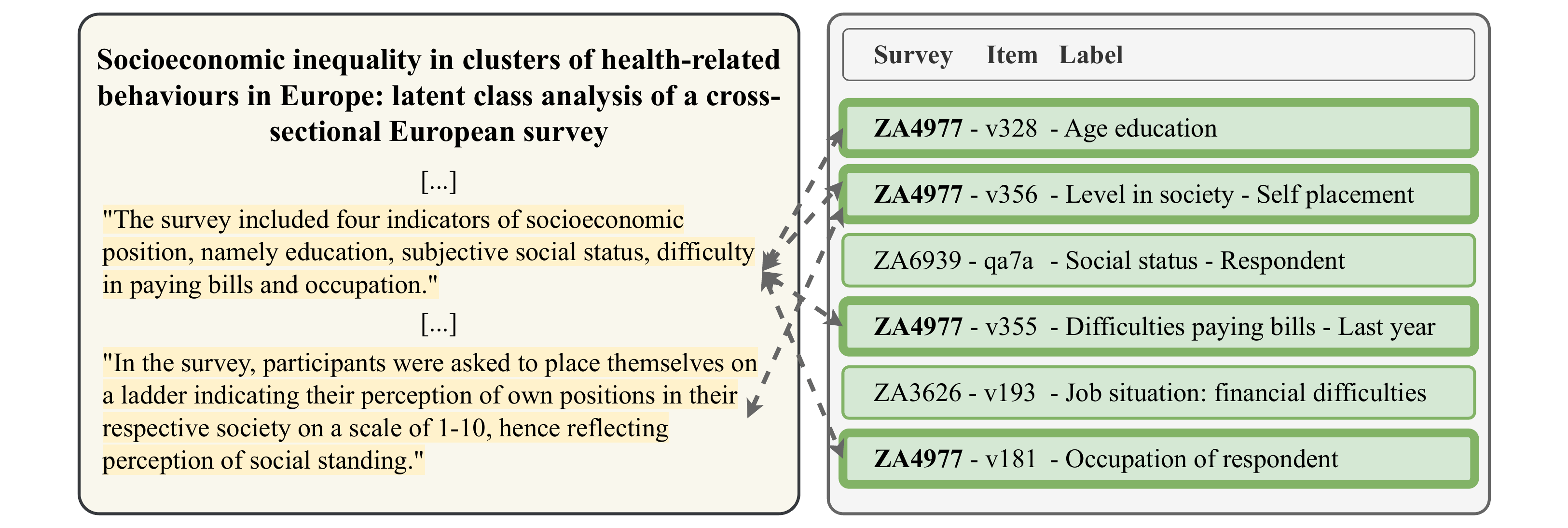}
    \caption{An example research publication (left) with sentences mentioning survey items~\citep{Kino2017} and a list of candidate survey items from different surveys (including the survey ID, the item ID, and the item label, which summarizes its content). The arrows indicate that a sentence mentions a survey item and should be linked. Some survey items from different surveys in the list are very similar to each other (e.g., {\fontfamily{cmtt}\selectfont v356} and {\fontfamily{cmtt}\selectfont qa7a}). These are difficult to disambiguate without document-level context or knowledge about relevant surveys}
    \label{figure:challenge-example}
\end{figure*}

Research in the social sciences has a long history of using surveys to understand societies by collecting information from individuals~\citep{d8fdf947-8b3a-3f20-8183-8309c4684f03,Kuczynski+1986+188+189,groves2011}.
For example, surveys are used to measure the influence of environmental factors on life satisfaction~\citep{Rajani2019}, to investigate the interaction between pro-environmental attitudes and support for environmental taxes~\citep{Davidovic2020}, and to explore the relationship between national identity and democracy~\citep{Gabrielsson2022}.
In these works, latent concepts, which do not appear in surveys explicitly, are first defined and then operationalized using one or more questions from surveys~\citep{6970150}, which are a collection of statements or questions, called \textit{survey items}, and their respective answer categories.
For example, the \textit{Eurobarometer 72.3} survey~\citep{ZA4977} includes over 450 survey items and is grouped into question sets, covering health-related topics.
One of the questions asks respondents to choose an answer from three categories related to the reason for going to a dental check-up recently.\footnote{\url{https://search.gesis.org/variables/exploredata-ZA4977_Varv91}}
~\citet{Kino2017} use this item, among twelve others, to measure health-related behaviors using the wording: ``Attendance for dental check-ups was indicated by use in the past 12 months either on own initiative, doctor's initiative or in a screening programme.''
Similar to this sentence, they paraphrase the other survey items they used in their methods section (as shown in the two yellow-highlighted sentences in Figure~\ref{figure:challenge-example}) and summarize the results for different items in tables.
However, due to a lack of standardization in citing individual survey items~\citep{doi:10.1177/20597991211026616}, no question numbers or unique identifiers are mentioned, which would be useful in automatically identifying the items.
Without such identifiers, finding relevant work based on survey items of interest, which is desired by social scientists~\citep{10.1007/978-3-319-24592-8_15}, is challenging~\citep{10.1162/qss_a_00264}.
The task becomes more difficult when the surveys used cannot easily or automatically be identified from the context of a paper~\citep{10.1007/978-3-642-33290-6_17}.
The green blocks without arrows in Figure~\ref{figure:challenge-example} show items from different surveys that are similar to those referenced in the publication.
Consequently, there is a need for identifying used survey items~\citep{10.1007/978-3-319-24592-8_15} to improve access to research along the FAIR (findable, accessible, interoperable,  re-usable) principles~\citep{Wilkinson2016}.

Previous work tackled this task using Natural Language Processing (NLP) techniques.
~\citet{10.1007/978-3-642-33290-6_17} proposed methods for identifying references to surveys within social science publications.
However, a publication rarely uses all questions of a survey~\citep{6970150}.
As such, more fine-grained linking on the level of survey items is required~\citep{10.1007/978-3-319-24592-8_15}.
Entity Linking (EL) is a method for enriching data by mapping mentions of entities to entries in a structured knowledge base (KB)~\citep{cucerzan-2007-large,HACHEY2013130}.
Similar to EL,~\citet{zielinski-mutschke-2017-mining,zielinski-mutschke-2018-towards} defined the task of detecting and disambiguating survey items by linking a set of survey items to sentences from social science publications mentioning them.
Their approach used machine learning methods based on static word embeddings~\citep{mikolov2013efficient}.
~\citet{10.1007/978-3-031-56069-9_22} extend the work of~\citet{zielinski-mutschke-2017-mining,zielinski-mutschke-2018-towards} by implementing an information system that links publications with survey items.
The underlying linking algorithms were developed during a shared task, which released novel data and evaluated neural approaches for detecting and disambiguating survey items mentioned in publication texts~\citep{tsereteli-etal-2022-overview}.
However, neither the initial work by~\citet{zielinski-mutschke-2017-mining,zielinski-mutschke-2018-towards} nor the shared task~\citep{tsereteli-etal-2022-overview} adequately define what a survey item mention looks like.
When going beyond explicit entities in EL, where instead of matching surface forms of entities, implicit descriptions of entities need to be matched, the decision of what to label is often left to the annotators without clear guidelines~\citep{bast-etal-2023-fair}.
Consequently, the previously proposed datasets have low inter-annotator agreement.
In addition, existing datasets are small because the task definition requires repetitively labeling vague references, resulting in slow annotation times.
These limitations make the existing datasets inadequate for rigorously benchmarking traditional and modern methods on this task.

In this work, we create a high-quality, bilingual benchmark dataset for \taskcapital{} (\taskabbrev{}), containing 20,454 sentences from 100 fully-annotated English and German social science publications, encompassing a diverse set of topics.
To create our dataset, we use a more precise definition of survey item mentions by differentiating between variable-level and question-level mentions~\citep{6970150}. 
In our definition, the former describes a concept whereas the latter introduces a specific survey item.
As a result, we reduce ambiguity during annotations and significantly increase inter-annotator agreement by over 35\% compared to our previously published dataset~\citep{tsereteli-etal-2022-overview}.
Each annotated mention additionally comes with semantic labels, allowing a more fine-grained evaluation for each mention type.
Our new dataset is larger, richer in labels, and more diverse than existing datasets for this task.
In addition, we frame identifying survey items in publications as a two-stage EL task: (i) \taskone{}, formulated as binary text classification (ii) and \tasktwo{}, formulated as retrieval over a KB of existing entities.
This framing is common in EL, where mentions are first detected and then disambiguated~\citep{guo-etal-2013-link,luo-etal-2015-joint,10.1145/2872427.2883061,ganea-hofmann-2017-deep}.
It allows us to evaluate each stage independently, and thus improve targeted aspects of the pipeline in the future.
While previous work has used this framing as well, so far, each of the stages has only been evaluated independently.
Thus, we contribute a more accurate evaluation of the performance on this task.

We conduct extensive experiments where we demonstrate the feasibility of automatically linking survey items to their in-text mentions when given access to a KB.
For each stage, we benchmark existing classical as well as modern techniques, and train specialized models that set a new state-of-the-art on our task.
For \taskone{}, we adapt multilingual transformer-based models~\citep{NIPS2017_3f5ee243} to the social science domain through continued pre-training~\citep{gururangan-etal-2020-dont} and compare against baselines that require few or no labeled data, such as nearest neighbor classification, data augmentation, and in-context learning with Large Language Models (LLMs).
For \tasktwo{}, we fine-tune multilingual sentence transformer models that were specialized on sentence-pair tasks using pseudo-labels and synthetically generated data.
We find that training sentence transformers on synthetic data can significantly improve performance on \tasktwo{}.
By evaluating each stage independently, we observe that a low recall in the first stage can degrade performance in the second stage.
An analysis of the fine-grained, diagnostic elements in our dataset shows that models in stage one struggle with mentions that require the context of multiple sentences, paraphrases, and sentences that mention a single survey item.
For the second stage, matching survey items with a sentence that mentions more than one is more challenging, as the in-text description of the survey items is usually short.
From the user-perspective, we show that providing the ten most-likely survey items for each sentence predicted to contain a mention results in over 50\% recall at the document-level.
Overall, we demonstrate that linking survey items is feasible when citations are provided for used surveys.
Future work could incorporate document-level context and model the two stages jointly. 
In addition, creating diverse synthetic data for long-tail entities and improving the construction of the knowledge base could further reduce errors.
The data and code are available at: \url{https://github.com/e-tornike/SIL}.

\paragraph{Contributions}
\begin{enumerate}
    \item We create a large bilingual benchmark dataset for \taskabbrev{}, containing 20,454 English and German sentences.
    \item We conduct an extensive and robust empirical study of classical and neural methods on two-stage \taskabbrev{}. To the best of our knowledge, our work is the first to evaluate the entire pipeline for \taskabbrev{}, including \taskone{}, which is usually ignored for implicit EL~\citep{Hosseini2024}.
    \item We show that domain-specific data, pseudo-labels, and synthetic data can be used to train state-of-the-art models on each of the stages for this task.
    \item We demonstrate the feasibility of the task by showing that a two-stage system can achieve a recall of over 50\% with our existing models by providing the top ten predicted survey items for relevant sentences.
\end{enumerate}

\paragraph{Outline}
The remainder of this paper is structured as follows.
We provide background information and describe related work in~\S\ref{section:background}.
We define each task for \taskabbrev{} in~\S\ref{section:task-definition}.
\S\ref{section:dataset} describes limitations of existing datasets and provide details on dataset construction.
\S\ref{section:task-one} and~\S\ref{section:task-two} introduce experiments and results for each of the tasks for \taskabbrev{}.
Finally, we present results for the full two-stage pipeline for \taskabbrev{} in~\S\ref{section:two-stage}, conduct a qualitative analysis in~\S\ref{section:qualitative-analysis}, provide concluding remarks in~\S\ref{section:conclusion}, and describe limitations and ethical considerations in~\S\ref{section:limitations}.

\section{Background}
\label{section:background}
This section provides necessary background information on related work by broadly describing relevant NLP applications and models for scientific texts, introducing and comparing the tasks of entity linking with survey item linking, and discussing related tasks that follow a similar two-stage pipeline to entity linking.

\paragraph{NLP for Scientific Texts}
NLP has been widely used in processing scientific publications, with applications ranging from summarizing to detecting generated scientific papers~\citep{sdp-2020-scholarly,sdp-2021-scholarly,sdp-2022-scholarly}.
Unstructured information in scientific texts can be turned into structured data using different methods, such as quantity extraction and citation function classification~\citep{wiesp-2022-information,wiesp-2023-information}.
Most modern NLP approaches rely on informative representations that are created out of raw data~\citep{Liu_2020}.
These representations are commonly learned vectors that are either static~\citep{mikolov2013efficient,pennington-etal-2014-glove,grave-etal-2018-learning} or contextual~\citep{peters-etal-2018-deep}, which can be used as general input features, and combined with learnable parameters, for many downstream tasks.
More recently, Pretrained Language Models (PLMs)~\citep{radford2019language,devlin-etal-2019-bert,delobelle-etal-2020-robbert,he2021deberta} have contributed to significant advances in NLP due to scaling and architectural improvements.
PLMs have been shown to perform better in scientific domains when trained on scientific texts~\citep{beltagy-etal-2019-scibert,cohan-etal-2020-specter}.
The growing number of open-access publications in different scientific disciplines~\citep{Piwowar2018} allows the adaptation of PLMs to more domains, including the social sciences.
Currently, however, the choice of PLMs for social sciences is limited.
In this work, we adapt a multilingual PLM~\citep{conneau-etal-2020-unsupervised} by training it on English and German full texts from social sciences publications (\S\ref{subsec:sosse}).
We show that this leads to a performance improvement on our task.

\paragraph{Entity Linking}
EL is used to map mentions of entities (e.g., companies or locations) to entries in a structured KB~\citep{HACHEY2013130}, such as Wikipedia pages.
It is often framed as a two-stage pipeline, where the first stage detects the location of an entity in a text, and the second disambiguates the entity using a KB~\citep{guo-etal-2013-link,luo-etal-2015-joint,10.1145/2872427.2883061,ganea-hofmann-2017-deep}.
End-to-end methods jointly optimize the two stages~\citep{li-etal-2020-efficient,ayoola-etal-2022-refined}, however, this requires large, labeled datasets.
Furthermore, many EL systems assume that entities are shared between training and test sets, require frequency statistics to estimate entity popularity, and need access to structured data, such as relation tuples or type hierarchies~\citep{logeswaran-etal-2019-zero}.
When these requirements cannot be fulfilled, zero-shot EL is an option, where entities that are not present in the KB are identified by text descriptions, which reduces reliance on such assumptions~\citep{logeswaran-etal-2019-zero}.
In this work, we rely solely on the descriptions of the entities for the second stage of our two-stage pipeline (\S\ref{section:task-two}).
EL approaches mostly focus on explicitly mentioned entities, however, additionally identifying implicit (or informal) entity mentions can help build more comprehensive information extraction tools~\citep{Hosseini2024}.
Implicit entity mentions lack a surface form and are mentioned using the descriptions of the entities~\citep{10.1007/978-3-319-34129-3_8,9560019,Hosseini2024}.
Because the level of specificity of the descriptions can vary, detecting and disambiguating such entities is more challenging for automatic systems as well as for humans.
Only a few datasets exist that include implicit entity mentions~\citep{10.1007/978-3-319-34129-3_8,bast-etal-2023-fair,Hosseini2024}.
Survey item mentions, the focus of this work, are often implicitly paraphrased.
This work creates a novel benchmark dataset for identifying survey item mentions in social science publications (\S\ref{section:dataset}), making a significant contribution towards implicit EL.

Different from two-stage or end-to-end EL, autoregressive EL proposes to generate unique entity names~\citep{cao2021autoregressive} and re-rank predicted entities during inference~\citep{mrini-etal-2022-detection}.
This approach, however, is not suitable for long entities that use general language, as the probability of generating a correct sequence of tokens decreases as the sequence length increases.
In our setting, descriptions can span an entire or even multiple sentences, where one sentence may describe one aspect of an entity (e.g., the question) and another sentence, a different aspect (e.g., the answers).

\paragraph{\taskcapital{}}
Our work is based on that of~\citet{zielinski-mutschke-2017-mining,zielinski-mutschke-2018-towards}, who first introduced the tasks of variable detection and disambiguation.
In our previous work~\citep{tsereteli-etal-2022-overview}, we re-formulated the original task from textual entailment, which requires a cross-encoder that takes as input the two sentences to be compared (a sentence containing a mention and an entity description), to retrieval, which uses a bi-encoder to produce sentence embeddings for each pair in the comparison.
These embeddings can be pre-computed once and stored for efficient search across a large set of entity descriptions.
Even though the learned embeddings in PLMs encode rich information, effectively deriving semantically meaningful sentence representations (e.g., by combining the token embeddings using mean-pooling) from PLMs is challenging, possibly due to the nature of the pre-training objectives (e.g., masked language modeling).
~\citet{reimers-gurevych-2019-sentence} showed that specializing LMs on sentence-pair tasks, such as semantic textual similarity (STS), leads to significant improvements in the embedding quality for retrieval and clustering applications.
In this work, we create two novel sentence-pair datasets and train sentence embeddings that are specialized on survey items, showing improvements for the second stage in our pipeline (\S\ref{subsec:sosse}).

Similar to the process a social scientist goes through, expert annotators have to search through a large KB of survey items to find those referenced in a publication.
Because of the difficulty of the task, the two available datasets for survey item EL are small.
The dataset by~\citet{zielinski-mutschke-2018-towards} contains less than 300 sentences mentioning items from 64 documents, covering a single cumulative survey conducted in Germany.
Our previous dataset contains only 44 documents~\citep{tsereteli-etal-2022-overview}, but nearly 6k sentences.
Half of the sentences are vague co-references of survey items.
The definitions used to identify survey item mentions in the previous datasets are imprecise and flawed (see~\S\ref{subsec:related-datasets} for more details).
To this end, we formulate a stricter definition, provide a more detailed guideline to annotators, and release a larger and more diverse dataset (\S\ref{section:dataset}).

Finally, there are considerable gaps in the evaluation in previous work~\citep{zielinski-mutschke-2018-towards,tsereteli-etal-2022-overview}.
For example, each task is evaluated independently, however, this does not portray a real-world setting, where errors from the first stage can affect the overall performance in the later stage.
Furthermore, previous work has not thoroughly benchmarked different methods and evaluated errors.
In this work, we benchmark supervised and unsupervised methods, evaluate real-world settings, and conduct an extensive ablation.

\paragraph{Related Tasks}
Apart from EL, related tasks also require the detection and disambiguation of mentions of text from a large reference set.
For example, in the task of skill recognition, sentences that mention skills are first classified and then clustered using sentence representations~\citep{escoet:escoe-tr-16}.
Political claim identification deals with classifying sentences containing claims and categorizing them into a predefined set of super- and subcategories~\citep{lapesa-etal-2020-debatenet,zaberer-etal-2023-political}.
Similar to our problem setting, datasets are often mentioned implicitly in scientific fields, such as in the biomedical domain~\citep{https://doi.org/10.1002/asi.24049}.
~\citet{https://doi.org/10.1002/asi.24049} identify dataset mentions and characterize their context as either using, creating, or sharing a dataset.
~\citet{10.1162/tacl_a_00592} use a similar two-stage EL approach, by first detecting dataset mentioning sentences and then linking them to a KB of 7.8K entities using retrieval models.
Finally,~\citet{cadavidsanchez2023evaluating} compare two-stage EL to end-to-end EL for linking non-named entity museum artifacts using descriptions of cultural heritage objects.
In comparison, our task comes with a unique challenge that requires KB filtering.
Different entities in our KB can have the exact same or very similar content (e.g., questions and answers), because surveys often get re-used and modified over the years.
As such, we filter the KB using references to surveys at the document-level.\footnote{We show the impact of filtering the KB and aggregating predictions at the document-level in Appendix~B.7 and B.8 in the supplementary material, respectively.}

\section{\taskcapital{}}\label{section:task-definition}
\begin{figure*}[!ht]
    \centering
    \includegraphics[width=\textwidth]{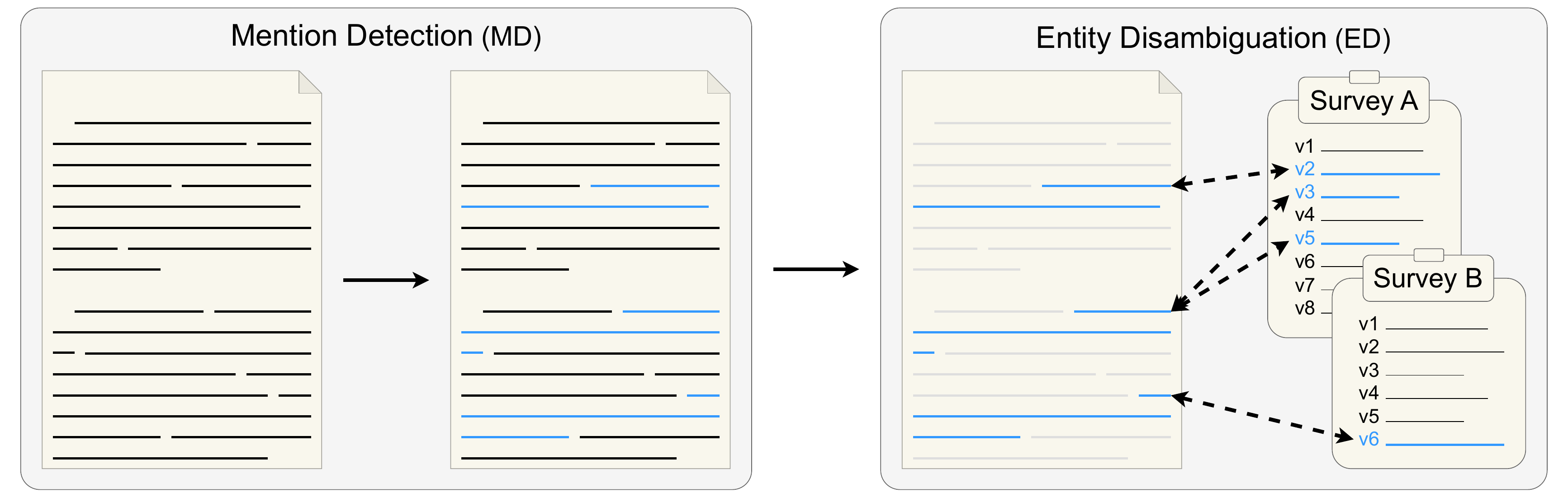}
    \caption{Two-stage pipeline for \task{}. First, the \taskoneabbrev{} stage identifies relevant sentences in a publication (left). The \tasktwoabbrev{} stage then links relevant survey items from surveys to each identified sentence (right)}
    \label{figure:SIL-pipeline}
\end{figure*}

\taskcapital{}~(\taskabbrev{}) is the task of detecting and disambiguating \textit{survey items} that are referenced in scientific texts.
A survey is made up of tens or hundreds of survey items that each contain a question or a statement and answers.
We treat each item as an entity, and the set of all items from all surveys as our KB.
\taskabbrev{} is divided into a standard two-stage pipeline (as shown in Figure~\ref{figure:SIL-pipeline}), where the first stage, \textit{\taskone{}} (MD), identifies the sentences that mention survey items, while the second stage, \textit{\tasktwo{}} (ED), links sentences containing mentions to entities in the KB.
Such a two-stage system has the benefit that each stage can be evaluated independently.
In the sequential model pipeline, we apply MD and ED sequentially.
We describe each stage and relevant metrics below.

\subsection{\taskonecapital{}}
\paragraph{Task Definition}
\taskonecapital{} is a binary text classification task that deals with identifying sentences in a document that mention survey items.
Given a set of documents $D$, the goal is to predict the binary label $l \in \{0,1\}$ for each sentence $s \in D$, where the label 1 marks sentences that mention a survey item, while all others are marked with label 0.

\paragraph{Evaluation Metrics}
Accuracy is a common metric to evaluate classification tasks, however, it is unable to quantify different error types and does not take class distribution into account.
Model errors such as type I (false positive) and type II (false negative) occur when an instance is classified as false when it is true and when an instance is classified as true when it is false, respectively.
Precision ($P$), recall ($R$), and $F$-measure (also called $F_\beta$ or $F_1$, when $\beta$ is set to 1) are more reliable metrics~\citep{info10040150} that account for these factors.
$P$ measures the percentage of correct positive predictions, while $R$ measure the percentage of positive instances that are found.
$F_1$ is computed as the harmonic mean of $P$ and $R$, where $\beta = 1$ gives equal weight to both components.
In binary classification for \taskone{}, the positive class (i.e., that a sentence mentions a survey item) is more relevant.
In the remainder of this work, unless states otherwise, we compute $P$, $R$, and $F_1$ for the positive binary case.

\subsection{\tasktwocapital{}}
\paragraph{Task Definition}
Entity disambiguation is formulated as an information retrieval task, where a given text is compared with a candidate set of survey items using cosine similarity, a vector-based similarity measure.
The goal is to retrieve a candidate set of the top $k$ most similar entities to the text.
These candidates can then be aggregated and refined at the document-level.
Formally, given a set of survey items $I$ and a set of sentences $S$ that mention a survey item, the goal is to predict the subset $I_s \in I$ that are mentioned in each $s \in S$.

\paragraph{Evaluation Metrics}
Before introducing the metrics used for this task, we describe the desiderata of a metric for \tasktwo{}.
First, survey items linked to a sentence are all equally relevant.
As such, a metric should not take the order of correct predictions into account with respect to other correct predictions, but with respect to incorrect ones.
Second, the number of relevant survey items varies per sentence.
Sentences most frequently mention a single survey item.
From the downstream application perspective, when survey items are predicted, a user reading a single sentence would likely be overloaded by having to process too many survey items per sentence.\footnote{This is similar to general search engines, which often provide no more than ten results per page.}
A suitable metric should thus evaluate a ranked set of predicted survey items.

Many metrics fall short of satisfying our desiderata.
For example, precision requires that the number of retrieved items are all relevant, $r$-precision assumes ground-truth knowledge about how many relevant items are necessary for each instance, and mean reciprocal rank (MRR) incorporates the rank of only the first relevant item into the score.
Recall ($R$), (mean) average precision (MAP), and normalized discounted cumulative gain (nDCG)~\citep{10.1145/582415.582418} satisfy our desiderata the best.
Recall is the simplest metric for \tasktwo{}, which ignores ranking position and relevance judgements.
In contrast, MAP takes ranking position into account by taking the mean of the average precision scores at each recall level over multiple instances.
Similar to recall, MAP considers the number of total relevant items, and it assumes that many retrieved item are desired.
It is defined as follows:

\begin{align}
    \text{MAP@}K &= \frac{1}{N}\sum^{n}_{n \in N} \text{AP@}K \\
    &= \frac{1}{N}\sum^{n}_{n \in N} \frac{1}{K}\sum^{k}_{k \in K} P\text{@}K,
\end{align}

\noindent where $K$ is the recall cutoff, $N$ the number of sentences in the evaluation set, and $P$ the precision.
While MAP takes ranking into consideration, it disregards relevance scores, which nDCG accounts for, as it considers both.
Discounted cumulative gain (DCG), the non-normalized formulation, is defined as follows:

\begin{align}
    \text{DCG@}K = \sum^{K}_{i = 1} \frac{r_i}{\log_2(i+1)},
\end{align}

\noindent where $r_i$ is the relevance score of the item at position $i$.
The normalized DCG score is then simply divided by the \textit{ideal} DCG (IDCG), which is the maximum DCG score with a perfect ranking:

\begin{align}
    \text{nDCG@}K = \frac{\text{DCG@}K}{\text{IDCG@}K}.
\end{align}

\noindent Each of the three scores provides a different insight into the evaluation.
Even though each score ranges between 0 and 1, they are not directly comparable.

\subsection{\taskthreecapital{}}
The \taskthreecapital{} (\taskthreeabbrev{}) applies \taskoneabbrev{} and \tasktwoabbrev{} sequentially.
This is different from \tasktwoabbrev{} alone, which we evaluate independently of \taskoneabbrev{}.
This allows us to not only measure the components of the pipeline independently, but also when they are combined.

\section{Dataset Construction}
\label{section:dataset}
This section describes the dataset construction for \taskabbrev{}.
First, we review existing datasets for the task and describe their shortcomings.
We then introduce our dataset by describing the document selection, annotation, and processing steps, and provide detailed dataset statistics. 
Finally, we describe the supplementary entity KB that we construct for the task.

\subsection{Limitations of Existing Datasets}\label{subsec:related-datasets}
~\citet{zielinski-mutschke-2018-towards} were first to create a dataset for \taskabbrev{}.
This dataset was released as trial data in our previous work, along with a novel dataset, SV-Ident~\citep{tsereteli-etal-2022-overview}.
The trial data contain 1,217 labeled sentences from 64 documents, of which 276 sentences mention entities.
In contrast, nearly half of the 5,972 sentences from the 44 documents in SV-Ident mention entities.
A reason for this is the imprecise definition of the entities that should be annotated.
~\citet{zielinski-mutschke-2018-towards} do not provide an exact definition for the term \textit{survey variable}, while~\citet{tsereteli-etal-2022-overview} define it as ``\textit{an item from a survey data set}''.
Neither work explicitly differentiates between concepts and survey items, both of which can be used to refer to a \textit{variable}.
This is a critical aspect that was not considered in the previous definitions, where variable-level mentions refer to the studied phenomena or concepts, while question-level mentions refer to the actual questions that have been asked to and answered by respondents of a survey~\citep{6970150}.
Based on the labeled sentences in the previous datasets, we conclude that~\citet{zielinski-mutschke-2018-towards} use the variable-level definition of an entity, which adds ambiguity.
For example, the sentence ``\textit{We also assess two aspects commonly associated with punitive views: political prioritization of law and order and a measure of racial attitudes specifically relevant for Germany in the form of anti-immigrant attitudes.}'' from the dataset by~\citet{zielinski-mutschke-2018-towards} describes two variable-level concepts, namely \textit{political prioritization of law and order} and \textit{racial attitudes}, however, no question-level survey items are mentioned.
In our previous work~\citep{tsereteli-etal-2022-overview}, we did not explicitly differentiate between variable-level and question-level mentions, resulting in the annotation of both.
In addition to this, both datasets label references to entities in the analyses of a publication.
Such references are problematic because they are repetitive and vague (e.g., ``\textit{The characteristics that are nearly as important as language are the respect for the constitution (91.5 percent) and not being a criminal (91.4 percent).}'').
This sentence references three variable-level entities that could each be operationalized using a single survey item.
Additional context is required in such cases.
This imprecise definition of entities results in the annotation of vague references, which are not informative for benchmarking models on the task.
Furthermore, we conclude that labeling such instances lead to the low inter-annotator agreement in SV-Ident.
In this work, we differentiate between variable-level and question-level entity mentions and demonstrate higher annotator agreement.
In addition, we introduce novel annotation elements that we later show ($\S$\ref{subsec:md-analysis}) provide a fine-grained view into model performance.

\begin{figure*}[!ht]
    \centering
    \includegraphics[width=\textwidth]{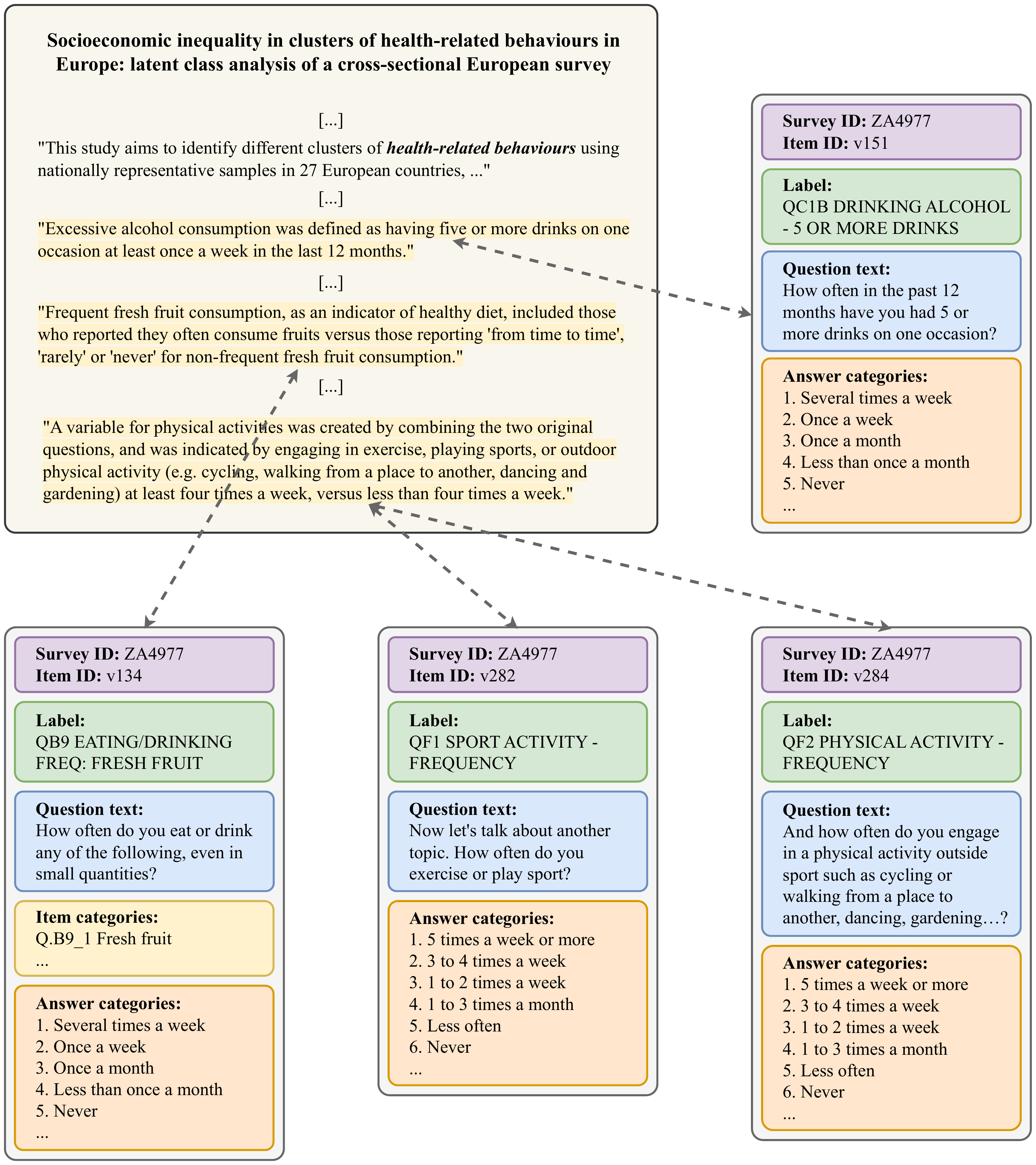}
    \caption{An example of sentences from a research publication~\citep{Kino2017} that studies the concept of \textit{health-related behaviour} using a number of survey items. The arrows indicate that a sentence mentions a survey item (highlighted in \hlyellow{yellow}) and should be linked. Each survey item contains metadata, such as the survey ID, item ID, label (summarizing its content), question (posed to participants), relevant item category from a set of categories (supplementary detail that is used in combination with the question), and answer categories (one of which is used as a response)}
    \label{figure:annotated-publication-example}
\end{figure*}

\subsection{\datacapital{}}\label{subsec:SILD-dataset}
We present the \textbf{S}urvey \textbf{I}tem \textbf{L}inking \textbf{D}ataset~(\data{}), which improves upon previous datasets by more accurately defining the concept of an in-text \textit{entity mention}, resulting in less ambiguous annotations and a higher inter-annotator agreement. 
Figure~\ref{figure:annotated-publication-example} provides an example publication that contains multiple references to survey items along with their corresponding metadata.
In the example, the concept \textit{health-related behaviour} is measured using survey items from the \textit{Eurobarometer 72.3} survey.
\data{} additionally contains coarse- and fine-grained labels, concepts, as well as sentence-to-sentence and concept-to-sentence relations.
In total, \data{} comprises 100 documents, which is 2.3 times larger than SV-Ident, and it contains 2.8 times as many entity mentions as the dataset by~\citet{zielinski-mutschke-2018-towards}.
\data{} is comparable in size to other EL datasets that contain fully-annotated scientific publications, such as SuperMat~\citep{doi:10.1080/27660400.2021.1918396} and GSAP-NER~\citep{otto-etal-2023-gsap}, which contain 142 and 100 documents, respectively. 

\paragraph{Document Selection and Processing}
Before labeling the data, we sampled 100 openly-accessible publications\footnote{The publications originate from the Social Science Open Access Repository (SSOAR) available under: \url{https://www.gesis.org/ssoar}} that cite surveys (of which 84 are in English and the remainder in German).
Links between publications and surveys are available on GESIS Search~\citep{8791137}, some of which are generated manually while others, semi-automatically.
At the time of downloading, there were over 4,000 publications linked to surveys.
Because not all surveys are linked to individual survey items, we only chose documents that were fully linked.
Finally, we extracted full-texts from the PDF files of the publications\footnote{We used GROBID~\citep{GROBID} to extract document content, such as title, sections, and paragraphs.} and manually filtered out documents with many parsing errors.
This resulted in over 500 documents, from which we manually selected 100 publications that mention at least one survey item.
This final set covers a broad range of subdomains, such as economics, politics, and psychology, varies in the methods used, such as empirical, descriptive, and theoretical, and contains documents published between the years 1986 and 2021 (see Appendix~A.2 in the supplementary material for more details).

Scientific literature is mainly published in the PDF file format, which removes all structural information from documents, making machine-processing difficult.
In this work, we use GROBID to extract full-texts from PDF files, but keep only sentences to simplify the annotation process, discarding section and paragraph information.
The sentences corresponding to each document are then converted into XML files, which are loaded directly into the INCEpTION platform~\citep{klie-etal-2018-inception}, where the annotation is carried out.
After annotations, each document is exported, parsed, and processed.
Due to PDF extraction errors during pre-processing, we delete noisy sentences, such as figure captions or tables, by applying a number of rule-based operations (e.g., splitting or merging texts).
We leave parsing captions and tables for future work.
Finally, we manually delete any noisy sentences that were not covered by our heuristics.

\paragraph{Annotation Procedure}
To fix the ambiguity problems in previous datasets (as explained in $\S$\ref{subsec:related-datasets}), we extend the annotation schema by~\citet{tsereteli-etal-2022-overview}.
In EL, entities are usually annotated and identified at the span-level~\citep{10.5555/3504035.3504698,kolitsas-etal-2018-end,wu-etal-2020-scalable}.
However, when entities are long sequences (as is the case in our setting), rather than single words or short phrases, annotating at the span-level is challenging, because entity boundaries are often ambiguous.
Thus, our work operates at the sentence-level, where a single sentence may contain multiple entities.
We explicitly define the term \textit{survey item mention} (with one exception described later) to mean any sentence that unambiguously \textit{defines} or states that a specific survey item is used to measure some latent phenomenon within the context of the publication.
This definition specifically targets question-level mentions, which are most informative for the purpose of our task, because they usually contain the largest overlap in terminology with survey items.
Using this definition, the most informative sentences that mention at least one survey item are annotated.
Example sentences are provided in Table~\ref{table:survey-item-mention-examples} (further examples are in Appendix~A.1 in the supplementary material).
Restricting the design to label only the most informative sentences in a document during annotation leads to less subjective labels, makes vague candidates less probable, and increases annotator agreement as a result.
Commonly, a survey item is associated with a single sentence in a document, however, in case of repetition or rephrasing, a survey item may be referenced in multiple informative sentences.
Sentences that may refer back to a previously defined survey item (e.g., in the analysis of the results) are not labeled, because such mentions are commonly vague, difficult to disambiguate, and unnecessary for the intended use-case of this work.
Such reference sentences are often less informative -- in many cases even vague -- limiting their usefulness, as models are unable to effectively learn to differentiate them from other sentences.
Given that survey items are typically defined once and later referenced with less detail within a document, it is sufficient for our downstream application to accurately link survey items to entire documents for which only a single, informative mention is necessary for disambiguation.
From an end-user perspective, given that we provide the source sentence that resulted in a survey item being linked to a publication, social scientists prefer to find sentences defining used survey items quickly rather than to read many vague references.

\begin{table}
    \centering
    \footnotesize
    \caption{Example sentences from publications that mention survey items. The associated \textit{type} and \textit{subtype} terms indicate the semantic and lexical classes the entities are labeled with. In contrast to explicit types, implicit types require additional context. A verbatim \textit{quotation} includes exact wording of a survey item, while a \textit{paraphrase} summarizes the content}
    \label{table:survey-item-mention-examples}
    \begin{tabularx}{\columnwidth}{l|X}
         \toprule
         \textbf{Sentence} & \textit{The wording of the questions used for the index was: 1. A working mother can establish just as warm and secure a relationship with her children as a mother who does not work} \\
         \textbf{Items} & {\fontfamily{cmtt}\selectfont \href{https://search.gesis.org/variables/exploredata-ZA7503_VarD056}{ZA7503\_VarD056}} \\
         \textbf{Type} & Explicit \\
         \textbf{Subtype} & Quotation \\
         \textbf{Doc. ID} & \href{https://www.ssoar.info/ssoar/handle/document/79550}{79550}~\citep{Fučík2020} \\         
         \midrule
         \textbf{Sentence} & \textit{While the first question captures attitudes to taxes explicitly, the second does so implicitly since taxes usually result in higher prices or fees.} \\
         \textbf{Items} & {\fontfamily{cmtt}\selectfont \href{https://search.gesis.org/variables/exploredata-ZA5500_Varv29}{ZA5500\_Varv29}}, {\fontfamily{cmtt}\selectfont \href{https://search.gesis.org/variables/exploredata-ZA5500_Varv30}{ZA5500\_Varv30}} \\
         \textbf{Type} & Implicit \\
         \textbf{Subtype} & Paraphrase \\
         \textbf{Doc. ID} & \href{https://www.ssoar.info/ssoar/handle/document/74431}{74431}~\citep{Davidovic2020} \\
         \botrule
    \end{tabularx}
\end{table}

Our annotation schema introduces additional detailed labels, referred to as \textit{diagnostic elements} or simply \textit{elements}.
These elements are connected to the sentences mentioning survey items and to the actual survey items.
They play an important role in conducting fine-grained evaluations of models and allow a more reliable measure of performance improvements (see the analysis in $\S$\ref{subsec:md-analysis}).
These elements (i) establish connections between concepts and sentences defining survey items, (ii) when necessary, link sentences together, and (iii) categorize sentences into fine-grained semantic and linguistic classes.
We define each element as follows:

\begin{enumerate}
    \item \textbf{\textit{Concepts}} are the abstract ideas that are \textit{operationalized} (i.e., defined) using survey items.\footnote{Note that \textit{explicit} and \textit{implicit} types are different from explicit and implicit entity mentions.} These are variable-level mentions.
    \item \textbf{\textit{Types}} are semantic classes for categorizing survey item mentions.
    These include:
    \begin{enumerate}[label=(\alph*)]
        \item \textit{Explicit} - contained in a single sentence.
        \item \textit{Implicit} - requires additional context (e.g., partial reference to an explicit type or citation to a different publication).
        \item \textit{Mixed} - contains more than one survey item of a different type.
        \item \textit{Other} - it does not fall into any of the above categories.
    \end{enumerate}
    \item \textbf{\textit{Subtypes}} are linguistic classes for categorizing survey item mentions. For simplicity, we only allow mapping a single subtype to each mention in a sentence. We define the following classes:
    \begin{enumerate}[label=(\alph*)]
        \item \textit{Quotation} - a direct quotation of (parts of) the metadata of a survey item. We consider this an explicit entity mention.
        \item \textit{Paraphrase} - a paraphrase of (parts of) the metadata of a survey item.
        \item \textit{Citation} - a citation of an unknown survey item from a different publication. This is the exception to our definition.
        \item \textit{Lexical inference} - a mention that requires world knowledge.
        \item \textit{Unspecified} - a mention that is not specified enough to disambiguate it from the context of the document. In this case, the \textit{Unk} (stands for \textit{unknown}) identifier is used for the survey item.
        \item \textit{Other} - mentions that do not fall into any of the above categories.
    \end{enumerate}
    \item \textbf{\textit{Relations}} are the semantic links between different elements, such as between explicit and implicit types or between concepts and sentences containing mentions.
    In the former case, an implicit sentence may require the context of an explicit one. In the latter case, a concept may be operationalized using one or more survey item mentions.
\end{enumerate}

\begin{figure*}[!ht]
    \centering
    \includegraphics[width=\textwidth]{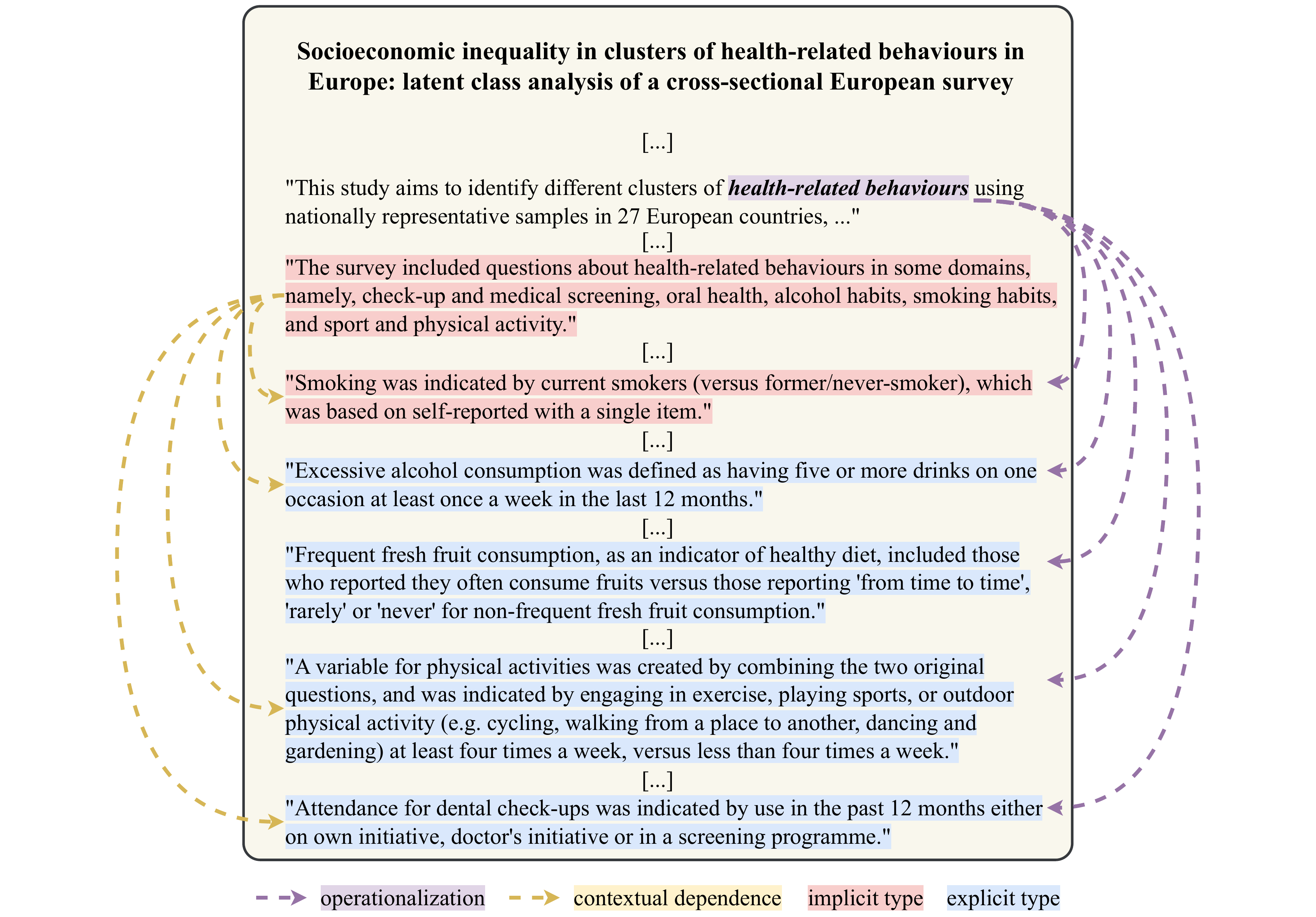}
    \caption{An example of sentences from a research publication~\citep{Kino2017} that studies the concept of \textit{health-related behaviour} using a number of survey items. The highlighted sentences cover different diagnostic elements: concept (\hlpurple{purple}), explicit sentence type (\hlblue{blue}), and implicit sentence type (\hlred{red}). The concept is linked with \hlpurple{purple} arrows to the sentences mentioning survey items that are used to operationalize it. \hlyellow{Yellow} arrows show the contextual dependence of implicit types, which require additional context to disambiguate}
    \label{figure:SILD-example}
\end{figure*}

\noindent The `type' and `subtype' columns in Table~\ref{table:survey-item-mention-examples} show different ways survey items are mentioned in scientific texts.
Implicit paraphrases are usually less specific and difficult to identify, whereas explicit quotations are the most informative.
We visualize the different elements in Figure~\ref{figure:SILD-example}, where the concept \textit{health-related behaviour} is operationalized using five survey items referenced in the implicit and explicit sentences highlighted in red and blue, respectively.
We visualize this relation using purple arrows.
All sentences mentioning survey items in the example can be categorized into the paraphrase subtype.
The first implicit sentences broadly summarize the survey items that are introduced later in the text, however, this sentence alone is insufficient for disambiguation.
As such, it has a contextual dependence relation to relevant sentences.
The second implicit sentence describes a survey item that is not available on GESIS, thus, we mark it as unknown.
Such relations allow us to evaluate whether adding appropriate context helps to identify implicit mentions.

Annotations are carried out on the INCEpTION platform.
One publication at a time, annotators read the sentences in a document and label relevant ones.
Concepts are labeled on the word-level, while survey item mentions are labeled on the sentence-level.
For sentences containing survey items, annotators access the surveys linked to each publication on GESIS Search, where individual survey items are listed.
Once a matching survey item is found, its unique identifier is copied into a text box on the platform.
Type and subtype categories are then selected and, if applicable, relations are drawn.
Finally, annotators include a score for how confident they are in each annotation.

We validate our annotation schema by having two trained annotators label 12 documents from the SV-Ident test set, which are also part of the documents in \data{}.
One of the annotators, a master's student in political science, who was paid and was also involved in labeling SV-Ident, while the other is the main author of this paper.
Compared to the inter-annotator agreement scores from SV-Ident, agreement using our guideline is significantly higher.
For agreement on sentences mentioning survey items, we report a Cohen's $\kappa$ value of 0.66 (+0.18 compared to SV-Ident), while for the multi-label task of listing which survey items are mentioned, we report a Krippendorff's $\alpha$ of 0.43 (+0.35 compared to SV-Ident).
Overall, the agreement is moderate to substantial for all reported elements (ranging between 0.43 and 0.63).
After consolidating disagreements, we list corner cases in our guideline.
Using the updated guideline, the main author of this paper annotated the remaining 82 documents in \data{}.

\subsubsection{Dataset Splits and Statistics}
To create training and evaluation datasets, we split the data at the document-level.
This prevents content overlap between the sentences of different splits, and it allows us to evaluate model generalization across documents.
Publications naturally have different referencing styles, resulting in some documents frequently quoting and others implicitly paraphrasing survey items.
In order to evaluate model performance on different difficulty levels, we create two different training, validation, and test sets.
We use all 16 German documents in the test set for both.
In the experiments, \taskoneabbrev{} models are trained on English data and evaluated on their cross-lingual transfer ability on the German data (\S\ref{section:task-one}).
The first test set, \data{}-Diff, encompasses 24 English documents that contain difficult survey item mentions, such as explicit and implicit paraphrases.
The training and validation sets are randomly sampled from the remaining documents, where 50 are used for training and 10 for validation.
The second set, \data{}-Rand, contains randomly sampled documents with equally-sized sets to \data{}-Diff.
In order to prevent future data contamination, all examples in this publication originate from the intersection of documents in the training sets of both splits.

\begin{table}
    \centering
    \footnotesize
    \caption{Dataset statistics for the \data{}-Diff dataset split. The first two rows describe the number of sentences with$^+$ and without$^-$ survey item mentions. The third and fourth rows list the number of total and unique survey items, respectively. The bottom two rows include the number of surveys referenced and papers in the dataset}
    \label{table:SILD-Diff-split-statistics}
    \begin{tabularx}{\columnwidth}{X|r|r|r|r|r}
        \toprule
         & \multirow[c]{2}{*}{\textbf{Train}} & \multirow[c]{2}{*}{\textbf{Dev}} & \multicolumn{2}{c|}{\textbf{Test}}  & \multirow[c]{2}{*}{\textbf{Total}} \\
        \\[-2ex]
         & & & \multicolumn{1}{c|}{\textbf{EN}} & \multicolumn{1}{c|}{\textbf{DE}} & \\
        \midrule
        \textbf{Sents.$^+$} & 364 & 75 & 237 & 107 & 783 \\
        \textbf{Sents.$^-$} & 9,666 & 1,900 & 5,523 & 2,582 & 19,671 \\
        \midrule
        \textbf{Items} & 810 & 195 & 654 & 162 & 1,796 \\
        \textbf{\{Items\}} & 520 & 154 & 536 & 126 & 1,283 \\
        \midrule
        \textbf{Surveys} & 45 & 7 & 46 & 16 & 97 \\
        \midrule
        \textbf{Papers} & 50 & 10 & 24 & 16 & 100 \\
        \botrule
    \end{tabularx}
\end{table}

The distributions for both dataset splits are comparable (Table~\ref{table:SILD-Diff-split-statistics} and, in the supplementary material, Table~A4).
\data{} contains 20,454 sentences, out of which 783 mention at least one survey item.
In total, 1,796 survey items are mentioned, out of which 1,283 unique items originate from 97 surveys.
Fine-grained statistics for \data{}-Diff are provided in Table~\ref{table:SILD-Diff-fine-grained-statistics}.
\textit{Explicit} and \textit{other} sentences are the most frequent types, whereas \textit{paraphrases} (implicit entity mentions) are twice as frequent as \textit{quotations} (explicit entity mentions).
For completeness, the statistics for \data{}-Rand are provided in Appendix~A.3 in the supplementary material.

\begin{table*}
    \centering
    \footnotesize
    \caption{Fine-grained statistics across diagnostic elements (type and subtype) for the \data{}-Diff dataset. The statistics are provided on both the level of sentences and of survey items. Sentences that contain only a single category are provided in the respective category rows. Sentences that contain more than one category are described in the \textit{mixed} rows. On the level of survey items, each mention is associated with a single type and subtype category. Statistics for training and development sets are aggregated}
    \label{table:SILD-Diff-fine-grained-statistics}
    \begin{tabular}{cl|r|r|r|r|r|r|r|r}
        \toprule
        & & \multicolumn{4}{c|}{\textbf{Sentences}} & \multicolumn{4}{c}{\textbf{Survey Items}} \\
        \\[-2ex]
        & & \textbf{Train +} & \multicolumn{2}{c|}{\textbf{Test}} & \multirow[c]{2}{*}{\textbf{Total}} & \textbf{Train +} & \multicolumn{2}{c|}{\textbf{Test}} & \multirow[c]{2}{*}{\textbf{Total}} \\
        & & \textbf{Dev} & \textbf{EN} & \textbf{DE} & & \textbf{Dev} & \textbf{EN} & \textbf{DE} & \\
        \midrule
        \parbox[t]{2mm}{\multirow{4}{*}{\rotatebox[origin=c]{90}{\textbf{Type}}}} & \textbf{Explicit} & 269 & 172 & 76 & 517 & 584 & 529 & 119 & 1232 \\
        & \textbf{Implicit} & 142 & 49 & 23 & 214 & 412 & 125 & 42 & 579 \\
        & \textbf{Other} & 19 & 8 & 5 & 32 & 8 & 0 & 0 & 8 \\
        & \textbf{Mixed} & 9 & 8 & 3 & 20 & - & - & - & - \\
        \midrule
        \parbox[t]{2mm}{\multirow{7}{*}{\rotatebox[origin=c]{90}{\textbf{Subtype}}}} & \textbf{Quotation} & 126 & 41 & 35 & 202 & 242 & 166 & 55 & 463 \\
        & \textbf{Paraphrase} & 231 & 158 & 51 & 440 & 675 & 433 & 87 & 1195 \\
        & \textbf{Lexical inference} & 11 & 5 & 3 & 19 & 26 & 15 & 10 & 51 \\
        & \textbf{Unspecified} & 11 & 4 & 0 & 15 & 13 & 25 & 0 & 38 \\
        & \textbf{Citation} & 7 & 2 & 7 & 16 & 7 & 2 & 7 & 16 \\
        & \textbf{Other} & 38 & 17 & 7 & 62 & 41 & 13 & 2 & 56 \\
        & \textbf{Mixed} & 15 & 10 & 4 & 29 & - & - & - & - \\
        \midrule
        & \textbf{Total} & 439 & 237 & 107 & 783 & 1004 & 654 & 161 & 1819 \\
        \botrule
    \end{tabular}
\end{table*}

\paragraph{Survey Item Metadata Knowledge Base}
Matching survey item mentions with existing entities requires access to a KB.
To this end, we create a large set of English and German survey items associated with their metadata, referred to as the \datasimcapital{} (\datasim{}) KB.
We download all available survey items from GESIS Search, resulting in 524,154 items from 1,571 surveys.
We use \datasim{} to generate synthetic data for training (\S\ref{subsec:md-data-augmentation}) and as a KB to disambiguate mentions (\S\ref{section:task-two}).
More statistics on \datasim{} are provided in Appendix~A.4 in the supplementary material.

\section{Mention Detection Experiments}
\label{section:task-one}
In this section, we benchmark different methods for the first stage of the pipeline, \taskone{}.
Linear text classification is often a good baseline against more sophisticated methods~\citep{lin-etal-2023-linear}.
However, we found that linear classifiers with lexical features considerably underperform compared to neural methods (see details in Appendix~B.1 in the supplementary material).
We describe the neural methods, which are based on fine-tuning transformers, below.
In addition, because collecting labeled data for our task is expensive, we also evaluate data augmentation methods.
Finally, we provide a fine-grained analysis using the diagnostic elements in our dataset.
Additional benchmarks, on which we found no additional benefits, such as k-NN, retrieval-augmented classification, and in-context learning, are provided in Appendix~B.3, B.4, and B.6 in the supplementary material, respectively.

We describe differences in experimental settings in each subsection.
Unless stated otherwise, models are trained on the training set of \data{}-Diff (which is comparable in performance to \data{}-Rand, as shown in Appendix~B.2 in the supplementary material), hyperparameters are tuned on the development set for 20 trials using an independent sampling algorithm, TPE~\citep{bergstra_etal_2011}, and performance is evaluated on the test set as the average across random seeds.
We report statistical significance according to the ASO test~\citep{del2018optimal,dror-etal-2019-deep,ulmer2022deep} using a confidence level of 0.95 and a threshold of $\epsilon_\text{min} \leq 0.05$ ($^{*}$).
The full list of models is provided in Appendix~C in the supplementary material.

\subsection{Baselines}\label{subsection:md-baselines}
Given the wide availability and diversity of PLMs, we pick a number of general and domain-specialized models to fine-tune on our task.
BERT~\citep{devlin-etal-2019-bert} and RoBERTa~\citep{DBLP:journals/corr/abs-1907-11692} are general PLMs that are common baselines.
As for models that are specialized for the scientific domain, only a few exist that include pre-training data from the social sciences~\citep{boyko2023interdisciplinary}.
We choose SciBERT~\citep{beltagy-etal-2019-scibert}, SSciBERT~\citep{10.1007/s11192-022-04602-4}, and SPECTER~\citep{cohan-etal-2020-specter} given their diverse pre-training corpora and optimization methods, of which only SSciBERT is specialized on the social science domain.
During fine-tuning, we update the weights of a linear classification head on top of each pre-trained model.
We train models for 20 epochs using the AdamW optimizer~\citep{loshchilov2018decoupled} and tune hyperparameters, such as learning rate and batch size.
We set the maximum sequence length to 64, which accounts for 88\% of sentences in the training set.
We present the performance for each of the fine-tuned transformer models in the \textit{Mono.} and \textit{Mul.} groups in Table~\ref{table:task1-transformer-results}.
Because the English and German test sets have different sizes (English is over two times larger), we compute the $F_1$-macro score, which adjusts for class-imbalance.
All PLMs, except for mBERT, have comparable performance for English, while \mbox{XLM-R$_{\text{large}}$} significantly outperforms all other models, improving on \mbox{XLM-R$_{\text{base}}$} by 11 points and reaching an $F_1$ score of 65.1 on German.
In our previous work~\citep{tsereteli-etal-2022-overview}, we trained separate models for each language.
Here, we show that models trained on English data can also generalize to German data.
This ability of cross-lingual transfer can allow models to transfer to more low-resourced languages in the future, improving the feasibility of the task in terms of costly labeling data.
The larger models also show improved recall, which is important for \taskone{}, given that this step only passes on relevant sentences to the proceeding pipeline.

\begin{table*}
    \centering
    \footnotesize
    \caption{Precision ($P$), recall ($R$), and $F_1$ scores for transformer-based classifiers trained on \data{}. Models are divided into monolingual (Mono), multilingual (Mul), and domain-specialized (DS) groups. Models are evaluated on English and German test sets independently. For each group, the highest score in each column is highlighted in bold and statistically significant results are marked with an asterisk. The subscript indicates the standard deviation across multiple seeds. 
    Multilingual models are able to transfer their learned knowledge from English to German.
    Domain-specialized models significantly outperform their base counterparts}
    \label{table:task1-transformer-results}
    \begin{tabular}{cl|lll|lll|l}
        \toprule
        & & \multicolumn{3}{c|}{\textbf{English}} & \multicolumn{3}{c|}{\textbf{German}} & \multicolumn{1}{c}{\textbf{Total}} \\
        & \textbf{Model} & \multicolumn{1}{c}{\textbf{$P$}} & \multicolumn{1}{c}{\textbf{$R$}} & \multicolumn{1}{c|}{\textbf{$F_1$}} & \multicolumn{1}{c}{\textbf{$P$}} & \multicolumn{1}{c}{\textbf{$R$}} & \multicolumn{1}{c|}{\textbf{$F_1$}} & \multicolumn{1}{c}{\textbf{$F_1$}} \\
        \midrule
        \parbox[t]{2mm}{\multirow{7}{*}{\rotatebox[origin=c]{90}{\textbf{Mono.}}}} & \textbf{BERT$_{\text{base}}$} & 73.4$_{\pm6}$ & 48.1$_{\pm3}$ & 57.9$_{\pm1}$ & 4.2$_{\pm0}$ & \textbf{86.4}$_{\pm10}$ & 8.0$_{\pm0}$ & 15.3$_{\pm1}$ \\
        & \textbf{BERT$_{\text{large}}$} & \textbf{76.3}$_{\pm2}$ & 49.3$_{\pm2}$ & 59.9$_{\pm2}$ & 4.9$_{\pm1}$ & 58.7$_{\pm12}$ & 9.0$_{\pm1}$ & 20.3$_{\pm3}$ \\
        & \textbf{RoBERTa$_{\text{base}}$} & 67.8$_{\pm2}$ & \textbf{54.1}$_{\pm2}$ & \textbf{60.1}$_{\pm2}$ & 6.0$_{\pm2}$ & 72.5$_{\pm11}$ & 10.9$_{\pm3}$ & 22.1$_{\pm6}$ \\
        & \textbf{RoBERTa$_{\text{large}}$} & 71.6$_{\pm5}$ & 51.4$_{\pm6}$ & 59.3$_{\pm2}$ & \textbf{42.8}$_{\pm20}$ & 40.0$_{\pm13}$ & \textbf{34.7}$_{\pm12}$ & \textbf{48.0}$_{\pm13}$ \\
        & \textbf{SciBERT} & 68.8$_{\pm3}$ & 51.0$_{\pm2}$ & 58.5$_{\pm1}$ & 27.6$_{\pm8}$ & 40.4$_{\pm8}$ & 31.1$_{\pm4}$ & 47.3$_{\pm3}$ \\
        & \textbf{SPECTER} & 71.0$_{\pm3}$ & 51.8$_{\pm2}$ & 59.8$_{\pm1}$ & 29.7$_{\pm15}$ & 37.2$_{\pm18}$ & 25.9$_{\pm5}$ & 45.4$_{\pm7}$ \\
        & \textbf{SsciBERT} & 72.6$_{\pm1}$ & 47.2$_{\pm1}$ & 57.2$_{\pm1}$ & 4.5$_{\pm0}$ & 80.0$_{\pm11}$ & 8.5$_{\pm0}$ & 16.5$_{\pm1}$ \\
        \midrule
        \parbox[t]{2mm}{\multirow{3}{*}{\rotatebox[origin=c]{90}{\textbf{Mul.}}}} & \textbf{mBERT$_{\text{base}}$} & 70.5$_{\pm5}$ & 42.8$_{\pm2}$ & 53.1$_{\pm1}$ & 60.6$_{\pm7}$ & 33.8$_{\pm8}$ & 42.4$_{\pm6}$ & 49.9$_{\pm2}$ \\
        & \textbf{XLM-R$_{\text{base}}$} & \textbf{71.8}$_{\pm2}$ & 49.5$_{\pm3}$ & 58.5$_{\pm2}$ & 69.5$_{\pm3}$ & 44.1$_{\pm1}$ & 53.9$_{\pm1}$ & 57.1$_{\pm1}$ \\
        & \textbf{XLM-R$_{\text{large}}$} & 69.9$_{\pm4}$ & \textbf{54.9}$_{\pm2}$ & \textbf{61.4}$_{\pm1}$ & \textbf{70.0}$_{\pm2}$ & \textbf{61.1}$_{\pm5}$ & \textbf{65.1}$^{*}_{\pm3}$ & \textbf{62.6}$^{*}_{\pm1}$ \\
        \midrule
        \parbox[t]{2mm}{\multirow{2}{*}{\rotatebox[origin=c]{90}{\textbf{DS}}}} & \textbf{\modelsoscibert{}$_{\text{base}}$} & \textbf{69.6}$_{\pm3}$ & 48.3$_{\pm5}$ & 56.8$_{\pm3}$ & 64.5$_{\pm12}$ & \textbf{57.6}$_{\pm9}$ & 59.1$_{\pm2}$ & 57.5$_{\pm2}$ \\
        & \textbf{\modelsoscixlmr{}$_{\text{base}}$} & \textbf{69.6}$_{\pm4}$ & \textbf{53.9}$_{\pm2}$ & \textbf{60.7}$_{\pm1}$ & \textbf{71.4}$_{\pm5}$ & 55.1$_{\pm6}$ & \textbf{61.8}$_{\pm3}$ & \textbf{61.0}$_{\pm1}$ \\
        \botrule
    \end{tabular}
\end{table*}

\subsection{Domain Adaptation}
PLMs can be adapted to a specific domain by pre-training on texts from that domain.
~\citet{beltagy-etal-2019-scibert} release SciBERT, a BERT model trained using texts from computer science and biomedical papers.
~\citet{10.1007/s11192-022-04602-4} continue pre-training SciBERT using social science abstracts.
~\citet{cohan-etal-2020-specter} encode paper titles and abstracts using SciBERT and incorporate citation information by pushing embeddings for papers that cite each other closer together and those that do not further apart.
Given that most social science publications rarely reference survey items within abstracts, it is unlikely that the pre-training corpus for SSciBERT contains many examples of survey item mentions, which is highly relevant for our application.
In contrast to certain scientific domains, such as computer science, research in the social sciences is often published in different languages.
As such, social science PLMs should be extended to the multilingual setting.
Due to these limitations of existing PLMs, we pre-train custom models on social science literature.
To the best of our knowledge, no model has yet been pre-trained on SSOAR full texts, which includes English and German publications from a diverse set of fields in the social sciences.
To this end, we download over 56k documents from SSOAR and remove those in \data{} to prevent data contamination.
We extract the texts using GROBID, and further preprocess them with custom heuristics, resulting in 44,741 clean documents in our full dataset, referred to as \datapt{}.
We split the documents in \datapt{} into sentences with a maximum of 512 tokens and set the batch size to 8.
We train mBERT, a multilingual BERT model, and \mbox{XLM-R}~\citep{conneau-etal-2020-unsupervised}, a multilingual RoBERTa model, using masked language modeling.
We refer to these models as \modelsoscibert{} and \modelsoscixlmr{}, respectively.
Because continued pre-training is expensive, we train models on a limited budget, namely for a total of 700k steps on four A100 GPUs.
Each model completed training in under 48 hours.
We leave more extensive experiments to future work.

Our domain-specialized PLMs, \modelsoscibert{} and \modelsoscixlmr{}, significantly improve upon their pre-trained counterparts in terms of recall (bottom group in Table~\ref{table:task1-transformer-results}).
Although continued pre-training of mBERT on \datapt{} does not outperform SciBERT, it is a viable option for adaption of multilingual models to the languages that are used in publishing social science literature.
Future work could more closely examine efficient hyperparameter tuning methods and ways to include publication-specific information, such as citations~\citep{10.1145/3529372.3530912}, to better adapt models to the scientific domain.

\subsection{Data Augmentation}\label{subsec:md-data-augmentation}
Although \data{} is the largest dataset for the task, identifying implicit mention types is challenging (in \$\ref{subsec:md-analysis} we report performance on different diagnostic elements in the data).
In order to collect more examples of implicit mentions, we turn to data augmentation, which is a commonly used technique to introduce data diversity and increase training data size~\citep{bayer22}.
As a starting point, we replace synonyms by randomly masking out a percentage of the words in each sentence in \data{} and use BERT to predict the missing words.
The augmented sentences are then combined with the original ones to produce the new training data.
During testing, in order to prevent data leakage (due to the high similarity between the source and the augmented sentences), we only evaluate on the source instances.
This method, however, creates instances that are very similar to those already in the training data.
Creating more diverse data is challenging, as generated data may introduce label noise.
To introduce realistic but diverse data, we simply train on the survey items in \datasim{}, as survey items mentions have a large overlap to \datasim{}. 
Recently, LLMs have shown that they can generate diverse texts, making them suitable for creating realistic examples.
For example, \citet{stavropoulos-etal-2023-empowering} used ChatGPT to generate synthetic examples of dataset and software mentions using available metadata.
Similarly, we use Vicuna-7B to generate synthetic texts given survey item metadata (see Appendix~B.5 in the supplementary material for the prompt).\footnote{We chose an open-source model over proprietary models for two reasons: first, open-source models are more interpretable, given that the training data and architecture are often known; second, there is no risk of data contamination running open-source models locally.}
In total, we generate over 400k synthetic sentences.
Due to computational constraints, we fine-tune XLM-R$_{\text{base}}$ for 5 epochs without hyperparameter tuning on a single seed and use the same dataset size as \data{} for the synonym-substituted data and different samples from \datasim{} and the LLM-generated examples.

Our results in Table~\ref{table:da-results} show models that are trained on 10k samples from \datasim{} and LLM-generated sentences.
Synonym substitution and examples from \datasim{} have a negative impact.
A suboptimal choice of initial hyperparamters and noisy texts in \datasim{} could explain these observations.
In contrast, LLM-generated data shows promising results, as it increases recall at the cost of precision.
For \taskoneabbrev{}, high recall is important, as it significantly impacts the quality of overall predictions in the second stage of the pipeline (see our results in~\S\ref{section:two-stage}).
We also evaluated a larger sample size, however, models were sensitive to the choice of hyperparameters.
We leave tuning hyperparameters on synthetic data to future work, where high-quality sentences could be sampled via active learning.

\begin{table*}[ht]
    \centering
    \footnotesize
    \caption{Data augmentation results using the original training data from \data{}, synonym substitution, and the metadata from \datasim{} as additional training data or as prompt input for LLM-generated sentences. 
    Training on LLM-generated data improves recall at the cost of precision}
    \label{table:da-results}
    \begin{tabular}{l|lll|lll|l}
        \toprule
        & \multicolumn{3}{c|}{\textbf{English}} & \multicolumn{3}{c|}{\textbf{German}} & \multicolumn{1}{c}{\textbf{Total}} \\
        \\[-2ex]
        \textbf{Data} & \multicolumn{1}{c}{\textbf{$P$}} & \multicolumn{1}{c}{\textbf{$R$}} & \multicolumn{1}{c|}{\textbf{$F_1$}} & \multicolumn{1}{c}{\textbf{$P$}} & \multicolumn{1}{c}{\textbf{$R$}} & \multicolumn{1}{c|}{\textbf{$F_1$}} & \multicolumn{1}{c}{\textbf{$F_1$}} \\
        \midrule
        \textbf{\data{}} & \textbf{69.7} & 48.9 & 57.5 & \textbf{68.3} & 41.1 & 50.9 & 55.6 \\
        \textbf{Synonym} & 41.6 & 24.1 & 30.5 & 60.0 & 16.4 & 25.7 & 29.2 \\
        \textbf{\datasim{}} & 49.8 & 9.9 & 16.3 & 55.1 & 13.3 & 20.5 & 17.7 \\
        \textbf{LLM-Gen} & 66.7 & \textbf{52.1} & \textbf{58.5} & 63.9 & \textbf{50.8} & \textbf{56.5} & \textbf{57.8} \\
        \botrule
    \end{tabular}
\end{table*}

\subsection{When does context matter?}\label{subsec:md-analysis}
To better understand which types of mentions models find challenging to identify, we evaluate the best model for \taskoneabbrev{}, namely XLM-R$_{\text{large}}$, using the diagnostic elements in \data{}, which we introduced in \$\ref{section:dataset}.
Due to the number of labels for each element in the test set, we restrict our evaluation to \textit{explicit} and \textit{implicit} types and \textit{paraphrase} and \textit{quotation} subtypes.
In line with our previous findings~\citep{tsereteli-etal-2022-overview}, \textit{implicit} types are more difficult to detect (we show recall scores in Table~\ref{table:diagnostic-results-md}).
Quotations, as describe in~\S\ref{section:dataset}, are explicit entity mentions, which most EL systems focus on.
Not surprisingly, identifying them is much easier than identifying implicit entity mentions (i.e., paraphrases).
Implicit types are the most challenging entity mentions, as they not only vaguely reference the entity but also require additional context.

\begin{table*}
    \centering
    \footnotesize
    \caption{Recall scores for the most-common type and subtype categories for \taskoneabbrev{}. The subscript indicates the standard deviation across multiple seeds. 
    Explicit and quotation are easier to detect than implicit and paraphrase}
    \label{table:diagnostic-results-md}
    \begin{tabular}{l|r|r|r|r}
        \toprule
        & \multicolumn{2}{c|}{\textbf{Type}} & \multicolumn{2}{c}{\textbf{Subtype}} \\
        \\[-2ex]
        \multicolumn{1}{c|}{\textbf{Model}} & \multicolumn{1}{c|}{\textbf{Explicit}} & \multicolumn{1}{c|}{\textbf{Implicit}} & \multicolumn{1}{c|}{\textbf{Quotation}} & \multicolumn{1}{c}{\textbf{Paraphrase}} \\
        \midrule
        \textbf{XLM-R$_{\text{base}}$} & 70.9$_{\pm2}$ & 38.8$_{\pm4}$ & 78.2$_{\pm1}$ & 61.2$_{\pm3}$ \\
        \textbf{XLM-R$_{\text{large}}$} & \textbf{78.9}$_{\pm2}$ & \textbf{46.4}$_{\pm4}$ & \textbf{85.0}$_{\pm1}$ & \textbf{69.4}$_{\pm2}$ \\
        \textbf{SoSci-XLM-R$_{\text{base}}$} & 77.2$_{\pm3}$ & 42.7$_{\pm5}$ & 84.5$_{\pm4}$ & 66.7$_{\pm3}$ \\
        \botrule
    \end{tabular}
\end{table*}

To this end, we take advantage of the relation annotations in \data{} to pick the appropriate context for each implicit sentence.
We append the contextually-dependent sentences either before or after the sentence containing the mention.
A limitation of this approach is the requirement for the relation label to a contextual sentence.
Given that links to contextually relevant sentences are difficult to generate, we also simply expand the context window of sentences by combining two neighboring sentences.
Table~\ref{table:context-results-md} shows that by including the relation information and using the neighbors as context significantly increases the performance for English.
The opposite is true for German, where performance degrades.
Looking at the sentence length, on average, English and German sentences that mention survey items contain the same number of characters in \data{}.
However, German sentences contain 1.5 survey items per sentence, while English have 2.8.
We investigate the performance difference between detecting sentences that mention one survey item and those that contain more than one below.

\begin{table}
    \centering
    \footnotesize
    \caption{Recall scores for XLM-R$_{\text{large}}$ fine-tuned on different target contexts that are concatenated to source sentences. Relation contexts require a relation label between two sentences, whereas neighbor contexts simply use the neighboring sentences as context.
    Recall improves for English but drops for German}
    \label{table:context-results-md}
    \begin{tabularx}{\columnwidth}{X|r|r|r}
        \toprule
        \multicolumn{1}{c|}{\textbf{Context}} & \multicolumn{1}{c|}{\textbf{English}} & \multicolumn{1}{c|}{\textbf{German}} & \multicolumn{1}{c}{\textbf{Total}} \\
        \midrule
        - & 44.9 & \textbf{33.5} & 41.3 \\
        Relation & 48.6 & 29.1 & 42.4 \\
        Neighbor & \textbf{57.6} & 25.8 & \textbf{47.4} \\
        \botrule
    \end{tabularx}
\end{table}

A sentence commonly mentions a single entity, however, our test data also contains 121 sentences that mention more than one item.
Surprisingly, recall for sentences with multiple items is 17 points higher than for those mentioning only a single item (Table~\ref{table:multi-item-results-md}).
Because multiple items are listed sequentially and frequently describe demographics of respondents using repetitive terms (e.g., gender or age), models likely exploit this pattern.
This could also explain the increase in precision when using additional context (Table~\ref{table:context-results-md}).
Sentences mentioning more than one survey item expand on the details of the items in the surrounding sentences, while those that only mention a single item, likely describe the item in the sentence alone.
Future work could further investigate data augmentation methods to generate difficult sentences.
Furthermore, document-level representations could incorporate wider context, which may help identify implicit paraphrases.

\begin{table}
    \centering
    \footnotesize
    \caption{Precision ($P$), recall ($R$), and $F_1$ scores for single- vs multi-item sentences for \taskoneabbrev{}. The subscript indicates the standard deviation across multiple seeds. The number of sentences of each category in the test set is provided in the second column. 
    Single-item sentences can more precisely be detected (higher $P$), but multi-item sentences have a higher $R$}
    \label{table:multi-item-results-md}
    \begin{tabular}{l|r|rrr}
        \toprule
        \multicolumn{1}{c|}{\textbf{Items}} & \multicolumn{1}{c|}{\textbf{Count}} & \multicolumn{1}{c}{\textbf{$P$}} & \multicolumn{1}{c}{\textbf{$R$}} & \multicolumn{1}{c}{\textbf{$F_1$}} \\
        \midrule
        \textbf{Single} & 223 & 73.5$_{\pm3}$ & 52.2$_{\pm3}$ & 61.0$_{\pm1}$ \\
        \midrule
        \textbf{Multi} & 121 & 65.5$_{\pm3}$ & 65.5$_{\pm3}$ & 65.3$_{\pm1}$ \\
        \botrule
    \end{tabular}
\end{table}

\subsection{Summary}
We conclude that for \taskoneabbrev{}, PLMs provide a strong baseline, but models can be adapted to the social science domain using a limited computational budget.
Recall, which is important for the task, can be improved through LLM-based data augmentation.
Furthermore, using the diagnostic elements we introduced in our dataset (\S\ref{section:dataset}), we found that implicit paraphrases and single-item sentences are the most challenging to detect.
Finally, incorporating context has mixed benefits for different languages.

\section{Entity Disambiguation Experiments}
\label{section:task-two}
In this section, we benchmark lexical and semantic Information Retrieval (IR) methods, including zero-shot systems, for the second stage of the pipeline, \tasktwo{}.
We define zero-shot retrieval as the task of retrieving relevant survey items without being trained on the task or the domain.
In contrast to text classification, lexical systems that use different weighting and normalization of exact term matching methods, such as BM25~\citep{10.1561/1500000019}, are strong baselines in IR~\citep{trec-fair-ranking-2021}.
However, such methods are unable to match semantically-related concepts, such as \textit{hot} and \textit{warm}.
We compare BM25 to semantic retrieval, which uses transformer-based models to encode texts into dense vectors.
In addition to using available models, we fine-tune models on two novel sentence-pair datasets.
In this section, \tasktwoabbrev{} is evaluated independent of \taskoneabbrev{} (i.e., assuming gold labels from the previous stage).
In~\S\ref{section:two-stage}, we evaluate the complete pipeline by using the predictions from the best model for \taskoneabbrev{}.
Similar to the analysis in $\S$\ref{section:task-one}, we provide a fine-grained perspective on performance for this stage of the pipeline.

As described in~\S~\ref{section:dataset}, \datasim{} contains rich metadata, such as label, question, sub-question, and item category.
We compare different metadata combinations to find the best-performing one by simply concatenating their values.
Combining the label, question, item category, and answers achieves the best results.
Because we evaluate models in a zero-shot setting, there is no source of randomness in our evaluation.
Hence, we run experiments using a single seed.

\subsection{Baselines}
We choose BM25 as our lexical baseline, given that it shows competitive performance compared to neural IR methods~\citep{trec-fair-ranking-2021}.
The lexical model is used in a zero-shot setting.
We encode each sentence $s$ in the collection of sentences $S$, where $s \in S$, into a sparse vector, where each element in the vector represents a term in $s$.
We then compute and store the vectors offline and filter them according to document-survey relations during inference.
Similarly, we encode each query sentence $q$ in the set of sentences $Q$ from the test set (i.e., each sentence that mentions a survey item) into a sparse vector.
Using cosine similarity, we compute the similarity of each query to each document in the subset $S_D$, which contains the survey items that are associated with the surveys linked to the publication the query originates from.
We then rank the documents by their scores and compute metrics for the top $k$ highest-scoring documents.
The first row in Table~\ref{table:ir-results} shows that BM25 is significantly better for English than for German.
This demonstrates that term-matching is a strong baseline for English.
A possible reason for the low scores for German may be the occurrence of compound words, where rare noun chains are single words (i.e., features), whereas in English, they are separated by spaces or hyphens.
Subword tokenization could mitigate the problem, which we leave to future work.

\begin{table*}
    \centering
    \footnotesize
    \caption{Recall ($R$), MAP, and nDCG scores for zero-shot retrieval models for \tasktwoabbrev{}. The table is split into a lexical model (L.), PLMs that have (Special.) and have not been (Base) trained on sentence-pair tasks.
    BM25 outperforms base XLM-R and most of the specialized models for English.
    For German, specialized models are better}
    \label{table:ir-results}
    \begin{tabular}{cl|rrr|rrr|rrr}
        \toprule
        & & \multicolumn{3}{c|}{\textbf{English}} & \multicolumn{3}{c|}{\textbf{German}} & \multicolumn{3}{c}{\textbf{Total}} \\
        & \textbf{Model} & \multicolumn{1}{c}{\textbf{$R$}} & \multicolumn{1}{c}{\textbf{$MAP$}} & \multicolumn{1}{c|}{\textbf{$nDCG$}} & \multicolumn{1}{c}{\textbf{$R$}} & \multicolumn{1}{c}{\textbf{$MAP$}} & \multicolumn{1}{c|}{\textbf{$nDCG$}} & \multicolumn{1}{c}{\textbf{$R$}} & \multicolumn{1}{c}{\textbf{$MAP$}} & \multicolumn{1}{c}{\textbf{$nDCG$}} \\
        \midrule
        \parbox[t]{2mm}{\multirow{1}{*}{\rotatebox[origin=c]{90}{\textbf{L.}}}} & \textbf{BM25} & 75.1 & 64.4 & 69.6 & 34.5 & 28.0 & 29.8 & 63.0 & 53.6 & 57.8 \\
        \midrule
        \parbox[t]{2mm}{\multirow{3}{*}{\rotatebox[origin=c]{90}{\textbf{Base}}}} & \textbf{XLM-R$_{\text{base}}$} & 30.4 & 20.1 & 24.1 & 23.9 & 15.0 & 17.6 & 28.5 & 18.7 & 22.2 \\
        & \textbf{XLM-R$_{\text{large}}$} & 24.1 & 12.0 & 16.5 & 18.3 & 7.6 & 10.2 & 22.4 & 10.7 & 14.7 \\
        & \textbf{SoSci-XLM-R$_{\text{base}}$} & 40.3 & 25.6 & 31.2 & 36.8 & 24.2 & 27.4 & 39.3 & 25.2 & 30.1 \\
        \midrule
        \parbox[t]{2mm}{\multirow{5}{*}{\rotatebox[origin=c]{90}{\textbf{Special.}}}} & \textbf{MiniLM} & 70.6 & 57.2 & 63.0 & 78.9 & 58.6 & 64.5 & 73.1 & 57.6 & 63.4 \\
        & \textbf{Cross} & 67.3 & 49.1 & 56.4 & 66.7 & 51.7 & 55.9 & 67.1 & 49.8 & 56.2 \\
        & \textbf{mE5$_{\text{small}}$} & 73.9 & 55.2 & 62.1 & 76.1 & 54.6 & 60.7 & 74.5 & 55.0 & 61.7 \\
        & \textbf{mE5$_{\text{base}}$} & 74.7 & 57.9 & 64.4 & 78.5 & \textbf{65.6} & 69.5 & 75.8 & 60.2 & 65.9 \\
        & \textbf{mE5$_{\text{large}}$} & \textbf{79.0} & \textbf{63.1} & \textbf{70.0} & \textbf{82.3} & 65.0 & \textbf{69.9} & \textbf{80.0} & \textbf{63.7} & \textbf{69.9} \\
        \botrule
    \end{tabular}
\end{table*}

Following a common approach in neural IR~\citep{10.1561/1500000061}, we encode sentences and survey items into dense semantic vectors using transformer-based bi-encoder models.
However, deriving embeddings from BERT, e.g., by averaging the output layer or using the output of the {\fontfamily{cmtt}\selectfont [CLS]} token, yields low-quality sentence embeddings.
Semantically meaningful sentence embeddings are commonly obtained by training on sentence-level objectives~\citep{reimers-gurevych-2019-sentence} or via contrastive learning~\citep{gao-etal-2021-simcse}, during which positive sentence pairs are pulled closer together in latent space while negative pairs are pushed apart using specialized loss functions~\citep{10.1145/3593590}.
While many sentence embedding models exist, we do not aim to benchmark all available models.
We choose three models that perform well on both English and German semantic textual similarity (STS) tasks on the MTEB leaderboard~\citep{muennighoff-etal-2023-mteb}.
Our chosen models are comparable, as all three use \mbox{XLM-R$_{\text{base}}$} as their base model.
\citet{reimers-gurevych-2020-making} distill a paraphrase teacher model (MiniLM), trained on seven paraphrase datasets~\citep{10.5555/3495724.3496209}, to an \mbox{XLM-R} student model.
A different approach, referred to as \textit{Cross}, continues fine-tuning MiniLM on English and translated German sentence pairs from the STSbenchmark dataset~\citep{cer-etal-2017-semeval}.
Finally, multilingual E5~\citep{wang2024text,wang2024multilingual} is first contrastively pre-trained with weak supervision on over 5 billion multilingual text pairs and later fine-tuned on 2 billion labeled text pairs.

As expected, deriving embeddings from \mbox{XLM-R} models that have not been fine-tuned on sentence-pair tasks results in poor performance (top half of Table~\ref{table:ir-results}).
In fact, BM25 significantly outperforms all \mbox{XLM-R} models.
The domain-specialized model, \mbox{SoSci-XLM-R$_{\text{base}}$}, outperforms both the base and large \mbox{XLM-R} counterparts.
This is promising, given that MiniLM, Cross, and mE5 could be retrained using the improved base model in future work.
Models trained on sentence-pair tasks achieve significant improvements (bottom half of Table~\ref{table:ir-results}).
The best model, (mE5$_{\text{large}}$), which was trained on a large set of sentences pairs, has a recall of 79 for English and over 82 for German.
Compared to the results from the data using the previous annotation guideline~\citep{tsereteli-etal-2022-overview}, performance using our newly annotated data shows that the task is feasible.
We attribute this to the new definition, which significantly reduces the ambiguity in the annotated sentences.
These results demonstrate that, while BM25 with lexical features is a competitive baseline, it is unable to compete with sentence embeddings due to their cross-lingual ability.

\subsection{Social Science Sentence-Embeddings}\label{subsec:sosse}
Although the sentence embeddings baselines above have been specialized on sentence-pair data, they many not be suitable for \tasktwoabbrev{} (e.g., post-comment or question-answer pairs).
Models that are trained on data that are closer to the intended use-case may produce more representative embeddings.
To this end, using the baselines above, we continue fine-tuning sentence embeddings on semantically related texts using a contrastive learning objective.
Two sentences that are topically related do not have to be semantically similar (e.g., an implicit entity mention and the entity description).
Sentence embeddings that have been trained on semantically related texts are not widely available for the scientific domain, as sentence embeddings are commonly optimized for STS benchmarks~\citep{cer-etal-2017-semeval}, where semantically similar pairs of texts are evaluated.
Recently, unsupervised contrastive learning methods for learning sentence embeddings, which use pseudo-labels~\citep{wang2024text} or synthetic data~\citep{wang2024improving}, have been shown to perform on-par or better than supervised methods.
In order to fine-tune sentence embeddings, we add a single linear layer with a skip connection head to the existing model and freeze the pre-trained layers.
The sentence embeddings are then extracted using mean token pooling.
We create two large semantically related sentence pair datasets by using the metadata in \datasim{} to create pseudo-labeled pairs of sentences.
In the \datasim{}, the title for a survey item typically summarizes the content of the \mbox{question} and item category, excluding function words.
Thus, we randomly pair the title with either the question or the item category.
The resulting dataset, referred to as metadata-pairs (MP), contains 215,000 randomly sampled sentence pairs.
While sentence pairs in MP arise organically, they do not accurately reflect the way survey items are mentioned in scientific texts.
Thus, we additionally re-use the more natural LLM-generated sentences from~\S\ref{subsec:md-data-augmentation} and pair them with randomly selected metadata (from title, question or item category) from the source survey item.
The resulting synthetic dataset, referred to as synthetic-pairs (SP), contains 430k randomly sampled sentence pairs.
We eliminate elements in pairs that consist of fewer than three words, as shorter texts tend to be noisy.
To avoid contaminating the data by training on texts identical to those in the \data{} evaluation set, we exclude near-duplicates with a Levenshtein distance of less than ten.\footnote{This approach helps address parsing issues as well.}
We split MP and SP datasets into training and validation sets with sizes of 200k/15k and 400k/30k, respectively.
The maximum sequence length is set to 128.
Models are trained separately on MP and SP for a maximum of 50 epochs using the multiple negatives loss~\citep{DBLP:journals/corr/HendersonASSLGK17}, and the results are averaged over three random seeds.
We train models using the \textit{quaterion} library,\footnote{\url{https://github.com/qdrant/quaterion}} which is built around the \textit{sentence-transformers}~\citep{reimers-gurevych-2019-sentence}.

Increasing the dataset size when training models on pseudo-labeled MP pairs or on synthetic SP pairs generally improves performance across all models (except for mE5 on German using MP).
Using SP, performance on English starts to degrade after 200k pairs.
The mE5 model benefits most from the synthetic data, reaching a MAP@10 score of over 68.
Table~\ref{table:sosse-vs-baseline} compares the MAP of the three baseline models to the respective SoSSE counterparts, trained on 200k samples from SP.
The best SoSSE model scores 5.3 and 2.5 points higher for English and German, respectively.
Overall, fine-tuning sentence transformers on semantically related survey item metadata pairs significantly improves performance on our task.
In future work, we would like to evaluate the sentence embeddings on other tasks, in order to understand whether the performance boost comes from the semantic relatedness aspect or from the similarity to our downstream task.
Prompt engineering methods could be investigated to generate more diverse and higher quality data.

\begin{table}
    \centering
    \footnotesize
    \caption{Comparison of MAP@10 scores between the best SoSSE models trained on 200k LLM-generated sentences (standard deviation across seeds is indicated in the subscript) and their respective base models (run with a single seed).
    Training on LLM-generated data generally improves performance}
    \label{table:sosse-vs-baseline}
    \begin{tabularx}{\columnwidth}{X|l|l|l}
        \toprule
        \multicolumn{1}{l|}{\textbf{Model}} & \multicolumn{1}{c|}{\textbf{Sample}} & \multicolumn{1}{c|}{\textbf{English}}  & \multicolumn{1}{c}{\textbf{German}} \\
        \midrule
        \textbf{MiniLM} & - & 57.2 & 58.6 \\
        \textbf{SoSSE-MiniLM} & 200k & 58.6$_{\pm0.1}$ & 61.7$_{\pm0.2}$ \\
        \midrule
        \textbf{Cross} & - & 49.1 & 51.7 \\
        \textbf{SoSSE-Cross} & 200k & 50.3$_{\pm0.3}$ & 58.2$_{\pm0.5}$ \\
        \midrule
        \textbf{mE5$_{\text{base}}$} & - & 57.9 & 65.6 \\
        \textbf{\mbox{SoSSE-mE5$_{\text{base}}$}} & 200k & \textbf{63.2}$_{\pm0.5}$ & \textbf{68.1}$_{\pm0.3}$ \\
        \botrule
    \end{tabularx}
\end{table}

\subsection{Survey Citations are Invaluable}\label{subsec:ed-analysis}
In this subsection, we evaluate the impact of reducing the penalty of retrieving thematically similar survey items, which are grouped in surveys, and the importance of filtering the KB using survey citations.
Survey item groups can be recognized through the suffix of the item identifier.
We use this information to extend the ground-truth labels.
Retrieved survey items that are thematically-similar to the ground-truth are more likely to be relevant than those which are thematically-dissimilar.
Recall improves by 7.6 points when expanding relevant survey items to at most 2 surrounding neighbors.
In practice, however, this means that rather than retrieving the top 10 items, we would retrieve the top 30 (in the worst case), which is unfavorable.

The documents we downloaded for \data{} are all linked to survey research.
However, when links are not available, the candidate set expands significantly.
Here, we evaluate the performance in this more challenging setting.
For efficiency, we only evaluate BM25, as it can provide a sufficient perspective on performance.
Recall drops from 75.1 to 30.8 for English and from 34.5 to 14.3 for German.
This suggests that citations to surveys a publication uses are essential for \taskabbrev{}.

\subsection{Summary}
For \tasktwoabbrev{}, we conclude that BM25 is a strong baseline for English, but more sophisticated tokenization methods are necessary for German.
Specializing transformer-based models on LLM-generated data specifically designed for the task is a viable and inexpensive option to boost performance.
\tasktwoabbrev{} is feasible as long as the links to surveys that a publication references are available.
Without such links, the task becomes significantly more challenging and results in poor performance.

\section{\taskthreecapital{}}\label{section:two-stage}
So far, we have examined the performance of \taskoneabbrev{} and \tasktwoabbrev{} independently.
In~\S~\ref{section:task-one}, we identified only the sentences that mentioned survey items.
Then, in~\S~\ref{section:task-two}, we assumed ground-truth knowledge about the sentences that mention survey items to evaluate \tasktwoabbrev{}.
In a real-world setting, the predicted sentences from the first stage would be passed on to the second stage.
Such two-stage pipelines, which we refer to as a \taskthreecapital{} (\taskthreeabbrev{}), have been used extensively in traditional EL~\citep{guo-etal-2013-link,luo-etal-2015-joint,10.1145/2872427.2883061,ganea-hofmann-2017-deep}.
However, the first stage is often ignored for implicit EL~\citep{Hosseini2024}.
Previous work on \taskabbrev{}~\citep{zielinski-mutschke-2018-towards,tsereteli-etal-2022-overview} did not evaluate the stages sequentially, hence leaving a gap in understanding the feasibility of the task.
We evaluate the true performance of \taskabbrev{} using Recall@10 for the \taskthreeabbrev{}.
For both BM25 and mE5, recall drops by 25 to 30 points (Table~\ref{table:two-stage-results}).
Errors from \taskoneabbrev{} get propagated to \tasktwoabbrev{}, given that \tasktwoabbrev{} has no filtering mechanism and simply retrieves the most similar candidates for each input.
This demonstrates the importance of the \taskoneabbrev{}, which significantly influences the final output that is used to enrich information systems.
The first stage does not directly access the KB, missing relevant mentions.
Future work could look at joint methods for end-to-end EL, which could directly incorporate information from the KB.

\begin{table}
    \centering
    \footnotesize
    \caption{Recall@10 scores for \taskthreeabbrev{}, which applies \taskoneabbrev{} and \tasktwoabbrev{} sequentially. Models with $^{\clubsuit}$ use an oracle instead of an \taskoneabbrev{} model. 
    Errors from the best model for \taskoneabbrev{} propagate to the second stage, resulting in an overall lower recall}
    \label{table:two-stage-results}
    \begin{tabularx}{\columnwidth}{X|r|r|r}
        \toprule
        \multicolumn{1}{c|}{\textbf{Model}} & \multicolumn{1}{c|}{\textbf{English}} & \multicolumn{1}{c|}{\textbf{German}} & \multicolumn{1}{c}{\textbf{Total}} \\
        \midrule
        \textbf{BM25}$^{\clubsuit}$ & 75.1 & 34.5 & 63.0 \\
        \textbf{BM25} & 42.4 & 24.4 & 37.1 \\
        \midrule
        \textbf{mE5}$_{\text{large}}^{\clubsuit}$ & \textbf{79.0} & \textbf{82.3} & \textbf{80.0} \\
        \textbf{mE5}$_{\text{large}}$ & 45.6 & 59.0 & 49.6 \\
        \botrule
    \end{tabularx}
\end{table}

\section{Qualitative Evaluation}\label{section:qualitative-analysis}
To gain a better understanding of which instances models consistently disambiguate correctly and incorrectly, we inspect predictions for selected instances from the validation set.
We count the number of times a model that was trained using different seeds with a fixed number of parameters retrieves an entity.
We then categorize the entities into those which are always predicted correctly (\textit{always}) or incorrectly (\textit{never}), and those which are predicted correctly by a majority of the models (\textit{majority}).
Here, predictions are based on the top 10 retrieved entities.
Overall, only 9 out of 451 entities are always predicted correctly, 213 fall into the \textit{majority} label, and the remaining 225 into \textit{never}.
We select four examples to showcase different properties of texts from the categories above (Table~\ref{table:qualitative-analysis-ed}).
The first sentence is an example of a cross-lingual reference from an English sentence to a German survey item that the multilingual models predict correctly most of the time.
Because multilingual models encode many languages into a shared latent space, semantically similar sentences can be retrieved across languages.
Furthermore, we observe that models often cannot predict entities for sentences that mention more than one survey item.
This is likely caused by the varying content of survey items that are mentioned in a single sentence (e.g., sentence 2).
Identifying mentions at the span-level could alleviate this problem, however, a challenge of implicit entity mentions is that mention boundaries are vague and costly to label.
Another solution might be to break query sentences that contain multiple mentions into shorter sequences during retrieval, rather than encoding the entire sentence.
We also observe that models struggle when entities are not clearly identifiable, when question content is missing in the KB (e.g., the entities linked to sentence 3), or when the KB contains multiple entries for an entity, such as the entities for sentence 4.
We attribute such errors to problems in the extraction of the survey items and the construction of the KB.
Future work could construct a high-quality KB of survey items using more sophisticated heuristics.

\begin{table*}
    \centering
    \footnotesize
    \caption{Examples of sentences that models frequently predict correctly or incorrectly for the \tasktwoabbrev{} task. 
    The first sentence shows that models are able to retrieve cross-lingual mentions. 
    The content of the survey items in sentence 2 vary, making it difficult for models to disambiguate. 
    Missing metadata (sentences 3 and 4) and duplicate entries in the KB (sentence 5) make the task challenging}
    \label{table:qualitative-analysis-ed}
    \begin{tabularx}{\textwidth}{l|l|X}
         \toprule
         \multirow{7}{1em}{\textbf{1}} & \textbf{Sentence} & A first step in this direction is a comparison between the assessments of the German economy and of the respondents' own economic situation. \\
         & \textbf{Items} & \href{https://search.gesis.org/variables/exploredata-ZA2800_VarV113}{ZA2800\_VarV113} - \textit{WIRTSCHAFTSLAGE IN DER BRD HEUTE} (\textit{Wie beurteilen Sie ganz allgemein die heutige wirtschaftliche Lage in Deutschland?})\\ 
         & \textbf{Doc. ID} & \href{https://www.ssoar.info/ssoar/handle/document/19984}{19984}~\citep{Terwey1998} \\
         & \textbf{Predictions} & Majority \\
         \midrule
         \multirow{8}{1em}{\textbf{2}} & \textbf{Sentence} & Demographic factors included gender, age, urbanisation and marital status. \\
         & \textbf{Items} & \href{https://search.gesis.org/variables/exploredata-ZA4977_Varv330}{ZA4977\_Varv330} - \textit{GENDER} \\
         & & \href{https://search.gesis.org/variables/exploredata-ZA4977_Varv331}{ZA4977\_Varv331} - \textit{AGE EXACT} (\textit{How old are you?}) \\
         & & \href{https://search.gesis.org/variables/exploredata-ZA4977_Varv334}{ZA4977\_Varv334} - \textit{TYPE OF COMMUNITY} (\textit{Would you say you live in a...? Rural area or village / Small or middle sized town / Large town}) \\
         & & \href{https://search.gesis.org/variables/exploredata-ZA4977_Varv326}{ZA4977\_Varv326} - \textit{MARITAL STATUS} \\
         & \textbf{Doc. ID} & \href{https://www.ssoar.info/ssoar/handle/document/74709}{74709}~\citep{Kino2017} \\
         & \textbf{Predictions} & v330 always, v326 majority, and v331/v334 never \\
         \midrule
         \multirow{7}{1em}{\textbf{3}} & \textbf{Sentence} & The first question has to do with the bright side of European integration, while the second question has to do with the consequences of a potential demise of the EU. \\
         & \textbf{Items} & \href{https://search.gesis.org/variables/exploredata-ZA3521_Varmembrshp}{ZA3521\_Varmembrshp} - \textit{MEMBERSHIP: GOOD/BAD} \\
         & & \href{https://search.gesis.org/variables/exploredata-ZA3521_Varregret}{ZA3521\_Varregret} - \textit{REGRET} \\
         & \textbf{Doc. ID} & \href{https://www.ssoar.info/ssoar/handle/document/74446}{74446}~\citep{Tselios2020} \\
         & \textbf{Predictions} & Never \\
         \midrule
         \multirow{10}{1em}{\textbf{4}} & \textbf{Sentence} & Self-placement and perceived party positions on a left-right scale also measure position issues, but in the more generalized form of ideological orientations. \\
         & \textbf{Items} & \href{https://search.gesis.org/variables/exploredata-ZA5320_Varc250}{ZA5320\_Varc250} - \textit{Left-right self assessment} \\
         & & \href{https://search.gesis.org/variables/exploredata-ZA5320_Vard250}{ZA5320\_Vard250} - \textit{Left-right self assessment} \\
         & & \href{https://search.gesis.org/variables/exploredata-ZA5320_Varc284a}{ZA5320\_Varc284a} - \textit{Left-right assessment parties - CDU} \\
         & & \href{https://search.gesis.org/variables/exploredata-ZA5320_Vard284a}{ZA5320\_Vard284a} - \textit{Left-right assessment parties - CDU} \\
         & \textbf{Doc. ID} & \href{https://www.ssoar.info/ssoar/handle/document/58744}{58744}~\citep{Wuttke2017}\\
         & \textbf{Predictions} & Majority: d250, d284a, ...; Never: c250, c284a, ... \\
         \botrule
    \end{tabularx}
\end{table*}

\section{Conclusion}
\label{section:conclusion}
In this work, we introduced a benchmark dataset, \data{}, to more reliably evaluate models on the task of \taskcapital{}.
We conducted an extensive study on different methods for the task, and showed that domain-specific and synthetic data can be used to train state-of-the-art models for the task.
In addition to standard evaluation methods, we provided a fine-grained analysis for different types of entity mentions.
We found that the first stage, \taskone{}, plays a critical role in the performance of the downstream application and that implicit paraphrases are the most challenging to identify.
We also observed that larger models showed better performance, even for implicit mentions, on \tasktwo{}.
Future work could generate more diverse mention types using LLMs to extend the training data.
Our proposed system could also be trained end-to-end, which could help reduce the errors that get propagated from the first stage.
Finally, providing a high-quality and clean KB of survey items could be a practical contribution that would significantly improve performance on our task.

\section{Limitations}\label{section:limitations}
Our work on \taskabbrev{} has limitations, which we describe below.
While we removed any overlapping questions and research data between \data{} and \datasim{}, there may be questions in \datasim{} that are very similar (i.e., use different wording or are semantically similar) to those annotated in \data{}.
However, the KB should be accessible to models in downstream applications.
As such, memorization of the entities in the KB could be beneficial.
In addition, \datasim{} only includes surveys that are indexed by GESIS.
We do not conduct EL for entities that are not present in the KB.
Furthermore, the dataset that we presented in this work, while covering many documents, topics, and surveys, may not be representative of all of social sciences.
It is important to further increase the dataset size and breadth.
Finally, given that the task is user-oriented, evaluation should ideally also consider the utility of the presented algorithms to users, which is not done in this work.

Since social science research is frequently published in different languages, methods should ideally be multilingual.
We aim to expand our dataset to more languages in the future.
Developers of language tools for social science research should carefully evaluate the fairness of their systems and their impact on research.
Fairness with respect to the identified survey items is important, because researchers, relying on the output of automated systems, could overlook less popular research data or survey items containing less common words.
Future work could evaluate these dimensions in more detail.

\backmatter

\bmhead{Supplementary information}
The annotation guideline is provided as supplementary material.

\bmhead{Acknowledgements}
We thank Sotaro Takeshita for their helpful feedback.
The authors acknowledge support by the state of Baden-W\"urttemberg through bwHPC and the German Research Foundation (DFG) through grant INST 35/1597-1 FUGG.

\bmhead{Funding}
This work is supported by the German Research Foundation (DFG) project VADIS, grant numbers ZA 939/5-1, PO 1900/5-1, EC 477/7-1, KR 4895/3-1.

\bmhead{Conflict of interest}
The authors have no relevant financial interests to disclose.
Two of the authors are involved in the VADIS Project (\href{https://vadis-project.github.io/}{https://vadis-project.github.io/}), which collaborates with GESIS, Fraunhofer ISI, and Hochschule Mannheim.

\bmhead{Data availability}
The benchmark dataset introduced in this work, \data{}, is available under an open license at \url{https://dx.doi.org/10.5281/zenodo.11397370}.

\bmhead{Code availability}
The code used in our experiments and the scripts to reproduce the \datasim{} knowledge base and the S44k corpus is available under an open license at \url{https://github.com/e-tornike/SIL}.

\bmhead{Author Contributions}
TT conceptualized, supervised, and conducted the annotation task; processed the datasets; conducted the experiments and the evaluation; and wrote the full paper.
SP supervised the entire project and helped with ideation.
DR supported with the writing.

\bibliography{sn-article}
\clearpage

\begin{appendices}

\section{Dataset Details}
\label{section:appendix}

\subsection{Entity Mention Examples}
\label{subsec:appendix-mention-examples}
Table~\ref{table:more-survey-item-mention-examples} provides additional examples of mentions.
The first sentence implicitly paraphrases a survey item measuring the justification of divorce, while the second sentence paraphrases three items explicitly and one implicitly (the item measuring the type of community a respondent lives in is paraphrased as \textit{urbanisation}.

\begin{table}
    \centering
    \footnotesize
    \caption{Example sentences from publications that mention survey items. The associated \textit{type} and \textit{subtype} terms indicate the semantic classes the mentions are labeled with. Implicit types require additional context, while explicit types do not. As for the subtypes, verbatim quotations include exact wording of survey items, while paraphrases summarize the content}
    \label{table:more-survey-item-mention-examples}
    \begin{tabularx}{\columnwidth}{l|X}
         \toprule
         \textbf{Sentence} & \textit{As the core indicator of the social acceptance of divorce, we used the item that focused on divorce within the set of questions measuring the justification of different kinds of social action.} \\
         \textbf{Survey Items} & {\fontfamily{cmtt}\selectfont \href{https://search.gesis.org/variables/exploredata-ZA7503_VarF121}{ZA7503\_VarF121}} \\
         \textbf{Type} & Implicit \\
         \textbf{Subtype} & Paraphrase \\
         \textbf{Document ID} & \href{https://www.ssoar.info/ssoar/handle/document/79550}{79550}~\citep{Fučík2020} \\
         \midrule
         \textbf{Sentence} & \textit{Demographic factors included gender, age, urbanisation and marital status.} \\
         \textbf{Survey Items} & {\fontfamily{cmtt}\selectfont \href{https://search.gesis.org/variables/exploredata-ZA4977_Varv330}{ZA4977\_Varv330}}, {\fontfamily{cmtt}\selectfont \href{https://search.gesis.org/variables/exploredata-ZA4977_Varv331}{ZA4977\_Varv331}}, {\fontfamily{cmtt}\selectfont \href{https://search.gesis.org/variables/exploredata-ZA4977_Varv334}{ZA4977\_Varv334}}, {\fontfamily{cmtt}\selectfont \href{https://search.gesis.org/variables/exploredata-ZA4977_Varv326}{ZA4977\_Varv326}} \\
         \textbf{Type} & Explicit, Explicit, Implicit, Explicit \\
         \textbf{Subtype} & Paraphrase \\
         \textbf{Document ID} & \href{https://www.ssoar.info/ssoar/handle/document/74709}{74709}~\citep{Kino2017} \\
         \botrule
    \end{tabularx}
\end{table}

\subsection{\data{} Document Details}
\label{subsec:appendix-SILD-statistics}
Each publication on SSOAR is tagged with keywords that summarize the fine-grained topics (e.g., public opinion) and classifications that cover broader topics (e.g., political process).
Tables~\ref{table:SILD-domain-keywords-statistics} and \ref{table:SILD-domain-classifications-statistics} show the distribution of the ten most common keywords and classifications that are associated with the documents in \data{} compared with the ratios of the terms in the full SSOAR collection.
The similarity between the distributions, based on the Jensen-Shannon divergence, is moderate ($0.32$) for keywords and high ($0.11$) for classifications.

\begin{table}[ht]
    \centering
    \footnotesize
    \caption{Distribution (in \%) of the ten most common paper classifications for the documents in \data{} compared with the ratios in SSOAR. Documents may have more than one classification. \data{} covers 45 of the 1,592 classifications in SSOAR}
    \label{table:SILD-domain-keywords-statistics}
    \begin{tabularx}{\columnwidth}{X|r|r}
        \toprule
        \textbf{Terms} & \textbf{\data{}} & \textbf{SSOAR} \\
        \midrule
               Attitude &  12 &    2 \\
      Fed. Rep. of Ger. &  10 &   15 \\
        Int. comparison &   7 &    1 \\
                     EU &   6 &    4 \\
        Voting behavior &   5 &    $<$1 \\
         Public opinion &   5 &    $<$1 \\
      Social inequality &   5 &    1 \\
     Political attitude &   4 &    $<$1 \\
             Perception &   4 &    $<$1 \\
        EU member state &   4 &    $<$1 \\
        \botrule
    \end{tabularx}
\end{table}

\begin{table}[ht]
    \centering
    \footnotesize
    \caption{Distribution (in \%) of the ten most common paper keywords for the documents in \data{} compared with the ratios in SSOAR. Documents may have more than one keyword. \data{} covers 502 unique keywords from the 59k in SSOAR}
    \label{table:SILD-domain-classifications-statistics}
    \begin{tabularx}{\columnwidth}{X|r|r}
        \toprule
        \textbf{Terms} & \textbf{\data{}} & \textbf{SSOAR} \\
        \midrule
        Political process &  32 &   12 \\
        Social psychology &  14 &    3 \\
Data collection \& analysis &  13 &    7 \\
        General sociology &   9 &    4 \\
        European politics &   8 &    4 \\
                  Ecology &   7 &    2 \\
         Family sociology &   7 &    3 \\
            Health policy &   7 &    2 \\
    Sociology of religion &   6 &    1 \\
           Macrosociology &   4 &    1 \\
        \botrule
    \end{tabularx}
\end{table}

\subsection{\data{}-Rand Statistics}
\label{subsec:appendix-SILD-rand}
We provide dataset statistics for the randomly-sampled version of the \data{} dataset (Tables~\ref{table:SILD-Rand-split-statistics} and \ref{table:SILD-Rand-fine-grained-statistics}).
The distribution for entity mentions is comparable to the difficult version, \data{}-Diff.

\begin{table}
    \centering
    \footnotesize
    \caption{Dataset statistics for the \data{}-Rand dataset split. The first two rows describe the number of sentences with$^+$ and without$^-$ survey item mentions. The third and fourth rows list the number of total and unique survey items, respectively. The bottom two rows include the number of surveys referenced and papers in the dataset}
    \label{table:SILD-Rand-split-statistics}
    \begin{tabularx}{\columnwidth}{X|r|r|r|r|r}
        \toprule
         & \multirow[c]{2}{*}{\textbf{Train}} & \multirow[c]{2}{*}{\textbf{Dev}} & \multicolumn{2}{c|}{\textbf{Test}}  & \multirow[c]{2}{*}{\textbf{Total}} \\
        \\[-2ex]
         & & & \multicolumn{1}{c|}{\textbf{EN}} & \multicolumn{1}{c|}{\textbf{DE}} & \\
        \midrule
        \textbf{Sents.$^+$} & 397 & 75 & 204 & 107 & 783 \\
        \textbf{Sents.$^-$} & 10,367 & 1,950 & 4,772 & 2,582 & 19,671 \\
        \midrule
        \textbf{Survey items} & 944 & 168 & 547 & 162 & 1,796 \\
        \textbf{Unique} & 697 & 88 & 434 & 126 & 1,283 \\
        \midrule
        \textbf{Surveys} & 45 & 8 & 45 & 16 & 97 \\
        \midrule
        \textbf{Papers} & 50 & 10 & 24 & 16 & 100 \\
        \botrule
    \end{tabularx}
\end{table}

\begin{table*}
    \centering
    \footnotesize
    \caption{Fine-grained statistics across diagnostic elements (type and subtype) for the \data{}-Rand dataset. The statistics are provided on both the level of sentences and of survey items. Sentences that contain only a single category are provided in the respective category rows. Sentences that contain more than one category are described in the \textit{mixed} rows. On the level of survey items, each mention is associated with a single type and subtype category. Statistics for training and development sets are aggregated}
    \label{table:SILD-Rand-fine-grained-statistics}
    \begin{tabular}{cl|r|r|r|r|r|r|r|r}
        \toprule
        & & \multicolumn{4}{c|}{\textbf{Sentences}} & \multicolumn{4}{c}{\textbf{Survey Items}} \\
        \\[-2ex]
        & & \textbf{Train +} & \multicolumn{2}{c|}{\textbf{Test}} & \multirow[c]{2}{*}{\textbf{Total}} & \textbf{Train +} & \multicolumn{2}{c|}{\textbf{Test}} & \multirow[c]{2}{*}{\textbf{Total}} \\
        & & \textbf{Dev} & \textbf{EN} & \textbf{DE} & & \textbf{Dev} & \textbf{EN} & \textbf{DE} & \\
        \midrule
        \parbox[t]{2mm}{\multirow{4}{*}{\rotatebox[origin=c]{90}{\textbf{Type}}}} & \textbf{Explicit} & 294 & 147 & 76 & 517 & 707 & 406 & 119 & 1232 \\
        & \textbf{Implicit} & 148 & 43 & 23 & 214 & 402 & 135 & 42 & 579 \\
        & \textbf{Mixed} & 15 & 2 & 3 & 20 & - & - & - & - \\
        & \textbf{Other} & 15 & 12 & 5 & 32 & 2 & 6 & 0 & 8 \\
        \midrule
        \parbox[t]{2mm}{\multirow{7}{*}{\rotatebox[origin=c]{90}{\textbf{Subtype}}}} & \textbf{Quotation} & 110 & 57 & 35 & 202 & 244 & 164 & 55 & 463 \\
        & \textbf{Paraphrase} & 279 & 110 & 51 & 440 & 767 & 341 & 87 & 1195 \\
        & \textbf{Lexical inference} & 11 & 5 & 3 & 19 & 30 & 11 & 10 & 51 \\
        & \textbf{Unspecified} & 12 & 3 & 0 & 15 & 34 & 4 & 0 & 38 \\
        & \textbf{Citation} & 8 & 1 & 7 & 16 & 8 & 1 & 7 & 16 \\
        & \textbf{Mixed} & 19 & 6 & 4 & 29 & - & - & - & - \\
        & \textbf{Other} & 33 & 22 & 7 & 62 & 28 & 26 & 2 & 56 \\
        \midrule
        & \textbf{Total} & 472 & 204 & 107 & 783 & 1111 & 547 & 161 & 1819 \\
        \botrule
    \end{tabular}
\end{table*}

\subsection{\datasim{} Statistics}
\label{subsec:appendix-gsim-statistics}
\begin{figure*}[!ht]
    \centering
    \includegraphics[scale=0.5]{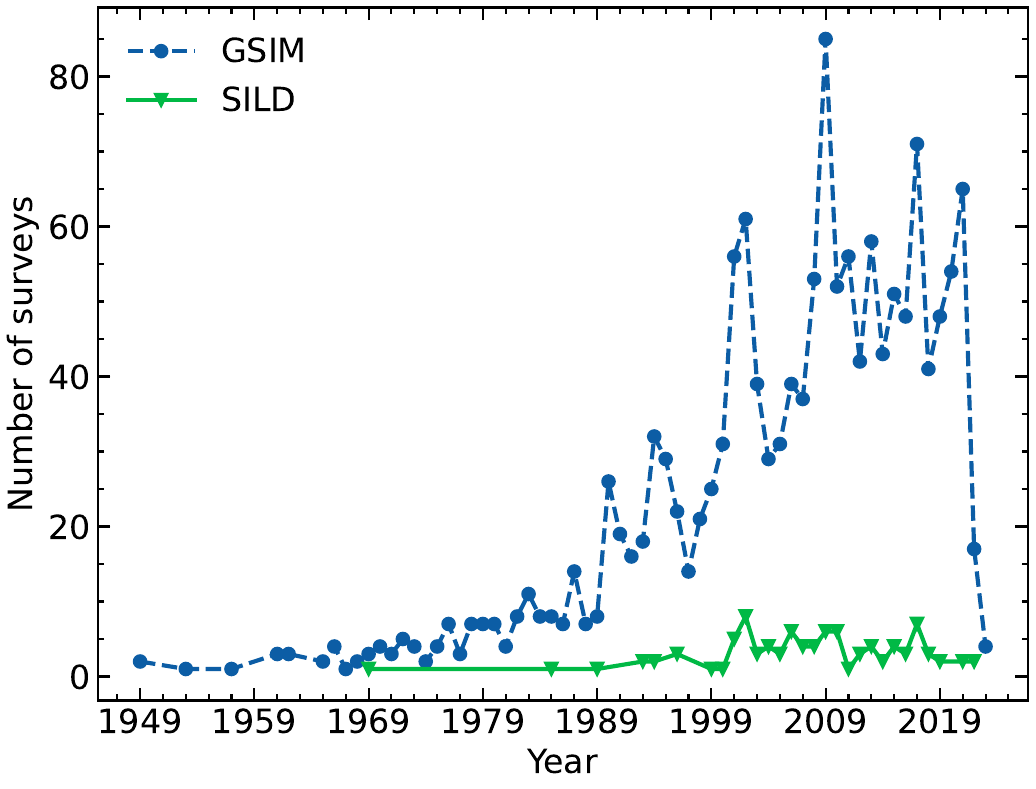}
    \caption{Comparison between \data{} and \datasim{} with respect to the number of surveys published each year}
    \label{figure:suveys-published-timeline}
\end{figure*}

\datasim{} contains many duplicates, which we filter across different metadata values.
After filtering, there are over 59k unique questions and 97k unique sub-questions that cover over 2k topics.
Most of the surveys in \datasim{} were published after 2000, but surveys published as early as 1949 are also included.
We plot the distribution of publication dates of the surveys in Figure~\ref{figure:suveys-published-timeline}.
By comparison, the surveys referenced in \data{} contain close to 8k questions and over 25k sub-questions, covering 510 topics.
Similar to \datasim{}, most survey data in \data{} is published after 2000, with the earliest publication in 1969.
While it appears that the number of surveys drops significantly for the years 2022 and 2023, this is an artifact of the processing and indexing pipelines used by GESIS, for which the most recent surveys are not yet linked.

\section{Additional Experiments}
\subsection{More baselines for \taskoneabbrev{}}
\label{subsec:appendix-md-results-full}
Logistic regression (LR)~\citep{RePEc:tin:wpaper:20020119} and support vector machine (SVM)~\citep{10.1023/A:1022627411411} are traditional classification models.
We compare performance for LR as well as linear and non-linear variants of SVM.
For both model families, we use term frequency–inverse document frequency (TF-IDF) lexical features, which measure the importance of a word in a collection.
As a preprocessing step, we lowercase all words in the dataset.
Because TF-IDF features are computed on the words in the training set, German words are never seen.
Thus, we do not evaluate these models on the German test set.
The first group in Table~\ref{table:md-results-full} shows the performance for each of the lexical models.
Even though LR has significantly higher precision, non-linear SVM outperforms the other variants in terms of recall and $F_1$.
The limited dataset size and the difference in topics (described in Section~4.2 of the source publication) between the training and test sets could be the cause of the low recall.
The table additionally includes DeBERTa and its multilingual version~\citep{he2021deberta}, which are larger than BERT and RoBERTa.
DeBERTA$_{\text{large}}$ shows the strongest performance for English and even outperforms mDeBERTa$_{\text{base}}$ on German.
Thus, the pre-training data for DeBERTa may include German data.

\begin{table*}
    \centering
    \footnotesize
    \caption{Precision ($P$), recall ($R$), and $F_1$ scores for logistic regression (LR), support vector machine (SVM), and transformer-based classifiers trained on \data{}. Models are divided into lexical (Lex.), monolingual (Mono), multilingual (Mul), and domain-specialized (DS) groups. Models are evaluated on English and German test sets independently (except for the lexical systems). For each group, the highest score in each column is highlighted in bold and statistically significant results are marked with an asterisk. The subscript indicates the standard deviation across multiple seeds. 
    Multilingual models are able to transfer their learned knowledge from English to German.
    Domain-specialized models significantly outperform their base counterparts}
    \label{table:md-results-full}
    \begin{tabular}{cl|lll|lll|l}
        \toprule
        & & \multicolumn{3}{c|}{\textbf{English}} & \multicolumn{3}{c|}{\textbf{German}} & \multicolumn{1}{c}{\textbf{Total}} \\
        & \textbf{Model} & \multicolumn{1}{c}{\textbf{$P$}} & \multicolumn{1}{c}{\textbf{$R$}} & \multicolumn{1}{c|}{\textbf{$F_1$}} & \multicolumn{1}{c}{\textbf{$P$}} & \multicolumn{1}{c}{\textbf{$R$}} & \multicolumn{1}{c|}{\textbf{$F_1$}} & \multicolumn{1}{c}{\textbf{$F_1$}} \\
        \midrule
        \parbox[t]{2mm}{\multirow{3}{*}{\rotatebox[origin=c]{90}{\textbf{Lex.}}}} & \textbf{LR} & \textbf{86.4}$^{*}_{\pm4}$ & 7.8$_{\pm4}$ & 14.0$_{\pm6}$ & - & - & - & - \\
        & \textbf{SVM$_{\text{lin.}}$} & 76.7$_{\pm1}$ & 15.3$_{\pm1}$ & 25.5$_{\pm2}$ & - & - & - & - \\
        & \textbf{SVM$_{\text{non-lin.}}$} & 66.7$_{\pm3}$ & \textbf{19.3}$^{*}_{\pm1}$ & \textbf{29.9}$^{*}_{\pm1}$ & - & - & - & - \\
        \midrule
        \parbox[t]{2mm}{\multirow{9}{*}{\rotatebox[origin=c]{90}{\textbf{Mono.}}}} & \textbf{BERT$_{\text{base}}$} & 73.4$_{\pm6}$ & 48.1$_{\pm3}$ & 57.9$_{\pm1}$ & 4.2$_{\pm0}$ & \textbf{86.4}$_{\pm10}$ & 8.0$_{\pm0}$ & 15.3$_{\pm1}$ \\
        & \textbf{BERT$_{\text{large}}$} & \textbf{76.3}$_{\pm2}$ & 49.3$_{\pm2}$ & 59.9$_{\pm2}$ & 4.9$_{\pm1}$ & 58.7$_{\pm12}$ & 9.0$_{\pm1}$ & 20.3$_{\pm3}$ \\
        & \textbf{RoBERTa$_{\text{base}}$} & 67.8$_{\pm2}$ & \textbf{54.1}$_{\pm2}$ & 60.1$_{\pm2}$ & 6.0$_{\pm2}$ & 72.5$_{\pm11}$ & 10.9$_{\pm3}$ & 22.1$_{\pm6}$ \\
        & \textbf{RoBERTa$_{\text{large}}$} & 71.6$_{\pm5}$ & 51.4$_{\pm6}$ & 59.3$_{\pm2}$ & 42.8$_{\pm20}$ & 40.0$_{\pm13}$ & 34.7$_{\pm12}$ & 48.0$_{\pm13}$ \\
        & \textbf{DeBERTa$_{\text{base}}$} & 71.9$_{\pm5}$ & 47.5$_{\pm4}$ & 56.9$_{\pm2}$ & 39.1$_{\pm24}$ & 44.9$_{\pm12}$ & 34.5$_{\pm10}$ & 46.3$_{\pm9}$ \\
        & \textbf{DeBERTa$_{\text{large}}$} & 72.2$_{\pm3}$ & \textbf{54.1}$_{\pm1}$ & \textbf{61.8}$_{\pm1}$ & \textbf{73.7}$^{*}_{\pm4}$ & 48.2$_{\pm7}$ & \textbf{57.8}$^{*}_{\pm4}$ & \textbf{60.7}$^{*}_{\pm2}$ \\
        & \textbf{SciBERT} & 68.8$_{\pm3}$ & 51.0$_{\pm2}$ & 58.5$_{\pm1}$ & 27.6$_{\pm8}$ & 40.4$_{\pm8}$ & 31.1$_{\pm4}$ & 47.3$_{\pm3}$ \\
        & \textbf{SPECTER} & 71.0$_{\pm3}$ & 51.8$_{\pm2}$ & 59.8$_{\pm1}$ & 29.7$_{\pm15}$ & 37.2$_{\pm18}$ & 25.9$_{\pm5}$ & 45.4$_{\pm7}$ \\
        & \textbf{SsciBERT} & 72.6$_{\pm1}$ & 47.2$_{\pm1}$ & 57.2$_{\pm1}$ & 4.5$_{\pm0}$ & 80.0$_{\pm11}$ & 8.5$_{\pm0}$ & 16.5$_{\pm1}$ \\
        \midrule
        \parbox[t]{2mm}{\multirow{4}{*}{\rotatebox[origin=c]{90}{\textbf{Mul.}}}} & \textbf{mBERT$_{\text{base}}$} & 70.5$_{\pm5}$ & 42.8$_{\pm2}$ & 53.1$_{\pm1}$ & 60.6$_{\pm7}$ & 33.8$_{\pm8}$ & 42.4$_{\pm6}$ & 49.9$_{\pm2}$ \\
        & \textbf{XLM-R$_{\text{base}}$} & \textbf{71.8}$_{\pm2}$ & 49.5$_{\pm3}$ & 58.5$_{\pm2}$ & 69.5$_{\pm3}$ & 44.1$_{\pm1}$ & 53.9$_{\pm1}$ & 57.1$_{\pm1}$ \\
        & \textbf{XLM-R$_{\text{large}}$} & 69.9$_{\pm4}$ & \textbf{54.9}$_{\pm2}$ & \textbf{61.4}$_{\pm1}$ & \textbf{70.0}$_{\pm2}$ & \textbf{61.1}$_{\pm5}$ & \textbf{65.1}$^{*}_{\pm3}$ & \textbf{62.6}$^{*}_{\pm1}$ \\
        & \textbf{mDeBERTa$_\text{base}$} & 71.3$_{\pm2}$ & 47.9$_{\pm4}$ & 57.2$_{\pm3}$ & 66.7$_{\pm4}$ & 42.4$_{\pm5}$ & 51.7$_{\pm5}$ & 55.5$_{\pm3}$ \\
        \midrule
        \parbox[t]{2mm}{\multirow{2}{*}{\rotatebox[origin=c]{90}{\textbf{DS}}}} & \textbf{\modelsoscibert{}$_{\text{base}}$} & \textbf{69.6}$_{\pm3}$ & 48.3$_{\pm5}$ & 56.8$_{\pm3}$ & 64.5$_{\pm12}$ & \textbf{57.6}$_{\pm9}$ & 59.1$_{\pm2}$ & 57.5$_{\pm2}$ \\
        & \textbf{\modelsoscixlmr{}$_{\text{base}}$} & \textbf{69.6}$_{\pm4}$ & \textbf{53.9}$_{\pm2}$ & \textbf{60.7}$_{\pm1}$ & \textbf{71.4}$_{\pm5}$ & 55.1$_{\pm6}$ & \textbf{61.8}$_{\pm3}$ & \textbf{61.0}$_{\pm1}$ \\
        \botrule
    \end{tabular}
\end{table*}

\subsection{\data{}-Diff vs \data{}-Rand}\label{subsec:appendix-diff-vs-rand}
Given that we split \data{} into two datasets of varying difficulty, we evaluate performance differences between the two (Table~\ref{table:diff-vs-rand}).
For \taskoneabbrev{}, performance is comparable, with slight differences in precision and recall across Languages.
A closer look at the German test set, which is the same for both splits, shows that the models trained on \data{}-Rand achieve a higher recall.
We hypothesize that this is due to the high quality data points that were introduced into the training dataset, which were originally in the test set.
We recommend using the \data{}-Diff split for evaluating challenging cases.

\begin{table*}
    \centering
    \footnotesize
    \caption{Precision ($P$), recall ($R$), and $F_1$ scores for XLM-R$_{\text{large}}$ fine-tuned on \data{}-Diff and \data{}-Rand. Models are evaluated on English and German test sets independently. The highest score in each column is highlighted in bold. 
    For English, there are no statistically significant differences. XLM-R$_{\text{large}}$ has the highest $F_1$ score overall, including for German. Furthermore, training models on social science publications significantly improves performance}
    \label{table:diff-vs-rand}
    \begin{tabular}{l|lll|lll|l}
        \toprule
        & \multicolumn{3}{c|}{\textbf{English}} & \multicolumn{3}{c|}{\textbf{German}} & \multicolumn{1}{c}{\textbf{Total}} \\
        \textbf{Data} & \multicolumn{1}{c}{\textbf{$P$}} & \multicolumn{1}{c}{\textbf{$R$}} & \multicolumn{1}{c|}{\textbf{$F_1$}} & \multicolumn{1}{c}{\textbf{$P$}} & \multicolumn{1}{c}{\textbf{$R$}} & \multicolumn{1}{c|}{\textbf{$F_1$}} & \multicolumn{1}{c}{\textbf{$F_1$}} \\
        \midrule
        \textbf{Diff} & 69.8$_{\pm3}$ & \textbf{62.2}$_{\pm3}$ & \textbf{65.6}$_{\pm2}$ & \textbf{73.5}$_{\pm6}$ & 55.7$_{\pm3}$ & 63.2$_{\pm3}$ & \textbf{64.8}$_{\pm1}$ \\
        \textbf{Rand} & \textbf{69.9}$_{\pm4}$ & 54.9$_{\pm2}$ & 61.4$_{\pm1}$ & 70.0$_{\pm2}$ & \textbf{61.1}$_{\pm5}$ & \textbf{65.1}$_{\pm3}$ & 62.6$_{\pm1}$ \\
        \botrule
    \end{tabular}
\end{table*}

\subsection{k-NN Classification}\label{subsec:appendix-kNN}
Because we have access to a large KB of survey item descriptions, we can use clustering-based methods to classify unseen instances based on the distance to the ones in the KB.
Clustering methods, which deal with finding clusters of similar objects in data, are commonly used for text classification~\citep{Aggarwal2012,DBLP:journals/corr/AllahyariPASTGK17a}.
The k-NN algorithm is a standard baseline, which uses vectors as features to find clusters based on nearest neighbors.
We use k-NN with sparse (TF-IDF) and dense (transformer-based embeddings) features.
We compare encoding the training data with encoding the KB, \datasim{}, as the retrieval corpus. 
During inference, we map each test instance to a cluster and its majority label.
Similar to the lexical baselines, we tune hyperparameters, such as the number of neighbors, the algorithm used to compute the nearest neighbors, the leaf size, the weight function, and the distance metric (either Manhattan or Euclidean) for 20 trials.
We observe that clustering alone cannot achieve high precision (Table~\ref{table:nnc-results}).
Using \datasim{} improves recall at the cost of precision.
Future work could incorporate the KB into \taskoneabbrev{} in end-to-end methods.

\begin{table*}
    \centering
    \footnotesize
    \caption{Precision ($P$), recall ($R$), and $F_1$ scores for nearest-neighbor classifiers trained on \data{} using different encoder models. Models use either \data{} (top) or \datasim{} (bottom) as the retrieval corpus, and are evaluated on English and German test sets independently. For each group, the highest score in each column is highlighted in bold. Nearest-neighbor classification alone is insufficient, but can achieve moderate recall. The retrieval-specialized model, multilingual E5 (mE5), performs best using both corpora}
    \label{table:nnc-results}
    \begin{tabular}{cl|lll|lll|l}
        \toprule
        & & \multicolumn{3}{c|}{\textbf{English}} & \multicolumn{3}{c|}{\textbf{German}} & \multicolumn{1}{c}{\textbf{Total}} \\
        \\[-2ex]
        & \textbf{Model} & \multicolumn{1}{c}{\textbf{$P$}} & \multicolumn{1}{c}{\textbf{$R$}} & \multicolumn{1}{c|}{\textbf{$F_1$}} & \multicolumn{1}{c}{\textbf{$P$}} & \multicolumn{1}{c}{\textbf{$R$}} & \multicolumn{1}{c|}{\textbf{$F_1$}} & \multicolumn{1}{c}{\textbf{$F_1$}} \\
        \midrule
        \parbox[t]{2mm}{\multirow{4}{*}{\rotatebox[origin=c]{90}{\textbf{\data{}}}}} & \textbf{TF-IDF} & 42.8 & 23.6 & 28.8 & - & - & - & - \\
        & \textbf{mBERT} & 58.6 & 28.4 & 37.1 & 47.8 & 34.7 & 35.1 & 35.9 \\
        & \textbf{MiniLM} & 47.6 & 21.7 & 27.2 & 31.9 & 12.2 & 15.8 & 23.7 \\
        & \textbf{mE5$_{\text{large}}$} & 55.5 & 28.0 & 36.2 & 45.5 & 32.0 & \textbf{35.7} & 35.9 \\
        \midrule
        \parbox[t]{2mm}{\multirow{4}{*}{\rotatebox[origin=c]{90}{\textbf{\datasim{}}}}} & \textbf{TF-IDF} & \textbf{62.7} & 13.5 & 22.2 & - & - & - & - \\
        & \textbf{mBERT} & 35.7 & 44.7 & 39.7 & \textbf{54.5} & 16.8 & 25.7 & 36.8 \\
        & \textbf{MiniLM} & 26.0 & 51.5 & 34.5 & 23.6 & \textbf{49.5} & 31.9 & 33.7 \\
        & \textbf{mE5$_{\text{large}}$} & 34.9 & \textbf{60.8} & \textbf{44.3} & 44.9 & 20.6 & 28.2 & \textbf{41.2} \\
        \botrule
    \end{tabular}
\end{table*}

\subsection{Retrieval-Augmented Classification}
\label{subsec:appendix-rac}
We next extend k-NN classification by complementing it with a fine-tuned transformer (XLM-R$_{\text{large}}$).
For simplicity, we train the k-NN and transformer separately, and linearly combine predictions during inference, giving equal weight to each classifier.
We observe that combining these classifiers leads to a slight improvement in precision (Table~\ref{table:rac-results}), however, recall significantly degrades.
Given that a high recall is desired for \taskone{}, this method is not viable.

\begin{table*}
    \centering
    \footnotesize
    \caption{Precision ($P$), recall ($R$), and $F_1$ scores retrieval-augmented classification using XLM-RoBERTa$_{\text{large}}$ fine-tuned on \data{} as the base classifier. Models are evaluated on English and German test sets independently. The highest score in each column is highlighted in bold. Augmenting the classifier with a retriever leads to a drop in performance for both models across all metrics (except precision for German)}
    \label{table:rac-results}
    \begin{tabular}{l|lll|lll|l}
        \toprule
        & \multicolumn{3}{c|}{\textbf{English}} & \multicolumn{3}{c|}{\textbf{German}} & \multicolumn{1}{c}{\textbf{Total}} \\
        \textbf{Model} & \multicolumn{1}{c}{\textbf{$P$}} & \multicolumn{1}{c}{\textbf{$R$}} & \multicolumn{1}{c|}{\textbf{$F_1$}} & \multicolumn{1}{c}{\textbf{$P$}} & \multicolumn{1}{c}{\textbf{$R$}} & \multicolumn{1}{c|}{\textbf{$F_1$}} & \multicolumn{1}{c}{\textbf{$F_1$}} \\
        \midrule
        \textbf{-} & 76.6 & \textbf{51.1} & \textbf{61.3} & 70.7 & \textbf{60.7} & \textbf{65.3} & \textbf{62.6} \\
        \textbf{mBERT} & 78.2 & 39.2 & 52.2 & \textbf{76.4} & 51.4 & 61.5 & 55.3 \\
        \textbf{mE5} & \textbf{78.5} & 43.0 & 55.6 & 71.4 & 42.1 & 52.9 & 54.7 \\
        \botrule
    \end{tabular}
\end{table*}

\subsection{Data Augmentation Prompt}
\label{subsec:appendix-da-prompt}
\begin{figure}[!ht]
    \centering
    \includegraphics[scale=0.3]{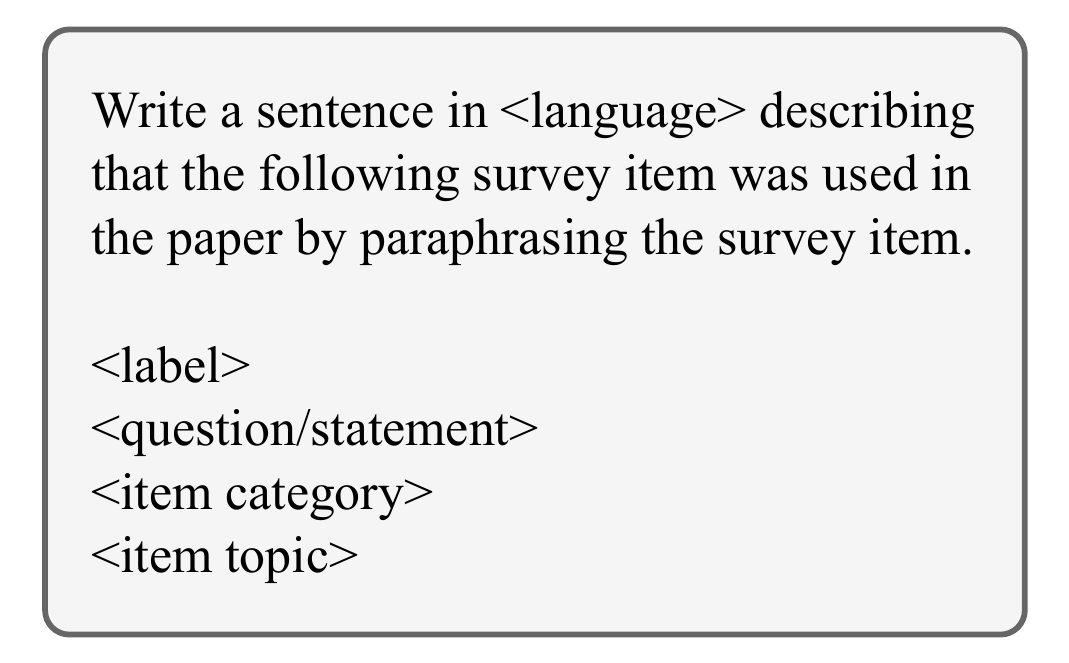}
    \caption{The prompt used for generating data using Vicuna-7B. The placeholder values indicate metadata from survey items that is replaced in the prompt}
    \label{figure:da-prompt}
\end{figure}

Figure~\ref{figure:da-prompt} shows the prompt used to generate 400k synthetic entity mentions.
The prompt includes label, question, item category, and topic metadata from survey items.
We randomly swap the language in the prompt to generate an equal number of English and German data.

\subsection{In-Context Learning}\label{subsec:appendix-icl}
With LLMs increasing in size and capabilities, ICL has recently become a promising paradigm for many NLP tasks~\citep{xie2022an,garg2023transformers,min-etal-2022-rethinking}.
Given the complexity and difficulty of identifying texts defining survey items, we investigate the potential benefits of ICL for \taskoneabbrev{}.
We formulate a prompt template that takes random positive examples (i.e., demonstrations) from \data{} and negative examples from \datasim{}.
We then prompt the model to generate binary answers for batches of examples.
We vary the number of demonstrations to evaluate the impact of the context.
We use the open-source models Mistral-7B (instruction fine-tuned) and Mixtral-8x7B due to their small size and strong performance compared to larger models~\citep{jiang2023mistral}.
Instruction fine-tuned models are only comparable to lexical baselines in our setting for \taskoneabbrev{} (Table~\ref{table:icl-results}).
Because our task does not have many demonstrations on the Internet, which LLMs can learn from during pre-training, we conclude that ICL is not a viable option for \taskoneabbrev{}.

\begin{table*}[h!]
    \centering
    \footnotesize
    \caption{Precision ($P$), recall ($R$), and $F_1$ scores for in-context learning using LLMs. Mistral-7B is an instruction fine-tuned model, while Mixtral8x7B is a mixture-of-experts model. The models are prompted with either 0, 10, 20, 50, or 100 examples (the subscript of each model), half of which contain survey item mentions. The highest score in each column is highlighted in bold. Increasing the number of examples in the prompt leads to mixed results with high variance. Mixtral-8x7B consistently outperforms Mistral-7B}
    \label{table:icl-results}
    \begin{tabular}{l|rrr|rrr|l}
        \toprule
        & \multicolumn{3}{c|}{\textbf{English}} & \multicolumn{3}{c|}{\textbf{German}} & \multicolumn{1}{c}{\textbf{Total}} \\
        \\[-2ex]
        \textbf{Model} & \multicolumn{1}{c}{\textbf{$P$}} & \multicolumn{1}{c}{\textbf{$R$}} & \multicolumn{1}{c|}{\textbf{$F_1$}} & \multicolumn{1}{c}{\textbf{$P$}} & \multicolumn{1}{c}{\textbf{$R$}} & \multicolumn{1}{c|}{\textbf{$F_1$}} & \multicolumn{1}{c}{\textbf{$F_1$}} \\
        \midrule
        \textbf{Mistral-7B$_{0}$} & 9.8$_{\pm0}$ & \textbf{80.7}$^*_{\pm0}$ & 17.5$_{\pm0}$ & 11.9$_{\pm0}$ & 48.6$_{\pm1}$ & 19.2$_{\pm0}$ & 17.8$_{\pm0}$ \\
        \textbf{Mistral-7B$_{10}$} & 18.2$_{\pm6}$ & 58.1$_{\pm15}$ & 26.0$_{\pm5}$ & 23.3$_{\pm8}$ & 35.3$_{\pm13}$ & 24.3$_{\pm5}$ & 25.6$_{\pm5}$ \\
        \textbf{Mistral-7B$_{20}$} & 18.3$_{\pm4}$ & 56.5$_{\pm16}$ & 26.2$_{\pm3}$ & 20.5$_{\pm5}$ & 38.7$_{\pm7}$ & 25.6$_{\pm2}$ & 26.1$_{\pm3}$ \\
        \textbf{Mistral-7B$_{50}$} & 14.6$_{\pm5}$ & 58.7$_{\pm11}$ & 22.5$_{\pm6}$ & 15.3$_{\pm6}$ & 49.7$_{\pm16}$ & 21.4$_{\pm5}$ & 22.1$_{\pm5}$ \\
        \textbf{Mistral-7B$_{100}$} & 8.7$_{\pm2}$ & 72.3$_{\pm5}$ & 15.4$_{\pm2}$ & 7.6$_{\pm2}$ & \textbf{73.5}$^*_{\pm6}$ & 13.6$_{\pm3}$ & 14.8$_{\pm3}$ \\
        \midrule
        \textbf{Mixtral-8x7B$_{0}$} & 9.6$_{\pm0}$ & 67.4$_{\pm1}$ & 16.9$_{\pm0}$ & 21.1$_{\pm1}$ & 35.1$_{\pm2}$ & \textbf{26.3}$_{\pm1}$ & 18.1$_{\pm0}$ \\
        \textbf{Mixtral-8x7B$_{10}$} & \textbf{38.4}$_{\pm14}$ & 36.1$_{\pm17}$ & 32.1$_{\pm2}$ & \textbf{57.4}$_{\pm25}$ & 22.1$_{\pm15}$ & 24.7$_{\pm9}$ & 32.9$_{\pm2}$ \\
        \textbf{Mixtral-8x7B$_{20}$} & 37.2$_{\pm8}$ & 36.6$_{\pm2}$ & \textbf{36.3}$_{\pm4}$ & 54.1$_{\pm14}$ & 16.8$_{\pm9}$ & 22.4$_{\pm7}$ & \textbf{35.2}$_{\pm4}$ \\
        \textbf{Mixtral-8x7B$_{50}$} & 38.1$_{\pm9}$ & 35.1$_{\pm9}$ & 35.0$_{\pm4}$ & 41.0$_{\pm9}$ & 20.6$_{\pm8}$ & 26.1$_{\pm8}$ & 33.4$_{\pm4}$ \\
        \botrule
    \end{tabular}
\end{table*}

\subsection{Diagnostic Elements for MD}\label{subsec:appendix-filter-kb}
In this subsection, we analyze the performance across diagnostic elements for \tasktwoabbrev{}, which we introduced in Section~4.2 of the source publication.
We use BM25 and the best model family for \tasktwoabbrev{}, namely mE5 and \mbox{SoSSE-mE5} trained on 200k samples from SP.
Similar to \taskoneabbrev{}, we observe that identifying implicit and paraphrase mentions is challenging for the base models for \tasktwoabbrev{} (Table~\ref{table:diagnostic-results-ed}).
Surprisingly, the large model variant has a higher recall for implicit types compared to explicit ones.
Future work could evaluate how the number of parameters influences performance with respect to different mention types and subtypes.
In contrast to the results from \taskoneabbrev{}, single-item sentences have significantly higher recall than those with multiple items (Table~\ref{table:multi-item-results-ed}).
The models used have no notion of partially matching a text.
As such, information may be lost when encoding multi-item sentences that describe different survey items.
Future work could examine partial sentence matching methods for this task.

\begin{table*}[h!]
        \centering
        \footnotesize
        \caption{Recall@10 scores for the two most-common type and subtype categories for \tasktwoabbrev{}. BM25 shows poor performance on implicit types, and is generally worse than mE5. The larger mE5 model is able to achieve a higher score for implicit types than for explicit ones. The gab between quotation and paraphrase subtypes decreases as the model size increases}
        \label{table:diagnostic-results-ed}
        \begin{tabular}{l|r|r|r|r}
            \toprule
            \textbf{Model} & \textbf{Explicit} & \textbf{Implicit} & \textbf{Quotation} & \textbf{Paraphrase} \\
            \midrule
            \textbf{BM25} & 68.4 & 38.6 & 67.1 & 64.8 \\
            \midrule
            \textbf{mE5}$_{\text{base}}$ & 79.3 & 64.9 & 89.4 & 74.8 \\
            \textbf{mE5}$_{\text{large}}$ & 81.2 & 82.7 & 89.8 & 80.4 \\
            \botrule
        \end{tabular}
\end{table*}

\begin{table}[h!]
    \centering
    \footnotesize
    \caption{Recall@10 scores for single- vs multi-item sentences for \tasktwoabbrev{}. The number of sentences of each category in the test set is provided in the third column. Single-item sentences have a significantly higher recall score, especially for \mbox{SoSSE-mE5$_{\text{base}}$}}
    \label{table:multi-item-results-ed}
    \begin{tabularx}{\columnwidth}{X|l|c|r}
        \toprule
        \multicolumn{1}{c|}{\textbf{Model}} & \multicolumn{1}{c|}{\textbf{Items}} & \multicolumn{1}{c|}{\textbf{Count}} & \multicolumn{1}{c}{\textbf{$R$}} \\
        \midrule
        \multirow{2}{1em}{\textbf{BM25}} & Single & 223 & 66.5 \\
        & Multi & 121 & 58.3 \\
        \midrule
        \multirow{2}{1em}{\textbf{\mbox{SoSSE-mE5$_{\text{base}}$}}} & Single & 223 & 87.8 \\
         & Multi & 121 & 69.6 \\
        \botrule
    \end{tabularx}
\end{table}

\subsection{Document-Level Aggregation}\label{subsec:appendix-document-level}
\begin{figure}[!ht]
    \centering
    \includegraphics[scale=0.4]{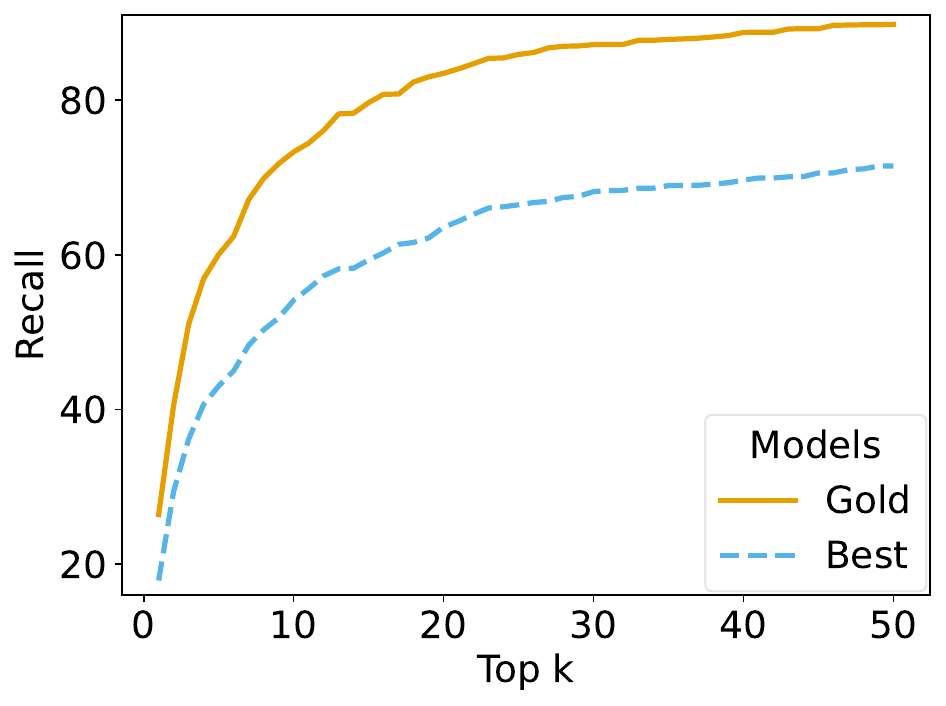}
    \caption{Recall scores for the top \textit{k} predictions, when aggregating the predictions on the document-level. \textit{Gold} assumes an oracle \taskoneabbrev{} model, while \textit{Best} uses XLM-R$_{\text{large}}$. Performance plateaus for both models at $k=30$. Using the best model for \taskoneabbrev{} achieves a recall of over 50\% at $k=10$}
    \label{figure:document-level-results}
\end{figure}

In the downstream application, a user may be provided with highlighted sentences that are linked to survey items.
In this scenario, it is reasonable to assume that the unique survey items linked to highlighted sentences in a document should, at best, cover all referenced survey items.
To measure how many survey items are correctly predicted per-document, we aggregate the predictions for all sentences in a document.
Recall is then computed for different cutoff values, which range between 1 and 50.
We compare the oracle model which uses the gold labels from \taskoneabbrev{} with the best model combination.
For both the gold and best model, we observe that performances plateaus around 30 retrieved items (Figure~\ref{figure:document-level-results}).
In practice, a platform could provide users with up to 10 predicted survey items, which can account for more than 50\% of the relevant survey items in a document.

\section{Resources}\label{subsec:appendix-resources}
We list the resources (i.e., models and software) used in this publications in Tables~\ref{table:appendix-resource-models} and~\ref{table:appendix-resource-software}.

\begin{table*}[h!]
    \centering
    \footnotesize
    \caption{A list of models used in this work}
    \label{table:appendix-resource-models}
    \begin{tabularx}{\textwidth}{l|l|X|l}
        \toprule
        \multicolumn{1}{c|}{\textbf{Model}} & \multicolumn{1}{c|}{\textbf{Params}} & \multicolumn{1}{c|}{\textbf{URL}} & \multicolumn{1}{c}{\textbf{License}} \\
        \midrule
        \textbf{BERT$_{\text{base}}$} & 110M & \href{https://huggingface.co/google-bert/bert-base-uncased}{https://huggingface.co/google-bert/bert-base-uncased} & Apache 2.0 \\
        \textbf{BERT$_{\text{large}}$} & 336M & \href{https://huggingface.co/google-bert/bert-large-uncased}{https://huggingface.co/google-bert/bert-large-uncased}  & Apache 2.0 \\
        \textbf{mBERT$_{\text{base}}$} & 168M & \href{https://huggingface.co/google-bert/bert-base-multilingual-uncased}{https://huggingface.co/google-bert/bert-base-multilingual-uncased}  & Apache 2.0\\
        \textbf{RoBERTa$_{\text{base}}$} & 125M & \href{https://huggingface.co/FacebookAI/roberta-base}{https://huggingface.co/FacebookAI/roberta-base}  & MIT \\
        \textbf{RoBERTa$_{\text{large}}$} & 355M & \href{https://huggingface.co/FacebookAI/roberta-large}{https://huggingface.co/FacebookAI/roberta-large}  & MIT \\
        \textbf{DeBERTa$_{\text{base}}$} & 183M & \href{https://huggingface.co/microsoft/deberta-v3-base}{https://huggingface.co/microsoft/deberta-v3-base}  & MIT \\
        \textbf{DeBERTa$_{\text{large}}$} & 434M & \href{https://huggingface.co/microsoft/deberta-v3-large}{https://huggingface.co/microsoft/deberta-v3-large}  & MIT \\
        \textbf{mDeBERTa$_{\text{base}}$} & 278M & \href{https://huggingface.co/microsoft/mdeberta-v3-base}{https://huggingface.co/microsoft/mdeberta-v3-base}  & MIT \\
        \textbf{SciBERT} & 110M & \href{https://huggingface.co/allenai/scibert\_scivocab\_uncased}{https://huggingface.co/allenai/scibert\_scivocab\_uncased}  & Apache 2.0\\
        \textbf{SPECTER} & 110M & \href{https://huggingface.co/allenai/specter}{https://huggingface.co/allenai/specter}  & Apache 2.0\\
        \textbf{SSciBERT} & 108M & \href{https://huggingface.co/KM4STfulltext/SSCI-BERT-e2}{https://huggingface.co/KM4STfulltext/SSCI-BERT-e2}  & Apache 2.0\\
        \textbf{XLM-R$_{\text{base}}$} & 279M & \href{https://huggingface.co/FacebookAI/xlm-roberta-base}{https://huggingface.co/FacebookAI/xlm-roberta-base}  & MIT\\
        \textbf{XLM-R$_{\text{large}}$} & 561M & \href{https://huggingface.co/FacebookAI/xlm-roberta-large}{https://huggingface.co/FacebookAI/xlm-roberta-large}  & MIT\\
        \textbf{MiniLM} & 118M & \href{https://huggingface.co/sentence-transformers/paraphrase-multilingual-MiniLM-L12-v2}{https://huggingface.co/sentence-transformers/paraphrase-multilingual-MiniLM-L12-v2}  & Apache 2.0 \\
        \textbf{Cross} & 278M & \href{https://huggingface.co/T-Systems-onsite/cross-en-de-roberta-sentence-transformer}{https://huggingface.co/T-Systems-onsite/cross-en-de-roberta-sentence-transformer}  & MIT\\
        \textbf{mE5$_{\text{small}}$} & 118M & \href{https://huggingface.co/intfloat/multilingual-e5-small}{https://huggingface.co/intfloat/multilingual-e5-small}  & MIT\\
        \textbf{mE5$_{\text{base}}$} & 278M & \href{https://huggingface.co/intfloat/multilingual-e5-base}{https://huggingface.co/intfloat/multilingual-e5-base}  & MIT\\
        \textbf{mE5$_{\text{large}}$} & 560M & \href{https://huggingface.co/intfloat/multilingual-e5-large}{https://huggingface.co/intfloat/multilingual-e5-large}  & MIT \\
        \textbf{Vicuna-7B} & 7B & \href{https://huggingface.co/lmsys/vicuna-7b-v1.5}{https://huggingface.co/lmsys/vicuna-7b-v1.5}  & Llama 2 \\
        \textbf{Mistral-7B} & 7B & \href{https://huggingface.co/mistralai/Mistral-7B-Instruct-v0.2}{https://huggingface.co/mistralai/Mistral-7B-Instruct-v0.2}  & Apache 2.0\\
        \textbf{Mixtral-8x7B} & 8x7B & \href{https://huggingface.co/mistralai/Mixtral-8x7B-Instruct-v0.1}{https://huggingface.co/mistralai/Mixtral-8x7B-Instruct-v0.1}  & Apache 2.0 \\
        \botrule
    \end{tabularx}
\end{table*}

\begin{table*}[h!]
    \centering
    \footnotesize
    \caption{A list of software used in this work}
    \label{table:appendix-resource-software}
    \begin{tabularx}{\textwidth}{l|X|l}
        \toprule
        \multicolumn{1}{c|}{\textbf{Software}} & \multicolumn{1}{c|}{\textbf{URL}} & \multicolumn{1}{c}{\textbf{License}} \\
        \midrule
        deep-significance~\citep{ulmer2022deep} & \href{https://github.com/Kaleidophon/deep-significance}{https://github.com/Kaleidophon/deep-significance} & GNU GPL v3.0 \\
        Faiss~\citep{douze2024faiss} & \href{https://github.com/facebookresearch/faiss}{https://github.com/facebookresearch/faiss} & MIT \\
        Grobid~\citep{GROBID} & \href{https://github.com/kermitt2/grobid}{https://github.com/kermitt2/grobid} & Apache 2.0 \\
        Haystack~\citep{Pietsch-Haystack-the-end-to-end-2019} & \href{https://github.com/deepset-ai/haystack}{https://github.com/deepset-ai/haystack} & Apache 2.0 \\
        INCEpTION~\citep{klie-etal-2018-inception} & \href{https://github.com/inception-project/inception}{https://github.com/inception-project/inception}  & Apache 2.0 \\
        nlpaug & \href{https://github.com/makcedward/nlpaug}{https://github.com/makcedward/nlpaug} & MIT \\
        Optuna~\citep{optuna_2019} & \href{https://github.com/optuna/optuna}{https://github.com/optuna/optuna}  & MIT \\
        PyTorch~\citep{10.5555/3454287.3455008} & \href{https://github.com/pytorch/pytorch}{https://github.com/pytorch/pytorch}  & BSD-style \\
        Qdrant & \href{https://github.com/qdrant/qdrant}{https://github.com/qdrant/qdrant}  & Apache 2.0 \\
        Quaterion & \href{https://github.com/qdrant/quaterion}{https://github.com/qdrant/quaterion}  & Apache 2.0 \\
        ranx~\citep{ranx} & \href{https://github.com/amenra/ranx}{https://github.com/amenra/ranx}  & MIT \\
        Scikit-Learn~\citep{JMLR:v12:pedregosa11a} & \href{https://github.com/scikit-learn/scikit-learn}{https://github.com/scikit-learn/scikit-learn}  & BSD 3-Clause \\
        Sentence-Transformers~\citep{reimers-gurevych-2019-sentence} & \href{https://github.com/UKPLab/sentence-transformers}{https://github.com/UKPLab/sentence-transformers}  & Apache 2.0 \\
        Transformers~\citep{wolf-etal-2020-transformers} & \href{https://github.com/huggingface/transformers}{https://github.com/huggingface/transformers}  & Apache 2.0 \\
        \botrule
    \end{tabularx}
\end{table*}

\clearpage
\section{Annotation Guideline}
This section of contains the annotation guideline.

\begin{figure*}[h!]
    \centering
    \includegraphics[width=0.95\textwidth]{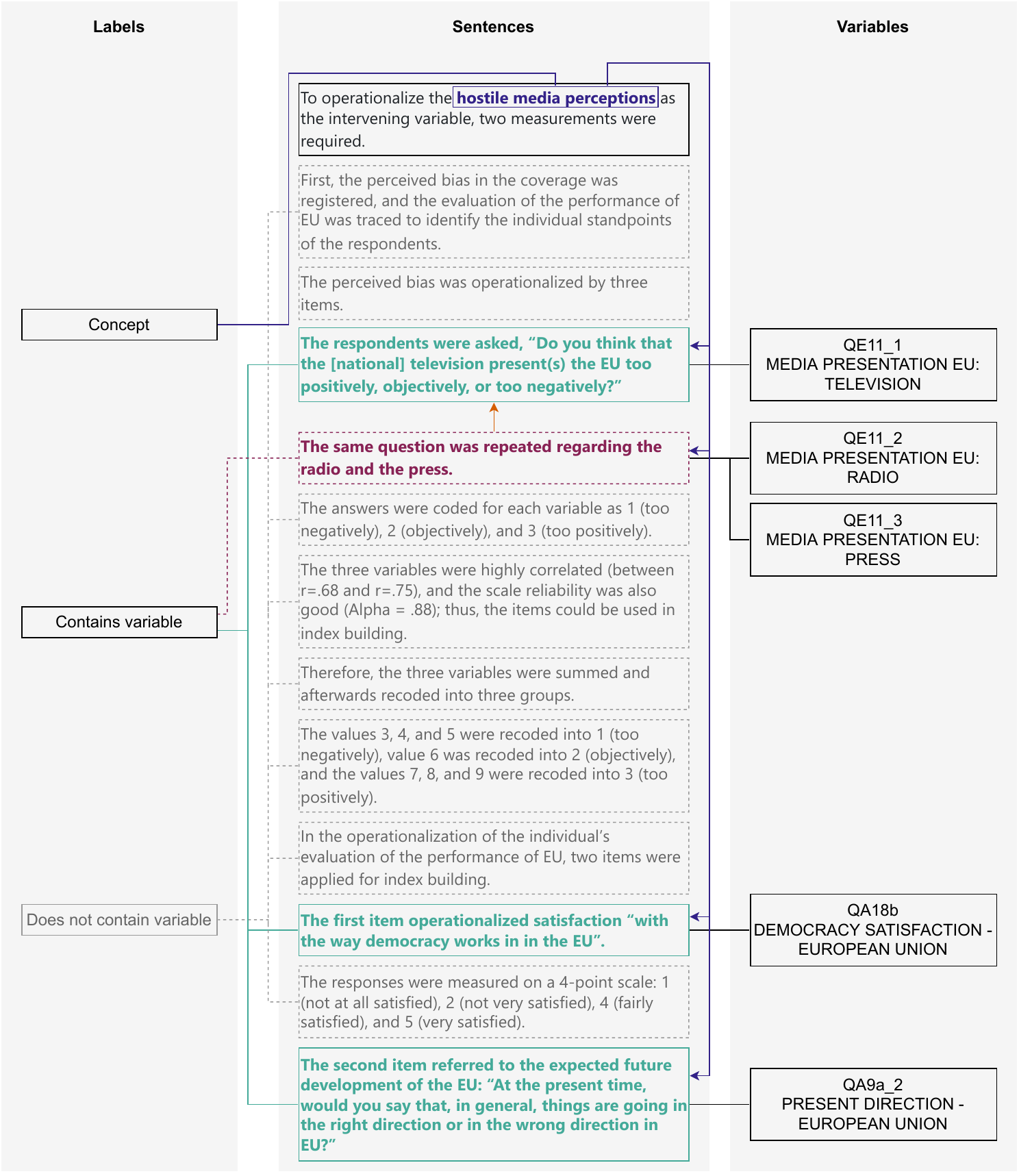}
    \caption{Example of {\color[HTML]{44AA99}explicit} ({\color[HTML]{44AA99}teal}) and {\color[HTML]{882255}implicit} ({\color[HTML]{882255}red}) survey item mentions and an operationalized {\color[HTML]{332288}concept} ({\color[HTML]{332288}purple}). Linked survey items: \href{https://search.gesis.org/variables/exploredata-ZA5876_Varqa9a_2}{QA9a\_2}, \href{https://search.gesis.org/variables/exploredata-ZA5876_Varqa18b}{QA18b}, \href{https://search.gesis.org/variables/exploredata-ZA5876_Varqe11_1}{QE11\_1}, \href{https://search.gesis.org/variables/exploredata-ZA5876_Varqe11_2}{QE11\_2}, and \href{https://search.gesis.org/variables/exploredata-ZA5876_Varqe11_3}{QE11\_3}. Text source: Ejaz et al., 2017.}
    \label{fig:example}
\end{figure*}

\subsection{Background}
In social science literature, survey question and answer pairs (often called \textit{survey items}\footnote{In this document, we will use the terms \textit{survey item}, \textit{survey variable}, and \textit{variable} interchangeably.}) play an important role in the discussion and analysis of observed social phenomenon.
As such, identifying which survey items are used in a research study is vital.
Commonly, used survey items are directly mentioned in the text of a research publication (at times in specific sections, e.g., the data or methods section).
To help improve evaluation of tools for automatically identifying and linking survey items to publications, such documents need to be \textit{annotated}\footnote{Annotation refers to the process of labeling data. This is commonly referred to as \textit{coding} in social science literature.} manually by humans.
This annotation guideline has been written in order to assist human annotators in the given task and provide users of the data, and derived models thereof, an understanding of how the data was collected.

\subsection{Terminology}
In order to promote a shared vocabulary, we use the following terminology for the remainder of this document:
\begin{itemize}
    \item \textit{Research data set} - a questionnaire consisting of multiple questions, called \textit{survey items}.
    \item \textit{Survey item} - an item from a \textit{research data set}.
    \item \textit{Survey item content} - the actual pieces of content of a \textit{survey item} from a \textit{research data set} (e.g., question, sub-question, or answers).
    \textit{Survey item} are also marked with IDs (and on the GESIS platform, with labels, as can be seen in the boxes on the right of Figure~\ref{fig:example}, where \textmd{QA9a\_2}, \textmd{QA18b}, \textmd{QE11\_1}, \textmd{QE11\_2}, and \textmd{QE11\_3} are the IDs and the remaining text the labels).
    \item \textit{Survey item mention/reference} - an in-text mention of an item from a \textit{research data set} (e.g., the {\color[HTML]{44AA99}teal} and {\color[HTML]{882255}red} colored text in the sentences in Figure~\ref{fig:example}).
    A mention can be in the form of a (partial) quotation, a paraphrase, a negative polarity item, lexical inference, or different fields (examples can be found in~\S\ref{subsec:appendix-examples}).
    Survey item mentions can further be divided into {\color[HTML]{44AA99}explicit} and {\color[HTML]{882255}implicit} mentions:
    \begin{itemize}
        \item {\color[HTML]{44AA99}Explicit} mention - a \textit{survey item mention} that can be linked to all relevant \textit{survey items} using only the sentence in which it occurs (e.g., the sentence with the {\color[HTML]{44AA99}teal} text in Figure~\ref{fig:example}).
        \item {\color[HTML]{882255}Implicit} mention - a \textit{survey item mention} that requires additional context in order to link it to all relevant \textit{survey items} (e.g., the sentence with the {\color[HTML]{882255}red} text in Figure~\ref{fig:example}).
    \end{itemize}
    \item \textit{Survey item mention definition} - a \textit{survey item mention} that is informative enough to accurately identify a mentioned survey item. While \textit{survey items} may be references ambiguously in different parts of a publication, a definition sentence makes an affirmative statement that a specific \textit{survey item} or a set of \textit{survey items} was used for a specific purpose (e.g., in the analysis of the study or studying a specific phenomenon). All colored sentences in the example in Figure~\ref{fig:example} are such definition sentences and some of the sentences in gray mention survey items, but do act as the defining sentence.
    \item \textit{Contextual dependence} - a relation between two sentences containing \textit{survey item mentions}.
    The relation is an unidirectional dependence (marked as a line with an arrow) from the source sentence (i.e., the {\color[HTML]{882255}implicit} mention) to the target sentence (i.e., a sentence with an {\color[HTML]{44AA99}explicit} mention).
    \item \textit{Concept} - an abstract term or phrase that is turned into a measurable observation (a.k.a. \textit{operationalization}) through the use of \textit{survey items} (e.g., the {\color[HTML]{332288}purple} phrase in the sentence in Figure~\ref{fig:example}).
    \item \textit{Annotated instance} (or just \textit{instance}) - a piece of text containing one or more \textit{survey item mentions} and the linked \textit{survey items} (e.g., each sentence with the associated connecting arrows and labels in Figure~\ref{fig:example} can be seen as such an instance).
\end{itemize}

\subsection{Task Description}
The goal of the annotation task is to create annotations in social science publications, which provide direct links to the corresponding survey items that the authors of the publications are referring to.
Given a document and a list of survey items, an annotator is asked to read the full document and highlight complete sentences that define a survey item.
While there may be many survey item mentions, only few will define them in enough detail to identify them accurately.
Figure~\ref{fig:example} shows an example with three sentences from a text, two of which contain and are linked to the respective survey items.

The annotation can thus be divided into five main tasks:
\begin{itemize}
    \item \textbf{Task 1} - Survey Item Detection: 
    \begin{itemize}
        \item Type: Sentence-level
        \item Description: For each document, identify and mark all full sentences that contain at least one survey item mention definition ({\color[HTML]{44AA99}explicit} and {\color[HTML]{882255}implicit}). In most cases, each survey item will only have a single defining sentence in the document.
    \end{itemize}
    \item \textbf{Task 2} - Survey Item Linking:
    \begin{itemize}
        \item Type: Sentence-to-survey-item-ID
        \item Description: For each sentence containing a survey item mention, select all relevant survey items from the provided research data sets.
    \end{itemize}
    \item \textbf{Task 3} - Survey Item Type Classification:
    \begin{itemize}
        \item Type: Sentence-to-sentence
        \item Description: Mark each survey item as either {\color[HTML]{44AA99}explicit} or {\color[HTML]{882255}implicit}.
        {\color[HTML]{44AA99}Explicit} mentions are self-contained (within a single sentence), while {\color[HTML]{882255}implicit} mentions require additional context, which may or may not be available.
        In addition, for each {\color[HTML]{882255}implicit} mention, link the relevant context(s) (i.e., {\color[HTML]{44AA99}explicit} mention(s)).
        In addition, label each mention into one or more linguistic subtype categories.
    \end{itemize}
    \item \textbf{Task 4} - Concept Detection:
    \begin{itemize}
        \item Type: Word-level
        \item Description: For each document, identify and highlight any abstract concepts that are operationalized using survey items to measure a phenomenon.
    \end{itemize}
    \item \textbf{Task 5} - Concept Linking:
    \begin{itemize}
        \item Type: Concept-to-sentence
        \item Description: Link each concept with its relevant survey item mentions ({\color[HTML]{44AA99}explicit} and {\color[HTML]{882255}implicit}).
    \end{itemize}
\end{itemize}

\noindent The annotation has three phases. In \textit{phase 1}, annotators get familiar with the annotation format and tools by labeling practice documents.
Annotators label real documents during \textit{phase 2} and review the annotations of all annotators and make necessary corrections in \textit{phase 3}.

\subsection{Data}
The dataset for annotation consists of a diverse set of social science publications in both English and German.
The publications range in size and topic.
Certain publications may contain many survey item mentions, while others contain very few or none at all.

\subsection{Annotation Procedure}
The annotation procedure will provide detailed explanations on how to annotate documents.
Use the example in Figure~\ref{fig:example} to guide you in making decisions about the annotations.
If uncertain, review the \textit{corner-cases} in the last section.

\paragraph{Step 0: Prepare environment}
\indent Log into the INCEpTION environment.
Open the ``annotation\_guideline.pdf``, which can be found by clicking the ``Guidelines`` button (book icon).
You are providing with relevant research data sets, which can be found under ``Guidelines``.

\paragraph{Step 1: Open document}
\indent View the current annotation project and open the next document in the list, which has not already been fully annotated. During \textit{phases 1} and \textit{2}, documents will be annotated in PDF format.
During \textit{phase 3}, sentences from the documents will be listed (one sentence per line), however, no tables, figures or additional content (e.g., appendices) will be visible.
The original PDFs will be available under ``Guidelines``.

\paragraph{Step 2: Get an overview}
\indent Read the title, abstract, and keywords (if available) to determine the topic of the text.
The keywords may mention concepts that are defined in the document.
Furthermore, some documents may list used survey items explicitly in tables or in the appendix.
If applicable, use these to help you with annotations.
During \textit{phase 3}, section headers are removed.

\begin{figure*}[h!]
    \centering
    \includegraphics[width=0.99\textwidth]{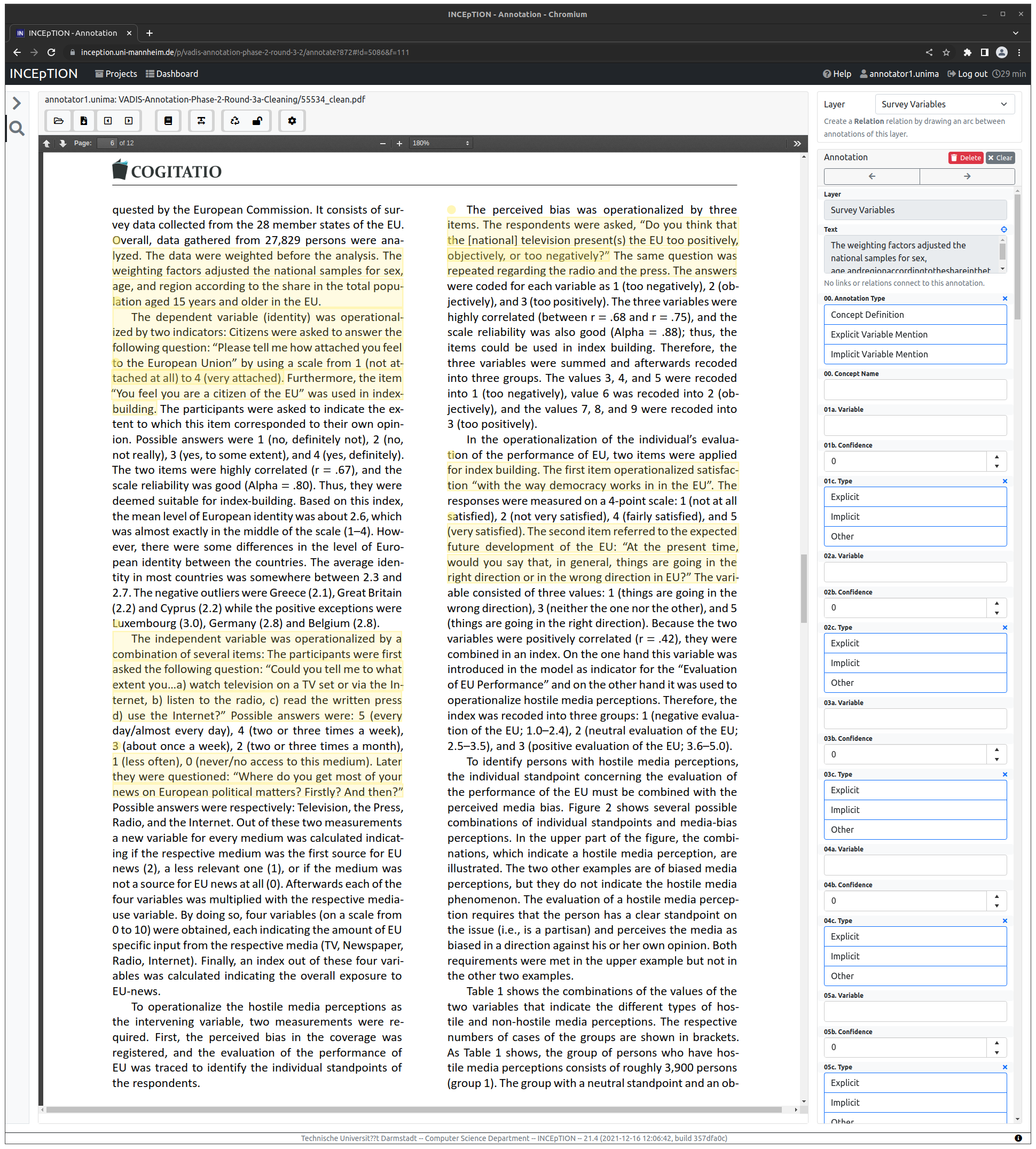}
    \caption{Example of Step 3: Identify survey item mentions. Here, the yellow sentences contain survey items.}
    \label{fig:step-3}
\end{figure*}

\paragraph{Step 3: Identify survey item mentions (Task 1)}
\indent Read the document sentence-by-sentence.
Identify and \textbf{highlight all sentences that define a survey item} using the ``Survey Variables`` layer (selected from the drop-down on the top of the right-hand sidebar).
Sentences may contain (part of) the survey item text literally, express the semantic content of the survey item in other words, or be narrower/broader.
To mark a sentence, highlight all words in the sentence.
If successful, this will create a box (henceforth, \textsc{hbox}) around the highlighted parts.
In case the \textsc{hbox} includes words from different sentences, try to delete and re-highlight the text to remain within the boundary of the sentence (it is better to highlight too few words than too many).
In case you want to delete an annotation, the red \textit{delete}-button on the right-hand sidebar can be used for each \textsc{hbox}.
See Figure~\ref{fig:step-3} for an example.

\paragraph{Step 4: Link survey items (Task 2)}
On the right-hand sidebar, you will see a number of repeating fields.
These fields come in pairs (e.g., ``01a. Variable``, ``01b. Confidence``, ``01c. Type``, and ``01d. Sub-Type`` go together).
For each survey item that is mentioned in a sentence (there may be more than one survey item), fill out all related fields.
During \textit{phase 3}, some fields and survey items will already be pre-annotated.
There may still be missing or incorrect values, which you should fill out or correct.
The pre-annotated survey item IDs can be combined (by appending the ID to the URL) with the URL of the survey item on the GESIS platform. 
Use the provided research data mapping (found under ``Guidelines`` in the ``Research Data Mapping`` file) to identify the research data IDs relevant for each document (e.g., document ``55534`` uses the research data with ID ``ZA5876``).
Using the research data IDs, search through the questionnaires provided under ``Guidelines`` (e.g., ``ZA5876\_q.pdf``) to find each survey item.
Once you find a survey item in the questionnaire, find the respective URL on the GESIS platform.
On GESIS, at the bottom of each research data page, the survey items are individually listed under the red ``Variables`` drop-down bar.
Click on one of the survey items to open it and expand its metadata.
Note that the question ID from the questionnaire is often (but not always) the ID listed in the metadata next to ``Item``.
Often, each question and sub-question pair is a single entry on the GESIS platform (e.g., ``ZA5876\_Varqa9a\_1`` and ``ZA5876\_Varqa9a\_2``).
Once you find the survey item that matches the one from the questionnaire, copy the URL of the survey item and paste it into the ``a. Variable`` field.
You may either copy the entire URL (e.g., \href{}{\url{https://search.gesis.org/variables/exploredata-ZA5876_Varqa9a_2}}) or just the part following ``exploredata-`` (e.g., ``ZA5876\_Varqa9a\_2``).
In case a sentence in the publication references a survey item not listed in the catalog, simply input an ``UNK`` variable tag (which stands for ``unknown``) in the variable text box.

\begin{figure*}[h!]
    \centering
    \includegraphics[width=0.99\textwidth]{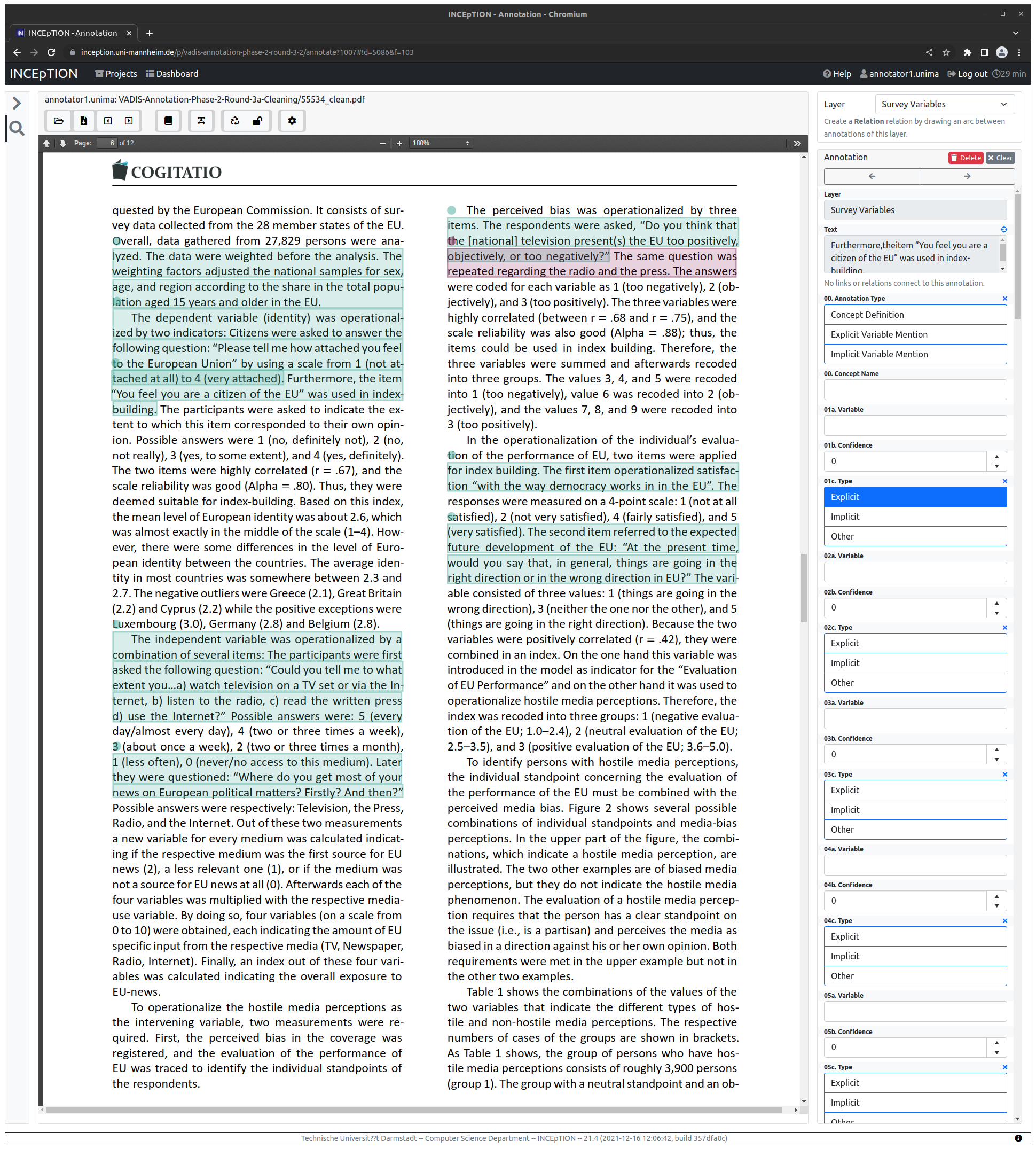}
    \caption{Example of Step 5: Label the survey item mention type. Here, the {\color[HTML]{44AA99}explicit} mentions are colored {\color[HTML]{44AA99}teal} and the {\color[HTML]{882255}implicit} mentions {\color[HTML]{882255}red}.}
    \label{fig:step-5}
\end{figure*}

\paragraph{Step 5: Label the survey item mention type (Task 3)}
For each survey item, select the type of mention from \textit{explicit}, \textit{implicit} or \textit{other}.
In addition, label the ``00a. Annotation Type`` for each sentence.
If a sentence contains multiple survey item mention types (i.e., both \textit{explicit} and \textit{implicit}), choose the ``Mixed Variable Mention`` type.
Explicit mentions are self-contained (within the sentence) and require no additional context (e.g., the {\color[HTML]{44AA99}teal} sentences in Figure~\ref{fig:example}).
Implicit mentions require additional context, world or background knowledge, which may not always be available in the document (e.g., the {\color[HTML]{882255}red} sentence in Figure~\ref{fig:example}).
If neither of these categories fit, use \textit{other}.
Lastly, label each mention into one or multiple subtype categories, such as ``Citation``, ``Lexical Inference``, ``Negation``, ``Paraphrase``, ``Quotation``, ``Other``, or ``Unspecified`` (refer to Table~\ref{table:annotation-examples} for examples for each of the categories).
See Figure~\ref{fig:step-5} for an example.

\begin{figure*}[h!]
    \centering
    \includegraphics[width=0.99\textwidth]{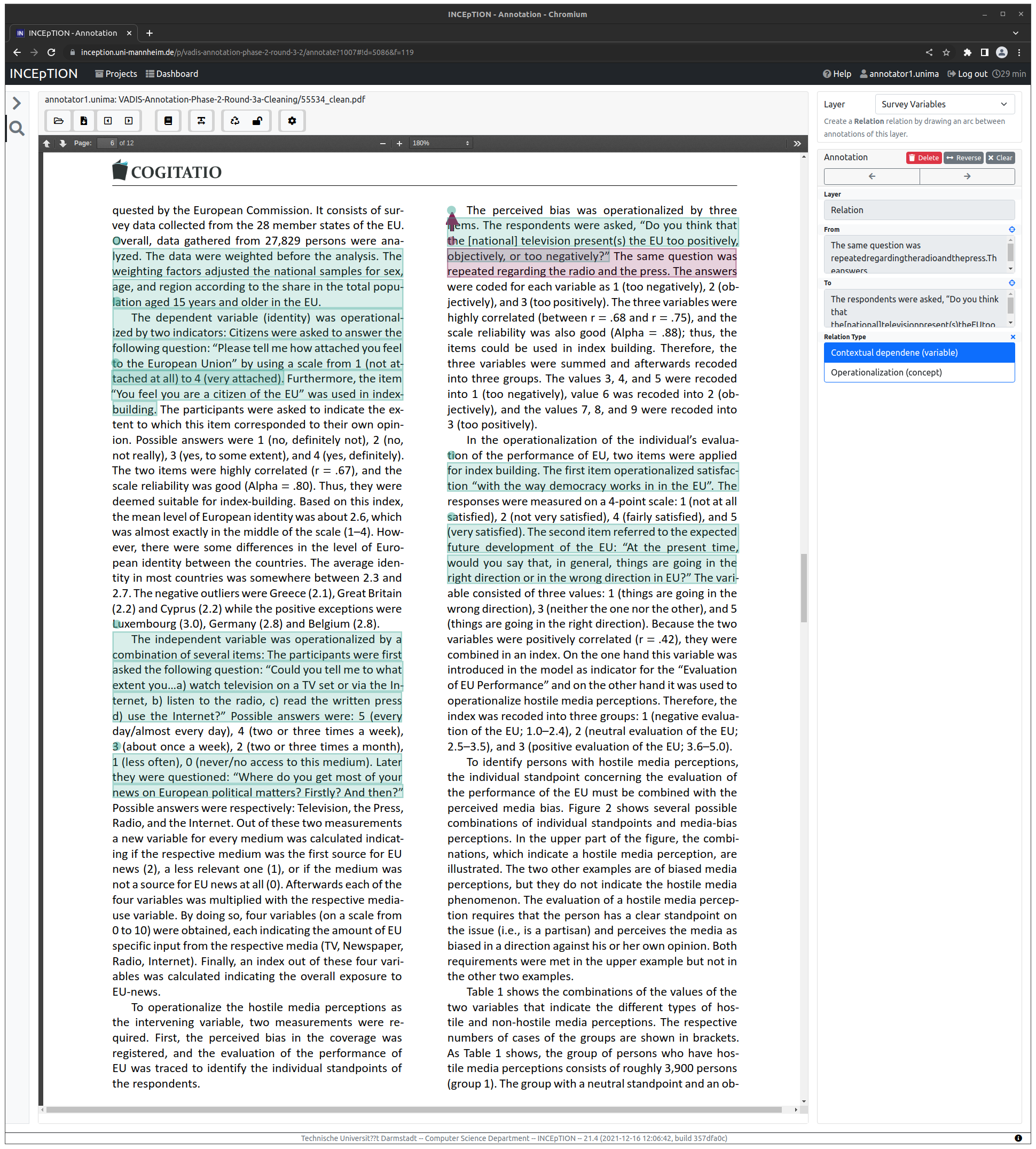}
    \caption{Example of Step 6: Link mentions to contexts. Here, the {\color[HTML]{882255}red} arrow is the relation from the {\color[HTML]{882255}implicit} to the {\color[HTML]{44AA99}explicit} mention.}
    \label{fig:step-6}
\end{figure*}

\paragraph{Step 6: Link mentions to contexts (Task 3)}
For each sentence that contains an {\color[HTML]{882255}implicit} or \textit{other} (if applicable) survey item mention type, select the context sentence(s) that are required to understand the mentioned survey item(s).
To do this, click and hold the \textsc{hbox} of the {\color[HTML]{882255}implicit} or \textit{other} mention and drag and release the arrow on top of the \textsc{hbox} of the sentence(s) with a relevant {\color[HTML]{44AA99}explicit} mention.
If successful, a dotted line should be drawn between the two \textsc{hbox}es.
Lastly, click the created line/arrow and mark the relation type from the right-hand sidebar as ``contextual dependence``.
See Figure~\ref{fig:step-6} for an example.

\begin{figure*}[h!]
    \centering
    \includegraphics[width=0.99\textwidth]{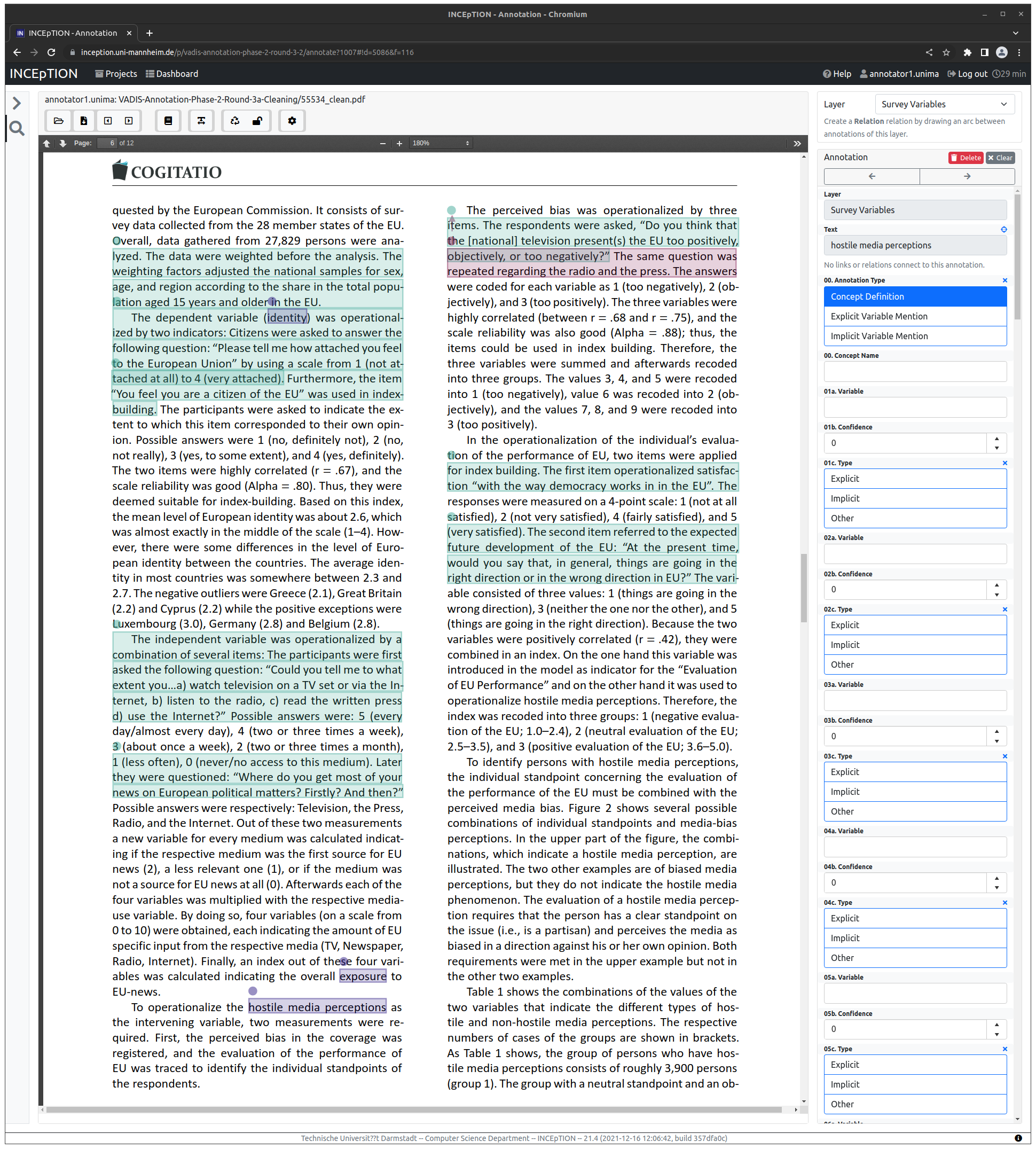}
    \caption{Example of Step 7: Identify concepts. Here, the {\color[HTML]{332288}purple} phrase is the concept.}
    \label{fig:step-7}
\end{figure*}

\paragraph{Step 7: Identify concepts (Task 4)}
Identify and highlight the word(s) of a concept in the sentence that defines how it is operationalized using survey items defined in the document.
Because a concept may occur frequently in a document, try to pick a sentence that most clearly states how it is operationalized.
From the drop-down menu under ``00. Annotation Type`` select the ``concept definition`` label.
Use the text box ``00. Concept Name`` to write out the full concept name.
See Figure~\ref{fig:step-7} for an example.

\begin{figure*}[h!]
    \centering
    \includegraphics[width=0.99\textwidth]{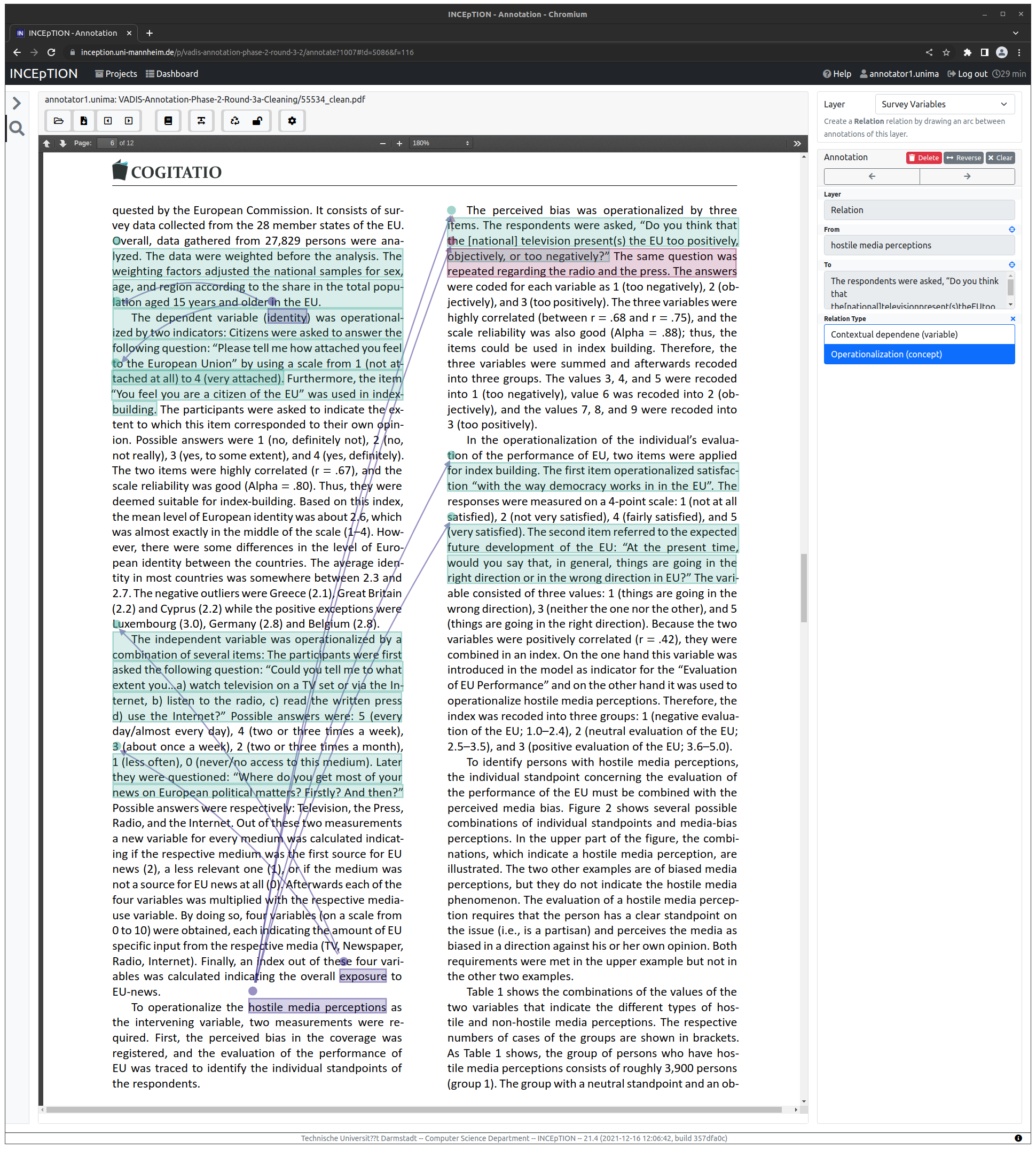}
    \caption{Example of Step 8: Link concepts. Here, the {\color[HTML]{332288}purple} arrows from the concepts to the survey item mentions imply the operationalization of the concepts through the linked survey items.}
    \label{fig:step-8}
\end{figure*}

\paragraph{Step 8: Link concepts (Task 5)}
Link each concept to the survey items that are used to operationalize it.
The relation should go from the concept to the survey items.
Similar to Step 4, select a reference type, but this time, select ``operationalization``.
In case the linked sentences contain more survey item mentions than are used to operationalize the concept, list the survey item names in the annotation for the concept (similar to Step~4).
See Figure~\ref{fig:step-8} for an example.

\paragraph{Step 9: Rate confidence}
\indent For each annotation of a survey item mention and concept, provide a rating for the confidence of the annotation on the provided scale (0=not very confident, 1=not confident, 2=neutral, 3=confident, 4=very confident).

\paragraph{Step 10: Label errors}
Some sentences may appear scrambled and contain parsing errors (e.g., missing spacing between words).
Mark such sentences with the error type using the ``Errors`` layer and by selecting the appropriate error type.
If none of the error types match the case, you may write a comment in the input box and hit ``Enter`` to save it.

\paragraph{Step 11: Lock the document}
Once you are done with a document, lock the document by clicking on the ``Finish document`` button (open lock icon).
After this step, no more changes to the document annotation are possible.
Continue annotating the next documents until there are no more documents to annotate.

\subsection{Corner-Cases}
\label{sec:corners}
\begin{itemize}
    \item When linking implicit mentions or concepts, if a survey item is mentioned more than once, create relations only to the first explicit survey item mention.
    
    \item If a survey item is defined in a table, footnote, or the appendix, label such mentions as well.

    \item If it is not clear which concept is operationalized with which survey items, identify and link the concepts, which you think could possibly have been operationalized by survey items.
    Make sure to pick the sentence that best describes how the concept is constructed.

    \item If the concept is not present in the sentence which clearly operationalized it, simply select the entire sentence and include the concept word/phrase in the ``00. Concept Name`` box.

    \item When linking concepts to survey item mentions, if a survey item mentioning sentence contains more variables than the concept uses, only link those sentences that contain all survey items used by the concept.
    For the remaining survey items which are used in the concept, include them in the concept by filling the field ``Variable``.
    You may only ignore the remaining fields for the survey item if you are certain that it is used in the concept.

    \item If you are confident that a survey item is used but not confident that another one is used in a concept, fill the ``Variable`` and ``Confidence`` fields for the survey item you are not certain about.
    This time, the confidence refers to how certain you are that the survey item is part of the concept.

    \item In the case that a sentence should contain both an explicit and an implicit survey item mention, use the "Mixed Variable Mention" Type for the Variable Type annotation. You may then continue to map relations between the implicit, and it's related explicit survey item mentions. If such a mixed sentence should contain more than one implicit survey item mention, please mark the sentence with the "Error" layer and the error type as "Multiple implicit variable mentions".
\end{itemize}

\subsection{Examples}\label{subsec:appendix-examples}
Examples for subtypes are presented in the Table~\ref{table:annotation-examples}.

\begin{table*}[h!]
    \centering
    \footnotesize
    \caption{Explanations for subtypes. Each subtype comes with a description and an example mention. For clarity, a few examples additionally show the content of the survey item(s) they mention.}
    \label{table:annotation-examples}
    \begin{tabularx}{\textwidth}[!ht]{l|X|X|X}
        \toprule
        \textbf{Sub-Type} & \textbf{Description} & \textbf{Survey Item Mention} & \textbf{Survey Item Content} \\
         \midrule

         Citation & A sentence that mentions a survey item from a different study. The exact survey item may not be always be identified. & ``In this regard, research has shown that individuals who perceive their society as rather unequal and ridden by inequality also perceive conflicts as more severe (Hadler 2003).`` & \\
         \midrule

         Lexical Inference & An implied meaning based on contextual clues combined with world/linguistic knowledge. & First, winners are more politically satisfied compared with losers, including those who voted for the Black–Red Grand Coalition. & ``Which party did you vote for with your second vote (`Zweitstimme`)?
         
         - Respondent didn't vote
         
         - The Christian Democratic/Christian Social Union CDU/CSU 
         
         - The Social Democratic Party SPD``\\
         \midrule

         Negation & A negation of the survey item content in quotation / paraphrased / etc. form. & Is there any place in the immediate vicinity in which you fear walking alone at night? & ``Is there any area in the immediate vicinity - I mean within a kilometer or so - where you would prefer not to walk alone at night?``\\
         \midrule

         Paraphrase & A paraphrase of the survey item content. & The second and the third questions come from the ISSP research, where respondents were asked about the influence of religious leaders on people’s votes and the government. & ``How much do you agree or disagree with each of the following: 
         
         - Religious leaders should not try to influence how people vote in elections.
         
         - Religious leaders should not try to influence government decisions.`` \\
         \midrule

         Quotation & A quote of the survey item content. & There is only one item measuring happiness which directly asks the respondents: `‘If you were to consider your life in general these days, how happy or unhappy would you say you are, on the whole...’` & ``If you were to consider your life in general these days, how happy or unhappy would you say you are, on the whole...`` \\
         \midrule

         Unspecified & A sentence, in which it is unclear whether a survey item being mentioned is used in research for a specific purpose or not. Sometimes, such sentences occur before the variable-defining sentence. The exact survey items may not be identifiable. & In addition, the regression of attitudes on demographic survey items such as age, education or income can be used to help clarify the meaning of an attitudinal question. &  \\
         \bottomrule
    \end{tabularx}
\end{table*}

\subsection{Screenshots}
The user interface of the annotation platform and the steps users must follow are shown in the screenshots below.

\end{appendices}

\end{document}